\documentclass[galaxies,review,accept,moreauthors,pdftex,10pt,a4paper]{mdpi}
\usepackage{amssymb}
\setitemize{parsep=6pt,itemsep=0pt,leftmargin=*,labelsep=5.5mm}
\setenumerate{parsep=6pt,itemsep=0pt,leftmargin=*,labelsep=5.5mm}
\setlist[description]{itemsep=0mm}

\firstpage{1}
\pubvolume{6}
\issuenum{1}
\articlenumber{1}
\pubyear{2018}
\copyrightyear{2018}
\history{Received: date; Accepted: date; Published: date}

\Title{A Review of the Theory of Galactic Winds Driven by Stellar Feedback}

\newcommand\pc{\;{\rm pc}}
\newcommand\kms{\; {\rm km}\;{\rm s}^{-1}}
\newcommand\Kel{\;{\rm K}}
\newcommand\Msun{\; M_{\odot}}
\newcommand{\pdif}[2]{\ensuremath{ \frac{\partial #1}{\partial #2}}}
\newcommand{\bnabla}{{\mbox{\boldmath$\nabla$}}}
\newcommand{\bkappa}{{\mbox{\boldmath$\kappa$}}}

\Author{Dong Zhang $^{1,2}$\orcidA{}}

\AuthorNames{Dong Zhang}

\address{$^{1}$ \quad Department of Astronomy, University of Virginia, 530 McCormick Road, Charlottesville, VA 22904, USA; dongzhz@umich.edu\\
$^{2}$ \quad Department of Astronomy, University of Michigan, 1085 S University Ave, Ann Arbor, MI 48109, USA}

\abstract{Galactic winds from star-forming galaxies are crucial to the process of galaxy formation and evolution, regulating star formation, shaping the stellar mass function and the mass-metallicity relation, and enriching the intergalactic medium with metals. Galactic winds associated with stellar feedback may be driven by overlapping supernova explosions, radiation pressure of starlight on dust grains, and cosmic rays. Galactic winds are multiphase, the growing observations of emission and absorption of cold molecular, cool atomic, ionized warm and hot outflowing gas in a large number of galaxies have not been completely understood. In this review article, I summarize the possible mechanisms associated with stars to launch galactic winds, and review the multidimensional hydrodynamic, radiation hydrodynamic and magnetohydrodynamic simulations of winds based on various algorithms. I also briefly discuss the theoretical challenges and possible future research~directions.}

\keyword{starbursts; galactic winds; supernovae; interstellar medium; dust; radiative transfer; radiation hydrodynamics; cosmic rays; magnetohydrodynamics}

\begin{document}


\section{Introduction}\label{section_Introduction}


The basic scenario of galaxy formation and evolution in the cosmological context can be described in the framework of the standard $\Lambda$ cold dark matter ($\Lambda$CDM) cosmology. However, many observed properties of galaxies remain difficult to explain. To obtain a realistic picture of cosmological galaxy formation and evolution which matches observations, astrophysical processes such as gas cooling, accretion, galaxy mergers, and feedback should be considered. Among them feedback is one of the most difficult ingredients to be understood. It is widely believed that feedback provides energy, mass, and metal to the interstellar medium (ISM) and the circumgalactic medium (CGM), and~drives galactic-scale winds.

Feedback from both stars and active galactic nucleus (AGN) activities has been proposed to explain galactic winds. The discovery of galactic winds can be traced back more than a half century ago, when a large-scale outflow in a starburst galaxy M82 was observed~\citep{Lynds63,Burbidge64}. At almost the same time quasars were discovered~\citep{Schmidt63}. Although the energy source of quasars was heavily debated in the beginning, the theory that quasars and other AGNs are powered by gas accretion onto supermassive black holes (SMBHs) was finally accepted by the end of 1960s~\citep{Hoyle63a,Hoyle63b,Salpeter64,Zeldovich64,Lynden69,Lynden71}. Since then, galactic winds have been observed ubiquitously in both the local and high-redshift Universe~\citep{Veilleux05,HT17, Naab17}. The theory for galactic winds developed over years is of outflows powered by the energy and momentum of multiple supernovae (SNe) from massive stars~\citep{DS86,Leitherer92,Heckman93,Tomisaka93,Suchkov94,Leitherer95,Heckman00}. By the end of the last century, AGN feedback and AGN-driven outflows have also gained momentum in the literature~\citep{Magorrian98,Silk98,Haehnelt98,Ferrarese00,Gebhardt00}.

AGN feedback is the astrophysical process to link the momentum and energy released by the central SMBHs to the surrounding medium, and drives outflows (see reviews~\citep{Fabian12,Kormendy13,Heckman14,King15} ). The energy from a growth of SMBH in galaxies is much more powerful than that from the host galaxy, and AGN feedback has a significant impact on the formation of massive galaxies. There are two modes of AGN feedback: the {\it quasar} mode, and the {\it radio} mode. The {\it quasar} (or {\it radiative}) mode is associated with high-luminosity AGNs, which are close to the Eddington limit. This is the case for typical young quasars at high redshift. Most of the energy is released by radiation from accretion disks around SMBHs, and radiation couples to the gas, transfers momentum to it, and drives it from the host galaxies e.g.,~\citep{Silk98,Matteo05,Zubovas13,Bieri17}. The {\it radio} (or {\it kinetic}) mode describes low-luminosity AGNs in which radio jets are the main source, and the energy is in a kinetic form. Typically, this is the case for local massive galaxies. The radio jets and cocoons heat the CGM and the intergalactic medium (IGM) e.g.,~\citep{Croton06,Bower06,McNamara07,Okamoto08,Gaibler11,Fabian12,Guo12,Guo16,Guo18,Duan18}. Radio~plasma lobes and X-ray cavities in gas-rich cool-core clusters show the evidence of AGN feedback~\citep{Salome06,Guo09,McNamara12,Gaspari12,Ciotti17}. Another strong evidence of AGN feedback is the AGN-powered outflows, including ultrafast outflows observed in X-ray and UV bands~\citep{Reeves09,Tombesi11,Tombesi12}, and high-velocity multiphase outflows observed in optical, infrared, and sub-millimeter bands~\citep{Mullaney13,Cicone14,Zakamska14}. A large number of references have explored the properties of AGN-driven outflows (\citep{Morganti17} and references therein), some best-studied objects include outflows from Mrk 231 e.g.,~\citep{Rupke11,Feruglio15,Morganti16}, and IC 5063 e.g.,~\citep{Morganti98,Dasyra16,Oosterloo17}.


On the other hand, stellar-driven galactic winds are more significant in low-mass galaxies, and~are dominated in starbursts without AGN activities such as M82, NGC 253, and Arp 220. Also,~stellar-driven feedback can work with AGN feedback together in some galaxies containing both starburst cores and central AGNs, such as Mrk 231. Stellar-driven feedback was probably also important in the high-redshift Universe after the first-generation stars formed but before the formation of SMBHs. Compared to AGN-driven galactic winds, stellar-driven galactic winds are also important from sub-galactic to cosmological scales. Next in this section I introduce the impact and observations of stellar-driven galactic winds. Hereafter, the terminology “galactic winds” in this review article emphasizes on the winds associated with stellar feedback.


\subsection{Impact of Galactic Winds Driven by Stellar Feedback}

Galactic winds driven by stellar feedback are important on various scales. On cosmological scales, the most significant effect of galactic winds is to shape the cosmic galaxy luminosity function, flattering~its low-mass end slope compared to that of the halo mass function. The observed galaxy luminosity function shows a flat low-mass-end slope and sharp exponential cut-off at the high-mass-end, which are different from the shape of the dark matter halo mass function predicted by the $\Lambda$CDM model~\citep{Benson03,Baldry08,Li09}. The high-mass-end can be explained by the AGN feedback~\citep{Bower06,Somerville08,Vogel14}, while~the flat low-mass-end of the galaxy luminosity is believed to be caused by stellar feedback, which~drives gas out of galaxies, and suppress the formation of low-mass galaxies~\citep{DS86,Benson03,Mo04,Puchwein13,Schaye15,Dave17}.



Also, galactic winds play a crucial role in determining the chemical evolution of galaxies and mass-metallicity relation. Galactic winds from stellar feedback are expected to be metal-rich, and~ massive galaxies with deeper gravitational potential are expected to retain more of the galactic wind ejecta than dwarf galaxies, therefore metal loss from galaxies is strongly anti-correlated with the mass of galaxies~\citep{Tremonti04,Finlator08,Baldry08,Peeples11,Andrews13,Sanders15,Dave17,Sanchez17}. Galactic winds are also suggested to be the primary sources to pollute the circumgalactic medium (CGM) and the intergalactic medium (IGM) with metals~\citep{Aguirre01a,Aguirre01b,Aguirre08,Prochaska17}.

Galactic winds are probably the most extreme manifestation of the feedback between star formation in a galaxy and its ISM~\citep{SH08a,SH08b,Schaye10,Muratov15}. On galactic scales, winds can heat gas in the ISM and prevent gas from cooling and forming stars in both dwarf and massive galaxies~\citep{Heckman00,Bertone07}. The unbound outflowing gas is injected into the CGM and IGM. On the other hand, if the velocity of a wind is below the galactic escape velocity, the wind eventually falls back onto the galaxy to form galactic fountain or later gas accretion inflows, which helps to establish the so-called “gas recycling” process in galaxies~\citep{Chisholm16,Tumlinson17,Voort17,Krumholz17}. Since galactic winds are from the central region of galaxies with low angular momentum, the~recycling of gas which falls back from the galactic inner region to outer region re-distributes angular momentum in the galaxies~\citep{Ubler14,Krumholz17,DeFelippis17}.

\subsection{ Multiphase Structure}

Multiwavelength observations from radio to $\gamma$-ray show that galactic winds in starbursts and star-forming galaxies have multiphase structure, which can be generally divided into five phases, e.g.,~\citep{HT17}: very hot ($\sim$$10^8\,$K), hot ($\sim$$10^6\,$--$10^7\,$K), warm (ionized, $\sim$$10^4\,$K), cool (neutral atomic, $\sim$$10^3\,$K), and cold (molecular and dust $ \lesssim$$100\,$K) phases.

\begin{figure}[H]
\centering
\includegraphics[width=13cm]{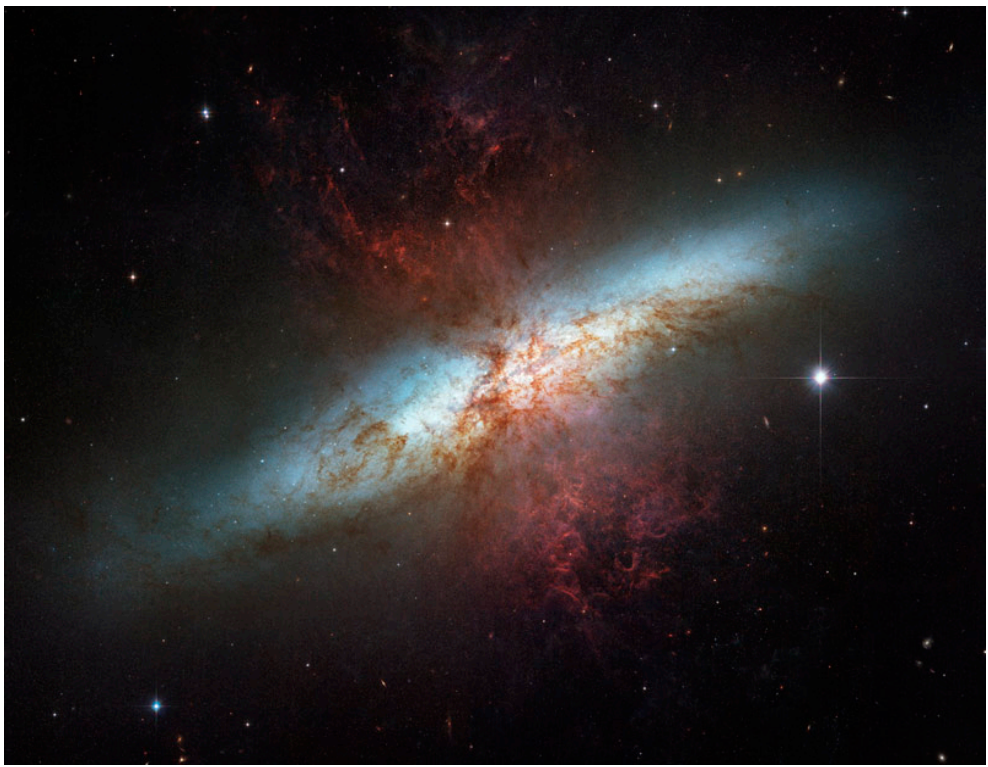}
\caption{The H$\alpha$ image (red component) of galactic wind in M82. Figure from the Hubble telescope galley (\url{http://hubblesite.org/image/1876/gallery}). H$\alpha$ image credit is from FOCAS, Subaru 8.3-m Telescope, NAOJ, and the disk image is from the Hubble telescope.}\label{fig_M82_1}
\end{figure}
\begin{figure}
\centering
\includegraphics[width=13cm]{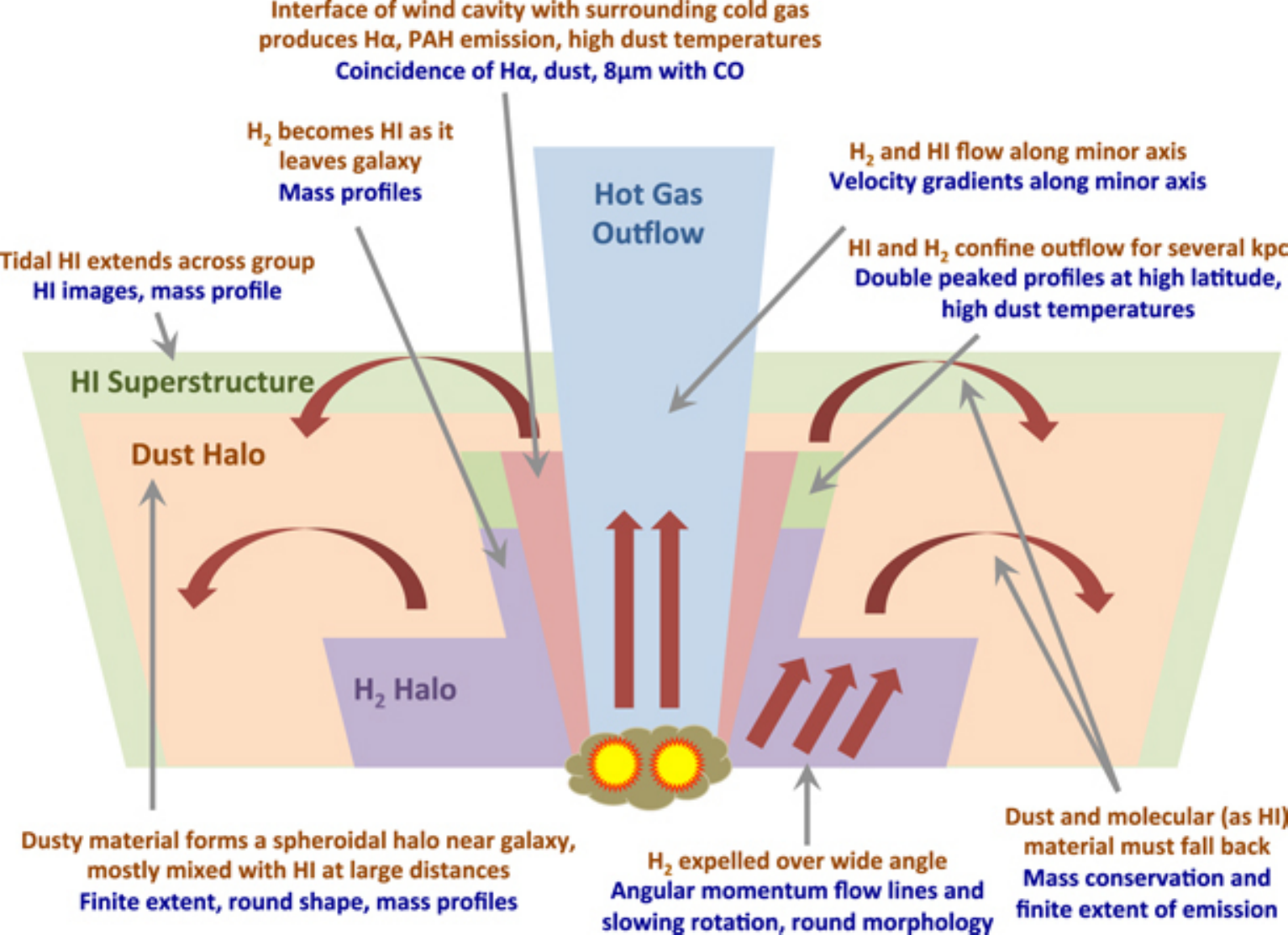}
\caption{A schematic of the multiphase wind (and fountain) in M82. The blue notes in the figure cite observational evidence for the red statements. Similar schematic has been made for NGC 253~\citep{Westmoquette13,Meier15}. Figure reproduced with permission from Leroy~et al. ApJ, 2015~\citep{Leroy15}.}\label{fig_M82_2}
\end{figure}

X-ray satellites including {\it Chandra} and XMM-Newton observatories have been used to observe highly ionized X-ray emitting hot phase of galactic winds, and also to constrain the hard X-ray emission directly from very hot winds driven by supernova (SN) explosions~\citep{SS00,McDowell03,SH07,SH09,Bravo17}. Emission lines such as H$\alpha$, N II, O II, O III, and absorptions such as Na I D, K I, Mg II have been used to trace morphologies and velocities of warm, cool and cold outflows in galactic winds, including warm ionized~\citep{SB98,Cecil02,Weiner09,Martin13,Tang14,Rubin14,Heckman15,Du16,Chisholm17}, neutral~atomic~\citep{Heckman00,Rupke02,Martin05,Rupke05a,Rupke05b,Rupke05c,Chen10,Kornei13,Rupke15}, molecular ~\citep{Sakamoto99,Veilleux09,Westmoquette13,Bolatto13,Salak13,Meier15,Leroy15,Walter17}, and dust~\citep{Roussel10,Hutton14,Melendez15,Hodges16,McCormick18,Jones18} outflows. 
As shown in the references, the multiphase outflowing gas can be accelerated to hundreds of km\,s$^{-1}$, even above 1000\,km\,s$^{-1}$. Taking the starburst M82 as an example: Figure~\ref{fig_M82_1} shows the H$\alpha$ traced warm phase of galactic wind in M82,  and Figure~\ref{fig_M82_2} demonstrates the schematic of the multiphase wind around M82. The H$\alpha$ emitting clouds reach $\sim$$600\,$km\,s$^{-1}$~\citep{McKeith95,SB98}, and the cool and cold outflows have velocities $\sim$$100$--$300\,$km\,s$^{-1}$ e.g.,~\citep{Walter02,Leroy15}. Also, the galactic wind in another starburst NGC 253 show similar properties~\citep{Bolatto13,Walter17}. A review of the recent observations of  multiphase stellar-driven galactic winds by David Rupke can be found in this special Issue. Previous research suggested that the warm phase of the wind is accelerated by the ram pressure of hot phase driven by SN explosions e.g.,~\citep{Heckman00,SH09}, but~recent simulations showed that such ram-pressure-acceleration scenario may be problematic~\citep{SB15, BS16, SR17, Zhang17, McCourt18}. From~a theoretical point of view, the driving mechanisms of multiphase winds driven by stellar feedback are far from clear, and the comparison between theory and observations is still incomplete.
Below I give a brief theoretical review of galactic winds driven by stellar feedback. In Section~\ref{section_SN}, I~review the current models of very hot galactic winds driven by SN explosions, the interaction between hot winds and the multiphase ISM, and the entrainment and acceleration of clouds in hot winds. In~Section~\ref{section_radiation}, I discuss the possibility of driving large-scale dusty winds by radiation pressure produced by the continuum absorption and scattering of starlight on dust grains. In addition to analytic models, I also show the radiation hydrodynamic simulations of the coupling between radiation, dust, and gas, which results depend on the numerical algorithms. In Section~\ref{section_CR}, I introduce cosmic rays as an alternative mechanism to launch and propagate galactic winds. Although the physics of cosmic-ray transport is still unclear, many theoretical models and numerical simulations have been carried out to explore cosmic-ray feedback in various galaxies. Conclusions and discussion are in Section~\ref{section_conclusions}.

\section{Galactic Winds Driven by Supernova Explosions}\label{section_SN}
\vspace{-6pt}
\subsection{ Impact of Supernovae and Supernova Remnants on the ISM}\label{section_individualSN}

The best-developed model of galactic winds invokes the heating of the ISM by overlapping core-collapse SN explosions. Before I discuss the impact of net SN explosions, I first briefly review the dynamics and feedback of an individual supernova remnant  (SNR). A SN from a massive star typically ejects $2$--$10\,M_{\odot}$ material into the ambient ISM and drive a shock into the ambient ISM. The interaction between a SN and its surrounding medium forms a SNR which can be observed by multiwavelength telescopes. The spherical expansion of a SNR in a uniform medium has been systematically studied and is divided into several stages:  the free expansion, Sedov-Taylor,  pressure-driven snowplow, and momentum-conserving stages (see review by Ostriker \& McKee~\citep{OM88}).

During the initial free expansion stage, the SNR expanses with nearly constant velocity until the swept-up mass of the ambient medium is comparable to the ejecta mass. Then the SNR enters the second stage--the adiabatic Sedov-Taylor stage, which was first studied by Sedov~\citep{Sedov59} and Taylor~\citep{Taylor50}. In this stage the information of SN ejecta mass is erased, and the properties of the SNR mainly depends on the ambient density and the initial SN energy. The structure of the SNR can be described by a self-similar solution. Let the radius of the forward shock front to be $r_s$, the self-similar structure gives $r_s \propto E^{\alpha}n^{\beta}t^{\eta}$ where $E,n,t$ are the initial SN energy, ambient density and time, and $\alpha,\beta,\eta$ are the self-similar indexes which can be solved by dimensional analysis. The solutions are~\citep{Draine11,Kim15}
\begin{eqnarray}\label{eq:ST}
r_s &=& 5.0 \pc \;E_{51}^{1/5}  n_0^{-1/5} t_3^{2/5},\\
v_s  &=& 1.95\times 10^3 \kms\; E_{51}^{1/5} n_0^{-1/5} t_3^{-3/5},\\
T_s &=& 5.25\times10^7 \Kel\; E_{51}^{2/5} n_0^{-2/5} t_3^{-6/5} ,
\end{eqnarray}
where $E_{51}=E/10^{51}\,$ergs, $t_3=t/10^3\,$yr and $n_0=n/$cm$^{3}$.

Once the radiative cooling in the SNR becomes important, the SNR leaves the Sedov-Taylor stage. According to Kim \& Ostriker \citep{Kim15}, the cooling time $t_{\rm cool}$, the radius of the forward shock front $r_{\rm cool}$, the shock front velocity,  the post-shock temperature $T_{\rm cool}$, and the swept-up medium mass $M_{\rm swept}$ are given by
\begin{eqnarray}\label{eq:cool}
t_{\rm cool} &=& 4.4\times10^4\,{\rm yr}\; E_{51}^{0.22} n_0^{-0.55},\\
r_{\rm cool} &=& 22.6 \pc\; E_{51}^{0.29} n_0^{-0.42}, \\
v_{\rm cool}  &=& 202\kms\; E_{51}^{0.07} n_0^{0.13},\\
T_{\rm cool}  &=& 5.67\times10^5 \Kel\; E_{51}^{0.13} n_0^{0.26}, \\
M_{\rm swept} &=& 1.68 \times 10^3 \Msun\; E_{51}^{0.87} n_0^{-0.26}.
\end{eqnarray}

For $t>t_{\rm cool}$, the SNR goes to the pressure-driven snowplow stage, in which the thermal energy of the SNR interior still cannot be neglected. The expansion of the SNR is due to thermal pressure of the SNR interior region, and the radius and velocity of the SNR shell can be approximately given by~\citep{Draine11}
\begin{eqnarray}\label{eq:snowplow}
r_{\rm slowplow} &\approx & r_{\rm cool}(t/t_{\rm cool})^{2/7},\\
v_{\rm slowplow} &\approx &\frac{2}{7}\frac{r_{\rm cool}}{t_{\rm cool}}\left(\frac{t}{t_{\rm cool}}\right)^{-5/7}.
\end{eqnarray}

\vspace{-6pt}
The SNR leaves the pressure-driven snowplow stage once the thermal energy of the SNR is radiated away, but the SNR still expands until its forward shock speed approaches the effective sound speed in the ambient medium ($c_s$). The SNR fade-away time and size can be estimated by~\citep{ZC18}
\begin{eqnarray}\label{eq:fade}
t_{\rm fade} &\approx & 1.83\times 10^6{\;\rm yr}\,E_{51}^{0.32} n_0^{-0.37}c_{s,10}^{-7/5},\\
r_{\rm fade} &\approx & 66\,\pc\; E_{51}^{0.32} n_0^{-0.37}c_{s,10}^{-2/5},
\end{eqnarray}
where $c_{s,10}=c_s/10\,$km/s.

The analytic estimate of SNR evolution in a uniform medium has been tested and confirmed by one- and multidimensional hydrodynamic simulations of spherical SNRs with radiative cooling~\citep{Chevalier74,Cioffi88,Thornton98,Blondin98,Kim15,Martizzi15,Walch15}. However, since the realistic ISM is highly inhomogeneous~\citep{Elmegreen04}, it is important to investigate the expansion of SNRs in an inhomogeneous medium~\citep{Martizzi15,Walch15,Iffrig15,ZC18}. \mbox{Martizzi et al.~\citep{Martizzi15}} initialized the turbulent medium with a lognormal density distribution~\citep{Ostriker01,Lemaster09} and performed 3D hydrodynamic simulations with adaptive mesh refinement treatment to compare SNR expansions in an inhomogeneous medium to that in a uniform medium. They found that no clear stages can be divided for SNRs in an inhomogeneous medium. At the same mean density, SNRs in an inhomogeneous medium expand faster along some preferred directions due to the existence of low-density channels compared to those in a uniform medium, and the momentum feedback provided by SNRs in an inhomogeneous medium is $\sim$$70\%$ of that in a uniform medium. Zhang \& Chevalier~\citep{ZC18}, on the other hand, did not observed low-density channels in an inhomogeneous medium, but found that the properties of remnants are mainly controlled by the mean density of the ambient medium.


Similar to galactic winds, the ISM is referred to have multiphase structure. Without SN feedback, the ISM medium can be separated into two phases: the cold neutral medium and warm medium as the consequence of thermal instability, while the warm phase can be further divided into warm neutral and warm ionized medium~\citep{Field65,Pikel68,Field69,Reynolds74,Haffner09}. The two-phase ISM model was later extended to include the feedback from SN explosions. Cox \& Smith~\citep{Cox74} first proposed that SNRs may create large-scale “tunnel” which is filled with hot and tenuous medium. McKee \& Ostriker~\citep{McKee77} then proposed the famous three-phase ISM model. The third phase is the “hot ionized medium”, which had been shock heated by overlapping SNRs to $\sim$$10^6\,$K and occupies most of the volume of the ISM. According to the three-phase model, the warm and cold medium is embedded in the hot medium as clouds, which~are supposed to be in thermal pressure equilibrium with the hot medium. However, observations have shown that the hot medium may be highly overpressured compared to other phases~\citep{Bowyer95}, thus the interstellar magnetic fields may play an important role in equilibrating the pressure balance and  suppressing the thermal evaporation of warm and cold clouds. Another problem of the three-phase model is that it cannot explain the substantial amount of warm H I in the ISM~\citep{Draine11}. Although many details are still under debate (see discussion by Cox~\citep{Cox05}), the three-phase ISM model has been the dominant conceptual framework of the ISM for four decades, and recent numerical simulations have provided better understanding of the multiphase ISM~\citep{Hennebelle14,Gatto15,Kim17b,Hill18}. The detailed discussion of the ISM model is beyond the scope of this review, but keep in mind that the theory of galactic winds is based on the theory of the ISM.


\subsection{ The CC85 model}\label{section_CC85}

According to McKee \& Ostriker~\citep{McKee77}, a galactic-scale “superbubble” forms if the SNR net energy cannot be efficiently radiated away by the multiphase ISM, otherwise a “galactic fountain” is produced to interact with the galactic halo. Although the physical processes of galactic wind launching and propagating by SN explosions are complex, a simple, clear analytic estimate can be applied to galactic winds. The classical SN-driven galactic wind model was first developed by Chevalier \& Clegg~\citep{CC85} (hereafter CC85 model).

Assuming a spherical symmetry with negligible radiative cooling, the total mass and energy injection rate into the wind are $\dot{M}_{\rm hot}$ and $\dot{E}_{\rm hot}$ respectively, the steady-state hydrodynamic equations for a hot wind driven by overlapping SN explosions are
\begin{eqnarray}
&&\frac{1}{r^{2}}\frac{d}{dr}\left(\rho u r^{2}\right)=q\label{CC85_eq1}\\
&&\rho u \frac{du}{dr}=-\frac{dP}{dr}-qu\label{CC85_eq2}\\
&&\frac{1}{r^{2}}\frac{d}{dr}\left[\rho u r^{2}\left(\frac{1}{2}u^{2}+\frac{\gamma}{\gamma-1}\frac{P}{\rho}\right)\right]=Q\label{CC85_eq3}.
\end{eqnarray}
where $r,u,\rho,P$ are the wind velocity, radial radius, density, and pressure, respectively. For $r<R$ the averaged injected efficiencies per unit volume $q$ and $Q$ are defined as $q=\dot{M}_{\rm hot}/V$ and $Q=\dot{E}_{\rm hot}/V$, where $V=4\pi R^3/3$ is the volume of the starburst/wind launching region with $R$ being the radius of this region. For $r>R$, we set $q=Q=0$. The solutions of Equations (\ref{CC85_eq1})--(\ref{CC85_eq3}) can be translated to the equations of the Mach number for the wind $M=u/\sqrt{\gamma P/\rho}$, which satisfies
\begin{equation}
\left[\frac{\gamma-1+2/M^{2}}{\gamma+1}\right]^{(1+\gamma)/[2(1+5\gamma)]}
\left(\frac{3\gamma + 1/M^2}{1+3\gamma}\right)^{-(1+3\gamma)/(1+5\gamma)}=\frac{r}{R}
\end{equation}
for $r<R$ and
\begin{equation}
M^{2/(\gamma-1)}\left(\frac{\gamma-1+2/M^{2}}{1+\gamma}\right)^{(\gamma+1)/[2(\gamma-1)]}=\left(\frac{r}{R}\right)^{2}
\end{equation}
for $r\geq R$. The solution of $M$ smoothly transfers from subsonic flow $M<1$  for $r<R$ to supersonic flow $M>1$ for $r>R$. The critical point $M=1$ is at $r=R$. The transonic solution is similar to the Parker solar wind model,   in which the velocity of the wind increases from subsonic to supersonic at a critical distance~\citep{Parker65}. However, in contrast to the Parker's model which is in an isothermal process, the CC85 model applies to adiabatic flows, while the wind structure changes for radiative flows (see Section~\ref{section_cooling}).

\begin{figure}[H]
\centering
\includegraphics[width=12cm]{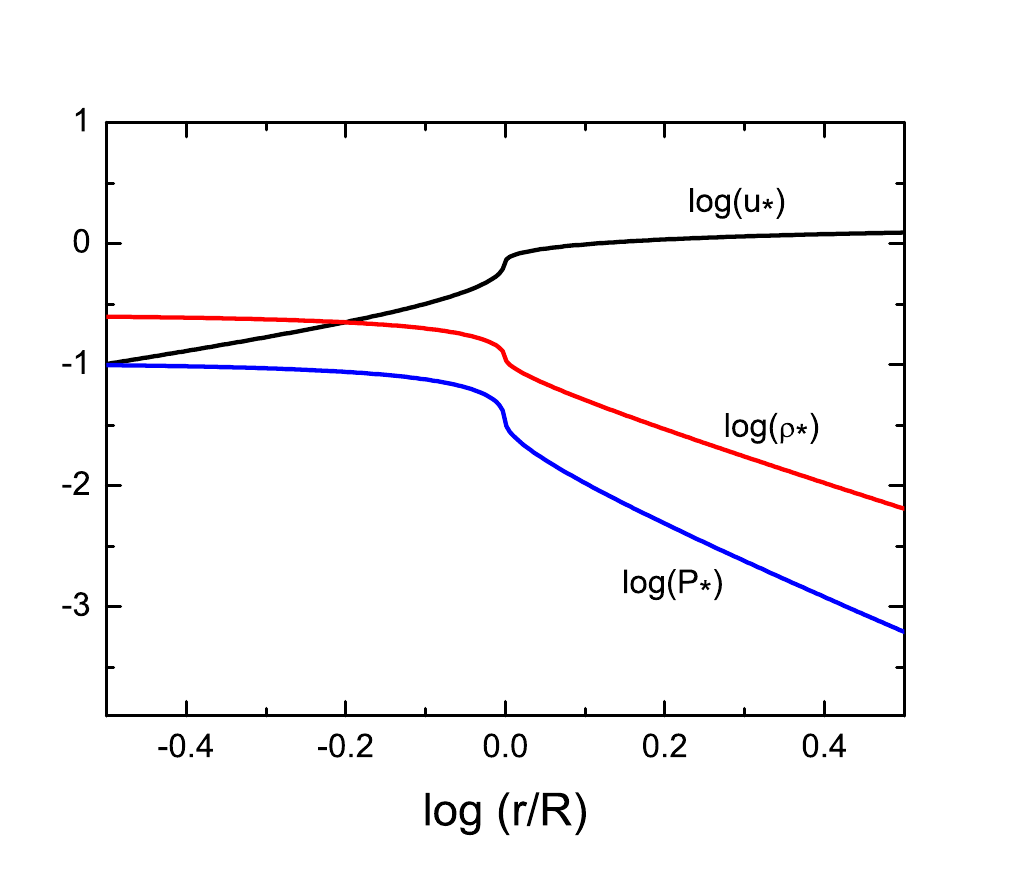}
\vspace{-12pt}
\caption{The steady-state wind solution as a function of radius $r/R$ for the CC85 model, where $R$ is the radius of the wind launching region, and the dimensionless variables are given by $u=u_{*}\dot{M}^{-1/2}\dot{E}^{1/2}$, $\rho=\rho_{*}\dot{M}^{3/2}\dot{E}^{-1/2}R^{-2}$, and $P=P_{*}\dot{M}^{1/2}\dot{E}^{1/2}R^{-2}$. Figure modified from Chevalier \& Clegg~\citep{CC85}. }\label{CC85}
\end{figure}

The wind solution is a function of radius $r$. We take $r_*=r/R$ as the dimensionless radius, and~dimensionless velocity $u_*$, density $\rho_*$ and pressure $P_*$ as
\begin{eqnarray}
&&u=u_{*}\dot{M}^{-1/2}\dot{E}^{1/2},\\\label{Appenu}
&&\rho=\rho_{*}\dot{M}^{3/2}\dot{E}^{-1/2}R^{-2},\\\label{Appenrho}
&&P=P_{*}\dot{M}^{1/2}\dot{E}^{1/2}R^{-2},\label{AppenP}
\end{eqnarray}
the solutions of these dimensionless variables are shown in Figure~\ref{CC85}. For $r \ll R$ ($r_* \ll 1$), we have $u_* \propto r_*$, and for $r \gg R$ ($r_* \gg 1$), the asymptotic velocity has $u_* \rightarrow \sqrt{2}$, $\rho_* \rightarrow 1/(4\sqrt{2}\pi r_*^{2}) \propto r_*^{-2}$, \mbox{and $P_* \propto r_* ^{-2\gamma}$. }

Next, we introduce two dimensionless parameters for the CC85 model, the thermalization efficiency $\alpha'$ and the mass-loading rate $\beta$ as
\begin{eqnarray}
\dot{E}_{\rm hot}&=&\alpha'\,\dot{E}_{\rm SN},\label{parameter1}\\
\dot{M}_{\rm hot}&=&\beta\,\textrm{SFR},\label{parameter2}
\end{eqnarray}
where SFR is the total star formation rate in a galaxy, the net energy rate provided by SNe $\dot{E}_{\rm SN}$ can be estimated by SFR that
\begin{equation}
\dot{E}_{\rm SN}=e \nu\;\textrm{SFR},
\end{equation}
where $e=10^{51}e_{51}$ ergs being the energy injected by an individual SN, and $\nu=(100M_{\odot})^{-1}\nu_{100}$ is the number of SNe per unit mass of star formation. Typically, for $100\,M_{\odot}$ of star formation one SN explodes, thus $\nu_{100}\simeq1$. More specifically, $\nu_{100}$ depends on the stellar initial mass function (IMF) and we have $\nu_{100}=1.18$ for Salpeter IMP~\citep{Leitherer99}. We define
\begin{equation}
\alpha = \alpha' e_{51} \nu_{100}
\end{equation}
as the new thermalization efficiency, thus we have
\begin{eqnarray}
&&n_{\rm hot}=10.0\;\textrm{cm}^{-3}\;\alpha^{-1/2}\beta^{3/2}\rho_* R_{200\rm pc}^{-2}\textrm{SFR}_{1},\label{winddensity1}\\
&&T_{\rm hot}=3.7\times10^{7}\;\textrm{K}\;\left(\frac{\alpha}{\beta}\right)\left(\frac{P_*}{\rho_*}\right)\label{WindT1},\\
&&V_{\rm hot}=710\;\textrm{km}\;\textrm{s}^{-1}\;\alpha^{1/2}\beta^{-1/2}u_*.\label{Vhot1}
\end{eqnarray}

Therefore, the asymptotic velocity of the wind is $V_{\rm hot}^{\infty} =10^3\,$km\,s$^{-1}$\,$\alpha^{1/2}\beta^{-1/2}$, the density and temperature of the wind at the center of the galaxy are $n_{\rm hot}=3\,$cm$^{-3}$\,$\alpha^{-1/2}\beta^{3/2}R_{\rm 200\,pc}^{-2}$SFR$_{1}$ and $T_{\rm hot} =1.5\times 10^7\,$K$(\alpha/\beta)$, where SFR$_1$=SFR/10$\,M_{\odot}\,{\rm yr}^{-1}$. Please note that for $\alpha$ $\sim$ $\beta$, the wind emits hard X-rays, which are hotter than the superbubble generated by the hot ionized medium according to the three-phase ISM model.

In addition to the CC85 model on galaxy scales, Cant\'{o}, Raga \& Rodr\'{i}guez~\citep{Canto00} developed an analytic model for winds from star clusters, where the multiple stellar winds interact with clusters of massive stars. Although the sources and scales of the cluster winds are different from the SN-driven galactic winds, the model is essentially the same as the CC85 model. Numerically, \mbox{Strickland \& Heckman~\citep{SH09}} performed a series of hydrodynamic simulations in 2D cylindrical symmetry to model the SN-driven winds from disk galaxies and compared to the results of the spherical CC85 model. They~ found that the wind from a disk-like starburst region of diameter 750\,pc and height 105\,pc has equivalent structure to that of a spherical CC85 model. The hot gas properties within a disk-like starburst region can be well predicted using the CC85 model. Also, recent 3D global simulations of SN-driven winds showed that the properties of SN-driven winds in the adiabatic phase are consistent with the CC85 model~\citep{Fielding17,Schneider18}. Therefore, it is widely believed that the CC85 model is a good approximation for adiabatic winds.

Importantly, there are two controlling parameters in the CC85 model: the thermalization efficiency $\alpha$, and the mass-loading rate $\beta$, both of which are difficult to be determined by direct observations. One~case study is for M82. Strickland \& Heckman~\citep{SH09}  used \textit{Chandra} X-ray observatory to constrain the hot wind from M82. The diffuse hard X-ray directly shows the existence of hot flows with the temperature of $\sim$$3$--$8\times 10^7\,$K, the thermalization efficiency and mass-loading rate can be constrained as $\alpha$ $\sim$ $0.3$--$1$ and $\beta$ $\sim$ $0.2$--$0.6$. More generally, Zhang~et al.~\citep{Zhang14} derived a general constraint on $\beta$ across a wide range of galaxies from dwarf starbursts to luminous infrared galaxies (LIRGs) and ultraluminous infrared galaxies (ULIRGs) using the observed linear relation between the X-ray luminosity and SFR~\citep{Grimm03,Ranalli03,Gilfanov04,Dijkstra12,Mineo12a,Mineo12b,Mineo14}. They found that for galaxies with SFR\,$\gtrsim 10\,M_{\odot}\,$yr$^{-1}$ the mass-loading rate is $\beta \lesssim 1$. This constraint means that the efficiency of converting the ambient ISM into the material of the hot wind fluid by overlapping SNe is low according to the CC85 model, which alone cannot explain the high mass-loading $\beta$ $\sim$ $1$--$10$  required by the integrated constraints on stellar feedback models in a cosmological context~\citep{Oppen06,Oppen08,Finlator08,Bower12,Puchwein13}. Bustard~et al.~\citep{Bustard16} modified the CC85 model including non-uniform mass and energy distribution in the starburst region, and radiative cooling in hot flows. They also used the $L_X-$SFR relation to constrain $\beta$ and found that $\beta \lesssim 5$ for SFR$\gtrsim\,10\,M_{\odot}\,$yr$^{-1}$ in general,  and $\beta \lesssim 2$ for SFR\,$\gtrsim\,10\,M_{\odot}\,$yr$^{-1}$ if the X-ray emission contributed by hot wind is less than 10\% of the total X-ray emission.

Please note that the above constraint on $\beta$ is only for mass-loading rate of the very hot phase of galactic winds (hereafter $\beta_{\rm hot}$). The constraint on mass-loading rates of other wind phases can be directly estimated by observations~\citep{Rupke05a,Martin05,Weiner09,Rubin14,Heckman15,Chisholm17}. For example, Martin~\citep{Martin05} estimated the mass-loading rate of neutral atomic outflows from ULIRGs with SFRs of $100$--$1000\,M_{\odot}\,$yr$^{-1}$, and~found that the mass-loading rate $\beta_{\rm cool}$$\sim\,$few, which is one order of magnitude higher than $\beta_{\rm hot}$. \mbox{Rubin~et al.~\citep{Rubin14}} found a more conservative estimate of the mass-loading rate of warm phase \mbox{$\beta_{\rm warm}\gtrsim 0.02$--$0.6$} from star-forming galaxies with SFRs $\lesssim 100\,M_{\odot}\,{\rm yr}^{-1}$ at redshift $z$ $\sim$ $0.5$, while~Weiner~\citep{Weiner09} found that the mass-loading of warm ionized phase  $\beta_{\rm warm}$ $\sim$ $1$ for star-forming galaxies at $z$ $\sim$ $1$. Heckman~et al.~\citep{Heckman15} had a higher estimate for the warm ionized phase $\beta_{\rm warm}$ $\sim$ $1$--$4$ from starburst galaxies. Chisholm~et al.~\citep{Chisholm17} found the warm phase of winds has scaling relations between mass-loading rates and the stellar mass of their host galaxies $\beta_{\rm warm}\propto M_*^{-0.4}$. This scaling relation is similar to recent numerical simulations, but the scalings obtained by observations are a factor of five smaller than some simulations~\citep{Muratov15}. In general, a large fraction of the mass of galactic winds are in other phases rather than the very hot phase. Then a question arises: what are the launching and acceleration mechanisms for the multiphase winds?

The mass-loading rate and thermalization efficiency of hot wind have also be constrained by hydrodynamic simulations~\citep{Creasey13, Girichidis16b,Martizzi16,Gatto17,Kim17a,Fielding17,Li17,Kim18}. Some work found low mass-loading rates $\beta \lesssim 1$ but others found higher. Usually these hydrodynamic simulations combined the very hot and hot components and defined hot winds with temperature $\gtrsim\,$a few\,$\times 10^{5}\,$K. Many simulations used local Cartesian boxes with vertical stratification for the ISM structure and injected SN explosions in the boxes to model the SN-driven winds. Creasey~et al.~\citep{Creasey13}'s simulations explored mass-loading rates as a function of galactic disk surface density and found that $\beta \lesssim 1$ and $\beta \propto \Sigma_{g}^{-0.82}$. Note~that $\beta$ is for all phases of winds, but~not only for hot winds. Martizzi~et al.~\citep{Martizzi16} performed local simulations and found low mass-loading rates $\beta \lesssim 1$. However, they showed that the local boxes cannot well predict the properties of hot winds because they do not capture the realistic global geometry and gravitational potential of galaxies. Fielding~et al.~\citep{Fielding17} used a global galactic disk setup with more realistic gravity and geometry than the vertically stratified structure to revisit the problems in Martizzi~et al., and~ found that $\alpha \lesssim 10^{-2}$ and $\beta \lesssim 1$, and $\alpha$ and $\beta$ decrease with the surface density $\Sigma_{\rm gas}$. However, $\beta$~decreases with spatial resolution while $\alpha$ is well converged with resolution. Li~et al.~\citep{Li17} performed simulations with fixed SN rates and more sophisticated treatment of radiative cooling, and found $\beta_{\rm hot} \simeq 2.1(\Sigma_{\rm gas}/1\,M_{\odot}\;{\rm pc}^{-2})^{-0.61\pm 0.03}$ with total mass-loading rate $\beta\simeq 7.4(\Sigma_{\rm gas}/1\,M_{\odot}\;{\rm pc}^{-2})^{-0.61\pm 0.03}$, which shows a shallower decline than that in Creasey~et al.~\citep{Creasey13}. The hot wind in their simulations is defined with flow temperature \mbox{$T > 3 \times10^{5}\,{\rm K}$}, then, for solar neighborhood Li~et al.'s model gives $\beta_{\rm hot} \simeq 0.8$ and $\beta \simeq 2$--$3$. \mbox{Kim \& Ostriker~\citep{Kim18}} gave a lower mass-loading rate with the averaged \mbox{$\beta\sim0.1$} for solar neighborhood. They defined hot winds as flows with temperature $T > 5 \times10^{5}\,{\rm K}$. Overall, the above numerical results basically support the constraint $\beta_{\rm hot}\lesssim 1$. On the other hand, higher value of $\beta$ was also obtained. Gatto~et al.~\citep{Gatto17} defined hot winds with temperature \mbox{$T > 3 \times10^{5}\,{\rm K}$}, and found that thermal pressure-driven outflows have $\beta \gtrsim 1$ at 1\,kpc if half of the volume near the galactic disk mid-plane can be heated to the hot phase. \mbox{Girichidis~et al.~\citep{Girichidis16b}} simulated winds for solar neighborhood condition and found large mass-loading rates $\beta$ $\sim$ $5$--$10$. According to Kim \& Ostriker~\citep{Kim18}, the reason for these large differences in mass and energy loading among many authors are mainly due to the different simulation settings in the vertical scale height of~ SNe.

\subsection{ The CC85 Model with Radiative Cooling}\label{section_cooling}

The traditional CC85 model is for adiabatic flows. Assuming the radius of a hot wind is much larger than the radius of wind launching/starburst region $r\gg R$, the dynamic timescale of a hot wind at radius $r$ is given by
\begin{equation}
t_{\rm dyn}\approx \frac{r}{V_{\rm hot}} \approx 1\times 10^{7}\,{\rm yr}\;\left(\frac{\beta}{\alpha}\right)^{1/2}r_{10},
\end{equation}
where $r_{10} = r/10\,$kpc. On the other hand, using an approximated power-law for radiative cooling, the~radiative cooling timescale of the wind at $r$ is (see Thompson~et al.~\citep{Thompson16} for details)
\begin{equation}
t_{\rm cool}\approx 3\times 10^6\,{\rm yr}\;\frac{\alpha^{2.20}}{\beta^{3.20}}\left(\frac{R_{\rm 200\,pc}}{r_{10}}\right)^{0.27}\frac{R_{\rm 200\,pc}^2 \Omega_{4\pi}}{\rm SFR_1}\label{eq_cool}.
\end{equation}

Remembering that the CC85 model is for a spherical wind, but Equation (\ref{eq_cool}) includes the factor of opening angle of the wind $\Omega_{4\pi} =\Omega/4\pi$ as a generalized case. Setting $t_{\rm cool}=t_{\rm adv}$, we can derive the critical radius beyond which radiative cooling is important in the flow:
\begin{equation}
r_{\rm cool}\approx 2\,{\rm kpc}\;R_{\rm 200\,pc}^{1.79}\frac{\alpha^{2.13}}{\beta^{1.92}}\left(\frac{\Omega_{4\pi}}{\rm SFR_1}\right)^{0.789}\label{rcool}.
\end{equation}

Equation (\ref{rcool}) shows that the cooling radius sensitively depends on both $\alpha$ and $\beta$. Setting $r_{\rm cool}=R$, we can also derive a critical $\beta$ which is given by
\begin{equation}
\beta_{\rm crit} \approx 0.55\,\alpha^{0.636}\left(\frac{R_{\rm 200\,pc}\Omega_{4\pi}}{\rm SFR_1}\right)^{0.364}.
\end{equation}

For example, for ULIRGs with SFR$\,=100\,M_{\odot}\,{\rm yr}^{-1}$ and $R=1$\,kpc, we have $\beta_{\rm crit}\sim 0.28 (\alpha/0.5)^{0.636}$. For M82-like starburst with SFR$\,\approx10\,M_{\odot}\,{\rm yr}^{-1}$ and $R=300$\,pc, $\beta_{\rm crit}$ $\sim$ $0.41 (\alpha/0.5)^{0.636}$. If $\beta \gtrsim \beta_{\rm crit}$, the~hot flow is radiative outside the starburst region $R$, and the CC85 model should be modified.

In contrast to the CC85 model, no analytic solutions can be obtained for the wind model with radiative cooling. Semi-analytic and numerical methods have been used to study the impact of radiative cooling on galactic or cluster winds, first by Wang~et al.~\citep{Wang95a,Wang95b}, then by others~\mbox{\citep{Efstathiou00,Silich03,Silich04,Tenorio07,Wunsch11,Thompson16,Bustard16}}. Compared to the solution of the CC85 model, some significant differences are as follows. The~temperature of a wind drops quickly from $\sim$$10^7\,$K to $\sim$$10^{3}$--$10^4\,$K due to rapid cooling, therefore the sound speed drops, and the Mach number increases. The sonic point is no longer at the starburst radius $R$. Since the hot wind transits from the very hot phase to the warm phase, Thompson~et al.~\citep{Thompson16} developed a model that the warm clouds in galactic winds are produced by the cooled hot winds due to thermal instability (see Section~\ref{section_acceleration1} for more details). The X-ray luminosity is also dimmer than that without radiative cooling, therefore the X-ray observations gave a less stringent constraint on $\beta_{\rm hot}$ than that from the CC85 model~\citep{Bustard16}.

In addition to radiative cooling, Scannapieco~\citep{Scannapieco17} found that Compton cooling due to scattering
between free elections and photons from the host galaxy can be more important than radiative cooling by atoms and ions in hot winds. The atomic and ionic radiative cooling can enhance density inhomogeneities in hot winds. Therefore, although the main cooling mechanism is Compton cooling, the hot wind can eventually generate warm clouds with a temperature of $\sim$$10^{4}\,$K, similar to the scenario given by Thompson~et al.~\citep{Thompson16}.

\subsection{ Acceleration of Warm, Cool and Cold Clouds in Hot Winds}\label{section_acceleration1}

Galactic winds have been observed to be multiphase, with clear evidence for cool, warm, and cold components from multiwavelength observations. A key question is how to accelerate the multiphase gas to the observed velocities with a typical value of a few hundred km/s, or even a thousand km/s. Since a hot wind driven by SNe is highly supersonic, the ram pressure of the wind is much stronger than the thermal pressure of it. A prevailing scenario is that the ISM material is advected into a hot wind driven by SNe and accelerated by the ram pressure of the wind~\citep{Heckman90,Heckman00,SS00,Martin05,Cooper08,SH09,Fujita09}.

Global simulations on galaxy scales have been carried out to model the acceleration of the ISM by a hot wind driven from the starburst region. In particular, the H$\alpha$ emitting ionized warm and Na I D cool phase of winds were modeled recently. Cooper~et al.~\citep{Cooper08} performed 3D simulations to study the interaction between a hot wind and an inhomogeneous ISM. They found that H$\alpha$ filaments are generated due to the breakup of clouds in the ISM in starburst region and accelerated by the ram pressure of the wind. In addition to the mass-loaded hot wind with big $\beta$, they proposed that soft X-ray also arise from bow shocks upstream of the dense clouds advected into the hot flow, and the interaction between these shocks. Cooper~et al.'s simulations showed that the soft X-ray emission region is basically associated with the H$\alpha$ emitting gas. Compared to the observation, the H$\alpha$ emitting clumps with a velocity of $\sim$$600\,$km\,s$^{-1}$ in M82 was considered to be accelerated by the hot wind~\citep{SH09}. Moreover, Fujita~et al.~\citep{Fujita09} carried out 2D simulations to study the origin of the Na I D-absorbing gas in ULIRGs. They found that cool gas can be accelerated to a mean velocity of $\sim$$320\,$km\,s$^{-1}$, and~some gas can be accelerated to $750\,$km\,s$^{-1}$. Therefore, they claimed that their model can explain the kinematics of cool gas seen in the Na I D observations. It seems that global simulations as a powerful tool have demonstrated the acceleration of a multiphase wind and matched observations. However, high-resolution simulations of cloud-hot wind interaction discovered a different story.

Motivated by global simulations which showed the formation and acceleration of filaments and clumps in hot winds, the dynamics of an individual cloud entrained in a hot wind has been studied. A more generalized problem--the interaction between clouds and hot medium or shocks in the context of ISM and hot winds have been studied for decades, both analytically~\citep{CM77,McKee78,Klein94, Zhang17}, and~numerically~\citep{Stone92,MacLow94,Schiano95,Vietri97,Marcolini05,Orlando05,Orlando06,Nakamura06,Orlando08,Shin08,Cooper09, Pittard09,Pittard10,Aluzas14,SB15, McCourt15, BS16, Banda16,Pittard16,McCourt18,Banda18,Goldsmith18}.  Although many works suggested that a cloud can survive in hot flows or shocks, recent numerical simulations focus on clouds on pc or sub-pc scales and hot flow parameters of galactic winds including radiative cooling and thermal conduction showed that a cloud is quickly shredded by the hot wind~\citep{SB15,BS16,McCourt18}. To describe the dynamics of clouds, an~important timescale needs to be introduced is the cloud crushing time $t_{\rm cc}$, which is the characteristic timescale for the cloud to be crushed by the hot wind passing through the cloud
\begin{equation}
t_{\rm cc} = \frac{R_{c}}{V_{\rm hot}}\left(\frac{\rho_{c}}{\rho_{\rm hot}}\right)^{1/2},
\end{equation}
where $R_{c}$ and $\rho_{c}$ are the radius and density of the cloud. Klein~et al.~\citep{Klein94} showed that the cloud destruction timescale is a few $t_{\rm cc}$. In the meantime, the cloud is being accelerated by the ambient hot wind. The timescale for the cloud to be comoving with the hot wind, i.e., the acceleration timescale is
\begin{equation}
t_{\rm acc} \approx\frac{4R_{c}}{3V_{\rm hot}}\left(\frac{\rho_{c}}{\rho_{\rm hot}}\right).
\end{equation}

For $\rho_{c}\gg \rho_{\rm hot}$, we always have $t_{\rm acc}\gg t_{\rm cc}$. If the destruction time of a cloud is comparable to $t_{\rm cc}$,  the~cloud can never be fully accelerated by the hot wind. However, numerical simulations showed that the cloud lifetime can be prolonged due to radiative cooling, thermal conduction, and magnetic fields.

Some more recent work suggested that clouds entrained in a hot wind are disrupted  by the Kelvin-Helmholtz (KH) instability on a timescale of $t_{\rm KH} \approx \kappa t_{\rm cc}$, with $\kappa \sim 10$ when radiative cooling in the cloud is efficient \citep{Cooper09,Andre12}. Scannapieco \& Br\"{u}ggen~\citep{SB15} used the cloud-following scheme to study the evolution of cold clouds entrained in a hot wind,  and found that the KH instability is strongly damped in supersonic hot flows, while clouds are shredded by shearing instability, which has a timescale $\propto t_{\rm cc}\sqrt{1+M_{\rm hot}}$ with $M_{\rm hot}$ being the Mach number of the hot wind. In particular, they found that the timescale for half of a cloud to be below 2/3 of the initial cloud density is $t_{\rm half}\approx 4t_{\rm cc}\sqrt{1+M_{\rm hot}}$, which is the typical lifetime of the cloud. Schneider \& Robertson~\citep{SR17} did similar simulations for both turbulent and spherical clouds but found a longer lifetime for spherical clouds. The difference is due to the different treatment of radiative cooling between two groups. Moreover, thermal conduction may also be important to evaporate the clouds~\citep{CM77,Krolik81,BS16}. Br\"{u}ggen \& Scannapieco~\citep{BS16} found that the electron thermal conduction can extend the lifetime of clouds by compressing clouds into dense filaments. They discussed that since the compression by thermal conduction causes the clouds to present a small cross section to hot flow, the terminal velocity of the clouds is still far below the hot wind velocity. Ignoring thermal conduction, McCourt~et al.~\citep{McCourt18} also observed the fragmentation of clouds interacting with a hot wind but argued that these fragments increase the total surface area and may cause rapid entrainment of cold gas. However, the fragments with a typical column density of $\sim$$10^{17}\,$cm$^{-2}$ is too small to be resolved by their simulations. Whether such small fragments can be accelerated to high velocity is still an open question.

\begin{figure}
\centering
\includegraphics[width=13cm]{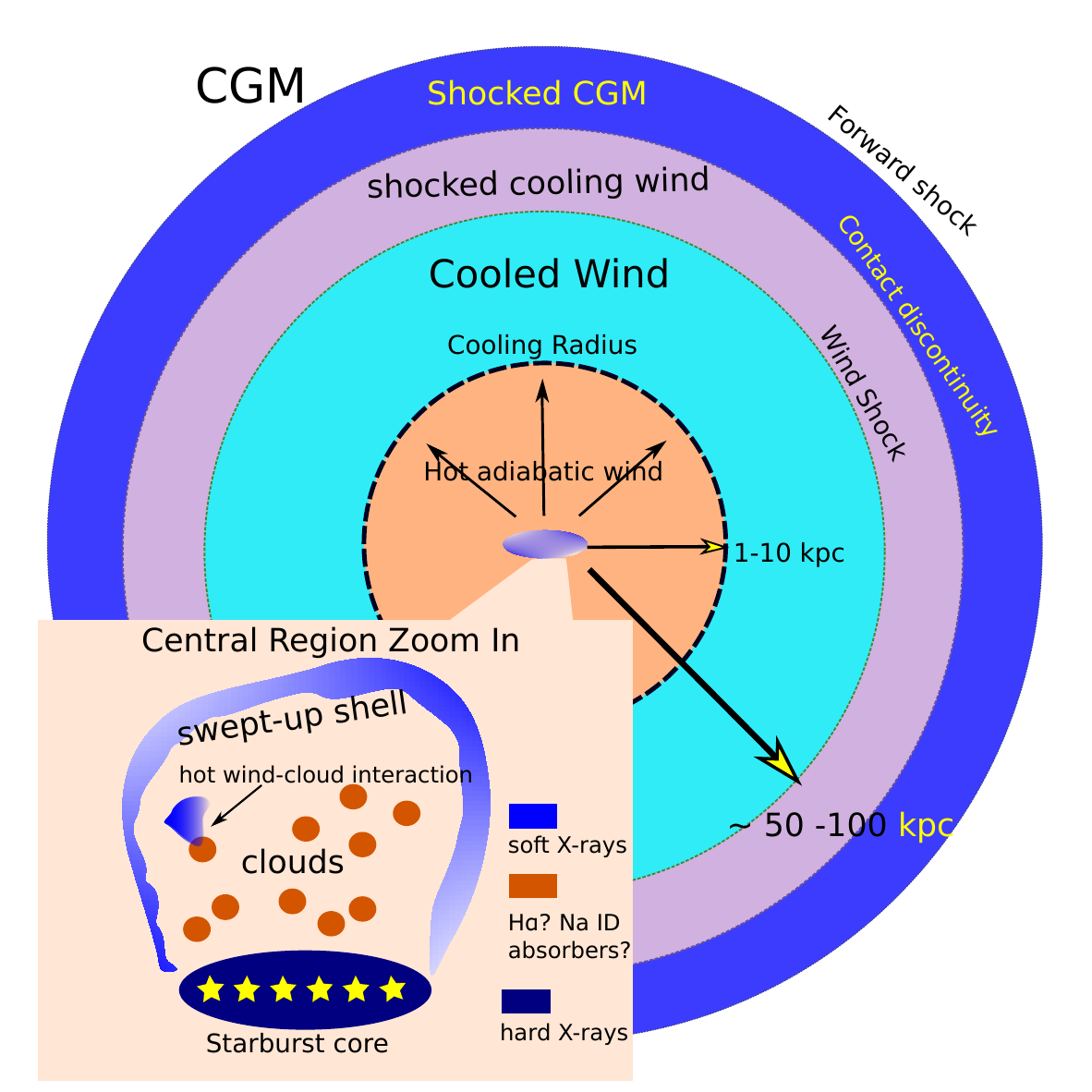}
\caption{The schematic of the evolution of a hot wind to a large distance of $\sim$$100\,$kpc. The hot wind launched from the starburst region (host) cools radiatively above the cooling radius at $\sim$$1$--$10\,$kpc, then the wind and the surrounding CGM are both shocked. Warm and cool clouds may be produced in the cooled wind by thermal instabilities, and advected to the halo. Please note that in this scenario the interaction between the hot wind and the ambient medium only provides mass-loading and increases $\beta$. The realistic process of the ISM and hot wind interaction needs to be further studied. The central region of the system is zoomed in (see the lower left schematic). The starburst region where the hot wind is launched contributes to the hard X-ray emission, while the interaction between the hot wind and clouds and the large-scale ISM contributes to the soft X-ray emission. If the entrained clouds are cold/cool/warm ($\sim$$10^{2}\,$--$10^{4}\,$K), the acceleration of clouds by the hot wind may explain the association between some soft X-rays, H$\alpha$ or molecular emission. However, it is still unclear whether clouds can be accelerated by the ram pressure of the hot wind to the observed velocities, or they may be completely shredded before fully accelerated due to hydrodynamic instabilities. Magnetic fields may be important to accelerating the entrained clouds. Figure modified from Thompson~et al.~\citep{Thompson16} and \mbox{Cooper~et al.
\citep{Cooper08}}.}\label{fig_hotwind}
\end{figure}

Some studies have shown that magnetic fields can delay the destruction of clouds~\mbox{\citep{MacLow94,Aluzas14,McCourt15,Banda16, Banda18}.} In addition to the ram pressure of hot winds, McCourt~et al.~\citep{McCourt15} showed that the magnetic field in a hot wind with $M_{\rm hot}=1.5$ enhances the drag force on an entrained cloud by a factor of $\sim$$(1+V_{A}^{2}/V_{\rm hot}^{2})$, where $V_A$ is the Alfv\'{e}n speed in the wind. According to their model, magnetic field is important only if its strength is above $\gtrsim$$100\,\mu{\rm G}\,\alpha^{1/4}\beta^{3/4}R_{\rm 200\,pc}^{-1}{\rm SFR_1}^{1/2}$. Banda-Barrag\'an~et al.~\citep{Banda16,Banda18} developed a model of magnetized clouds entrained in a hot flow with higher Mach number $M_{\rm hot}\simeq 4$--$5$, and found that spherical clouds may be accelerated to $\sim$$0.1$--$0.2$ of the hot wind velocity for the plasma beta $\beta_{\rm A}\simeq 10$--$100$, which corresponds to a magnetic field strength of $\sim$$3$--$10\,\mu{\rm G}\,\alpha^{1/4}\beta^{1/4}R_{\rm 200\,pc}^{-1}{\rm SFR_1}^{1/2}$. Turbulence also helps the cloud expansion and increases the cross section of the cloud, leading the cloud to reach $\sim$$0.3$--$0.4$ of the hot wind velocity. Magnetic fields seem to be the key ingredient to accelerate the ISM clouds in a hot wind, and the realistic magnetic field structure constrained by observations need to be used for magnetohydrodynamic (MHD) simulations (e.g., M82 magnetic fields~\citep{Adebahr17}). Please note that even magnetic fields are important to  suppress the shredding and evaporation of the clouds, the spatial extent and acceleration profile should still be tested and compared to observations of resolved systems such as M82 and NGC 253~\citep{Leroy15,Walter17}.

Another possible origin of warm/cool gas in galactic winds arises from thermal instability in the winds. Returning to the radiative cooling of hot winds (Section~\ref{section_cooling}), Thompson~et al.~\citep{Thompson16} proposed that the multiphase gas from the ISM is initially accelerated by the ram pressure of a hot flow, then~rapidly shredded, thereby increasing the mass-loading of the hot flow, and then the thermally driven hot flow with radiative cooling produces warm/cool gas again with high velocities by thermal instability. The cooled wind interacts with the CGM and deposit warm/cool gas into the halo of the host galaxy at $\sim$$50$--$100\,$kpc. This result may explain the observed warm/cool outflows in the halos of galaxies~\citep{Steidel10,Prochaska11,Tumlinson13,Werk14,Werk16,Borthakur16}. Figure \ref{fig_hotwind} shows the schematic of the evolution of a hot wind with radiative cooling. This picture was studied numerically by Schneider~et al.~\citep{Schneider18}, who~found multiphase wind generated in their simulations. However, note that this picture assumes that the ISM near the starburst/wind launching region is destroyed completely and only contributes to the mass-loading of the wind. It is important to explore the interaction between the ISM and the hot wind on galaxy scales more carefully in the future. As mentioned in Section~\ref{section_cooling}, Scannapieco~et al.~\citep{Scannapieco17} observed inhomogeneities and fragments formation in hot flows due to Compton and atomic/ionic cooling. This picture may provide another explanation of warm/cool gas in the halos, but also neglects the interaction between the wind and the ISM and CGM.

\section{Galactic Winds Driven by Radiation Pressure}\label{section_radiation}
\vspace{-6pt}
\subsection{ Radiation Feedback, and Radiation Pressure on Dust}

Radiation from individual or multiple stars may also be important in regulating star formation and launching galactic winds.

For massive stars, photoionization and radiation pressure are the dominant feedback mechanisms to affect their surrounding medium before the final SN explosions. Young massive stars emit mostly in the UV band and create HII regions by heating and ionizing the neutral gas up to a temperature of $T\sim 10^4\,$K~\citep{Whitworth79,McKee84,McKee89,Matzner02,Dale12,Walch12,Emerick08}. On the other hand, radiation pressure from massive stars acting on dust grains which absorb the direct UV photons and re-emit in the infrared (IR) band may play a more significant role than photoionization to affect the surround ISM~\citep{Krumholz09b,Kuiper11,Kuiper12,Klassen16,Kuiper16,Rosen16,Crocker18}, although this conclusion is still under debate~\citep{Sales16,Ishiki17,Ishiki18,Haid18}. In the galactic context, radiation pressure on dust grains has been discussed as a mechanism for launching galactic winds in starbursts and rapidly star-forming galaxies~\citep{Harwit62,Chiao72,Chiao73,Ferrara90,Murray05,Thompson05,Murray11,Hopkins12b}, and disrupting the giant molecular clouds (GMCs)~\citep{Harwit62,Odell67,Scoville01,Krumholz09b,Murray10,Skinner15,Raskutti16,Gupta16,Raskutti17,Tsang18}. Photoionization may be more important in Milky Way-like galaxies~\citep{Hopkins12b,Renaud13,Kannan14,Tremblin14}.

According to the CC85 model, the momentum deposition rate of a SN-driven wind is
\begin{equation}
\dot{M}_{\rm hot}V_{\rm hot} \approx 6.3\times 10^{34}\,{\rm g}\,
{\rm cm}\,{\rm s}^{-2}\,\alpha^{1/2}\beta^{1/2}{\rm SFR_1}\label{equ:SNmomentum}
\end{equation}

Assuming the bolometric luminosity of the host galaxy is $L=\epsilon  c^2 {\rm SFR}$ with a typical value of $\epsilon \sim 10^{-3}$~\citep{Leitherer99,Bruzual03}  for the Salpter initial mass function (IMF), Equation (\ref{equ:SNmomentum}) can be rewritten as
\begin{equation}
\dot{M}_{\rm hot}V_{\rm hot} \approx 3.3\,\alpha^{1/2}\beta^{1/2} \epsilon_{-3}^{-1}\frac{L}{c},
\end{equation}
where $\epsilon_{-3}=\epsilon/10^{-3}$. For the typical value of star-forming galaxies $\alpha \beta \sim 1$, we have $\dot{M}_{\rm hot}V_{\rm hot} \sim L/c$. In general, the momentum injection by the SN-driven hot wind is comparable to that provided by the radiation field. Therefore, we expect that radiation pressure on dust grains may be important to driving galactic winds. Please note that a radiation-pressure-driven wind is a kind of momentum-driven wind (see the discussion in Murray~et al.~\citep{Murray05}). We also call these winds as dust-driven winds in order to distinguish them from winds driven by radiation pressure on special lines (i.e., line-driven galactic winds e.g.,~\citep{Proga98,Proga99a,Proga99b,Proga02,Proga04,Risaliti10,Dyda18a,Dyda18b}).

\begin{figure}
\centering
\includegraphics[width=12.cm]{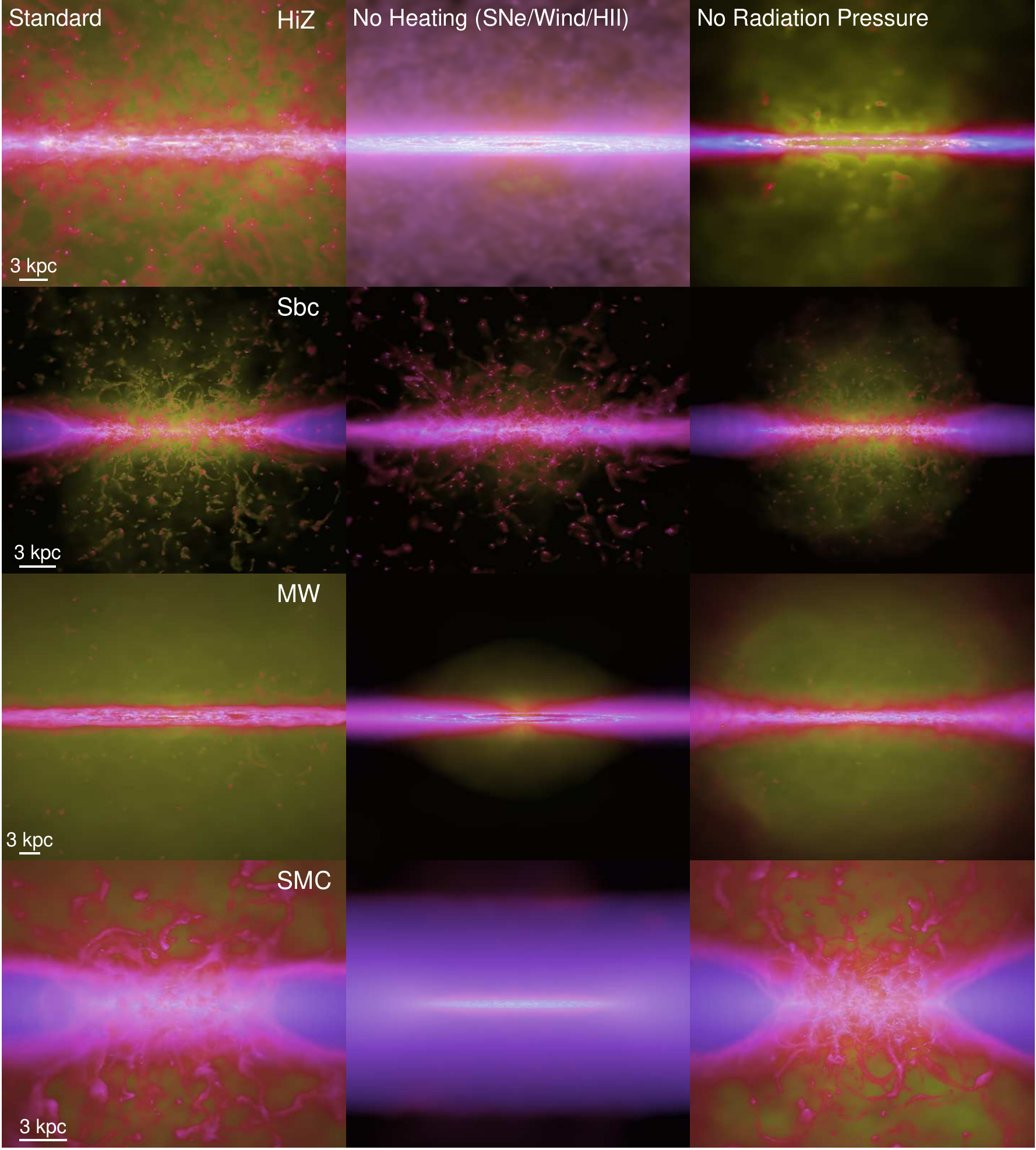}
\caption{The relative importance of SN, stellar wind and HII heating, and radiation pressure on dust in galactic winds. Edge-on gas morphologies of winds from: High-redshift (Hiz), starburst~(Sbc), Milky~Way-like (MW) and Small Magellanic Cloud (SMC)-like galaxies, with different feedback implemented. {\it \textbf{Left}}: standard model with all feedback mechanisms turning on. {\it \textbf{Middle}}: turning off heating from SN, stellar winds and HII photoionization. {\it \textbf{Right}}: turning off radiation pressure momentum injection. Figure reproduced with permission from Hopkins~et al. MNRAS, 2012~\citep{Hopkins12b}.}\label{fig_Hopkins}
\end{figure}

The relative importance of the SN and radiation feedback have been explored both analytically and numerically. A good example is the numerical simulations by Hopkins~et al.~\citep{Hopkins12b}, who combined SN (including stellar winds and HII) heating and radiation feedback together to model galactic winds in a wide range of galaxies (see Figure~\ref{fig_Hopkins}. They did not include cosmic rays, see Section~\ref{section_CR}.) Figure~\ref{fig_Hopkins} shows that the radiation pressure on dust is important in high-redshift and starburst galaxies, while SN heating and momentum injection are dominated in the Milky Way- and Small Magellanic Cloud (SMC)-like galaxies. This is a good demonstration, but more details need to be considered. For example, they used the Tree-SPH code \textsc{gadget-III}~\citep{Springel05}, later hydrodynamic simulations of SN-driven winds using sub-grid models for individual SNe have revealed more details of the SN feedback (see Section~\ref{section_CC85}). Also, it is assumed that momentum injection by radiation pressure is $\dot{P}_{\rm rad}\approx (1+\tau_{\rm IR})L_{\rm incident}$, which needs to be studied more carefully by radiation hydrodynamics.

Returning to the theory of stellar winds, the dust-driven winds in the stellar context has been well studied for decades~\citep{Gilman72,Salpeter74,Kwok75,Deguchi80,Kozasa84,Gail85,Bowen88,Fleischer92,Sedlmayr95,Krueger97,Ferrarotti06,Takigawa17}. It is believed that dust grains can survive and grow around the cool, luminous red supergiants, and AGB stars. Radiation pressure on dust grains can drive stellar winds from these stars. Lamers \& Cassinelli~\citep{Lamers99} reviewed the necessary condition for driving winds by dust. First of all, dust grains need to survive against sublimation. If the radiative equilibrium temperature $\approx (F/a_r c)^{1/4}$ with $F$ being the radiation flux onto the dust and $a_r$ being the radiation constant is above the “condensation temperature”, then the grains will sublimate rather than grow. Different dust grains with different sizes have various sublimation temperature, with a typical value between $\sim$$500$--$3000\,$K~\citep{Kobayashi09,Kobayashi11,Baskin18}. In addition to dust sublimation, dust sputtering also occurs due to high-energy atom/ion incident on a grain surface and could destroy a fraction of dust mass in the environment of hot gas~\citep{Draine79a,Draine79b,Covatto00,Gray04,Nath08,Everett10b,Jones11,Draine11b}. Another necessary condition is that the dust is coupled with the gas. The dust grains obtain momentum from the absorption or scattering of photons, and they collide with the gas~\citep{Gilman72},  then the drag force between the dust grains and the gas causes the coupling of the dust to the gas. The requirement of the momentum coupling sets up the lower limit of the mass loss rate and velocity of a dust-driven wind~\citep{Berruyer83,Mastrodemos96}.

Similar necessary conditions also need to be considered in the galactic context. Murray~et al.~\citep{Murray05} estimated that the typical length below which the dust is dynamically coupled with the gas is a few hundred kpc, which is significantly larger than the typical scales on which the galactic winds are launched and accelerated. Although the dust and the gas are dynamically coupled, they may not be in thermal balance. The dust is thermally in equilibrium with the radiation field, but the gas may have different temperature. Goldsmith~\citep{Goldsmith01} showed that the dust and gas are only well coupled in regions denser than $\sim$$10^{4}$--$10^{5}\,$cm$^{-3}$. Please note that the inner few hundred pc of ULIRGs has mean gas density $\sim$$10^3$--$10^4\,$cm$^{-3}$. Narayanan~et al.~\citep{Narayanan11,Narayanan12} found that the dust and gas are thermally coupled in ULIRGs. However, the density of the dust-driven winds is much lower than that in the galactic disks, therefore the dust and gas in the winds are dynamically coupled but not thermally coupled (i.e., $V_{\rm dust}=V_{\rm gas}$ but $T_{\rm dust} \neq T_{\rm gas}$). The dust temperature can be estimated by $T_{\rm dust}\approx (F/a_r c)^{1/4}\approx 64\,$K\,$(F/10^{13}\,L_{\odot}\,{\rm kpc}^{2})^{1/4}$, and the dust is still possible to survive in the hot medium with temperature much higher than the dust sublimation temperature.

\unskip

\begin{figure}[H]
\centering
\includegraphics[width=12cm]{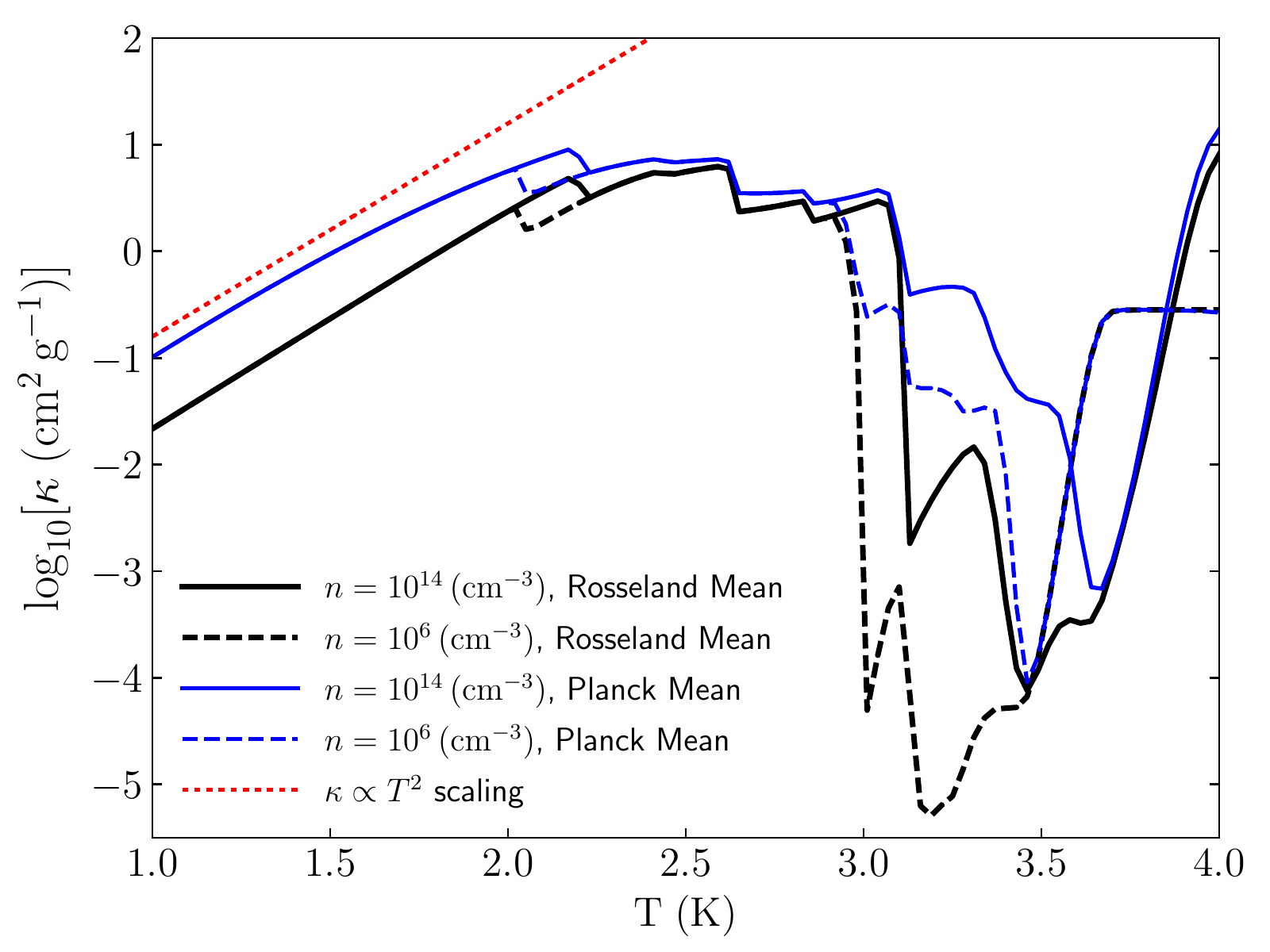}
\caption{The Rossenland mean ({\it black lines}) and Planck mean ({\it blue lines}) opacities of dust from the temperature range $10\,{\rm K}\leqslant T  \leqslant 10^4\,{\rm K}$. The gas density $n=10^{14}$\,cm$^{-3}$ and $n=10^{6}$\,cm$^{-3}$. Figure~reproduced using the publicly available code of Semenov~et al.~\citep{Semenov03}. Other opacities~\citep{Sharp92,Seaton94,Iglesias96,Bell94,Pollack94,Alexander75,Alexander94} were also plotted and compared to Semenov~et al.'s opacity in~\citep{Semenov03}. Please note that the Rossenland mean opacities are scaled as $\kappa \propto T^2$ while the Planck mean opacities are slightly different from  $\kappa \propto T^2$  at low temperature. The opacities decrease significantly at the sublimation temperature of dust at $T$ $\sim$ 1000 K. The rise in the opacity at higher temperature is due first to H scattering, then to bound- and free-free interactions, then to scattering off of free electrons.}\label{fig_opacity}
\end{figure}

Since the dust is coupled with the gas, the dust opacity not only depends on the composition and size of the dust grains, but also depends on the dust-to-gas ratio. Some widely used dust opacities include the frequency-averaged Rosseland and Planck mean opacities~\citep{Semenov03,Ferguson05}. Figure~\ref{fig_opacity} shows these two opacities in the literature, most of which was discussed in Semenov~et al.~\citep{Semenov03}. For dust temperature $\lesssim$$150\,$K, the opacity is scaled as $\kappa \approx \kappa_0 T^2$. Thompson~et al.~\citep{Thompson05} discussed that the $T^2$ scaling follows from the fact that the dust absorption cross section scales as $\lambda^{-2} \propto T^2$ in the Rayleigh limit. The normalization $\kappa_0$ depends on the properties of grains and dust-to-gas ratio. For the Galactic environment $\kappa_0$ $\sim$ $(2$--$3)\times 10^{-4}$\,cm$^2$\,g$^{-1}$ for the Rosseland mean $\kappa_R$ and $\kappa_0$ $\sim$ $10^{-3}$ \,cm$^2$\,g$^{-1}$ for the Planck mean. For dust temperature between $150\,{\rm K}\lesssim T \lesssim 1500$--$2000\,{\rm K}$, the dust opacities are almost flat around $5$--$10$\,cm$^{2}$\,g$^{-1}$; and for dust temperature $\gtrsim$1500--$2000\,$K, the opacities significantly drop, because the temperature has reached the dust sublimation temperature, but still below the hydrogen ionization temperature $\sim$$10^4\,$K. Then the gas dominates over the opacities at $T\gtrsim 10^4\,$K.

Considering gravity, a basic requirement for radiation pressure to drive winds is that the radiation flux of a galaxy near or approach the Eddington limit for dust. A galactic disk which is optical thick to the UV radiation ($\tau_{\rm UV}>1$) requires the gas surface density \mbox{$\Sigma_{g}> 1/\kappa_{\rm UV}$}, \mbox{or $\Sigma_{g}> 10^{-3}\,{\rm g}\,{\rm cm^{-2}}$ $\sim$ $5\,M_{\odot}\,{\rm pc^{-2}}$,} where $\kappa_{\rm UV}$ $\sim$ $10^{3}\,{\rm cm^{2}}\,{\rm g^{-1}}$ is a characteristic UV opacity~\citep{Thompson05}. The~ typical surface density of rapidly star-forming galaxies and starbursts is above this value and are optical thick to UV. For galaxies which are still optically thin to IR, we can derive the Eddington limit for single-scattering limit, which means that all photons are scattered or absorbed once and then escape the system. Therefore $\kappa$ $\sim$ $2/\Sigma_g$, and the Eddington limit flux for these galaxies (see details in Andrews \& Thompson~\citep{Andrews11}) is
\begin{equation}
 F_{\rm Edd} \approx \frac{4 \pi G c \Sigma}{\kappa} \sim 10^{8} \; {\rm L_{\odot} \; kpc^{-2}} \;
 \left(\frac{\Sigma_{\rm g}}{10 \; {\rm M_{\odot}\;pc^{-2}}} \right)^2 \, f_{\rm
 g}^{-1},
\end{equation}
where $f_{\rm g}$ is the mass fraction of gas. For galaxies (mostly starbursts) which are optical thick to IR ($\Sigma_{\rm g}>5000\,M_{\odot}\,{\rm pc^{-2}} \, f_{\rm dg, \, 150}^{-1}$) with $f_{\rm dg, \, 150}=f_{\rm dg} \times 150$ being the dust-to-gas ratio, other two Eddington limit fluxes can be derived, one is for warm starbursts ($T_{\rm dust}\lesssim 150\,$K) that
\begin{equation}
 F_{\rm Edd}  \sim 10^{13} \; {\rm L_{\odot} \, kpc^{-2}} \, f_{\rm
g}^{-1/2} \, f_{\rm dg, \, 150}^{-1},
\end{equation}
another is for hot starbursts ($T_{\rm dust}\gtrsim150\,$K but still below the dust sublimation temperature) that
\begin{equation}
 F_{\rm Edd} \sim 10^{15} \; {\rm L_{\odot}\;kpc^{-2}} \; \left
 (\frac{\Sigma_{\rm g}}{10^6 \; {\rm M_{\odot}\;pc^{-2}}} \right ) \; f_{\rm
 g}^{-1/2} \, f_{\rm dg, \, 150}^{-1}.
\end{equation}

Andrews \& Thompson~\citep{Andrews11} compared the above Eddington limit fluxes to observations and found that normal galaxies are sub-Eddington in general, so the impact of radiation pressure is not significant. Coker~et al.~\citep{Coker13} found that M82 is sub-Eddington on kpc scales, but the single-scattering Eddington ratio is close to be unity on smaller scales $\sim$$250$--$500\,$pc. The starburst cores of Arp 220 may near or reach the Eddington limit for dust~\citep{Wilson14,Barcos18}. Most LIRGs and ULIRGs are most likely to be sub-Eddington to the IR radiation~\citep{Barcos17,ZD17,Zhang18}, but they can still be locally super-Eddington due to the density inhomogeneity of the disks~\citep{TK16}. Please note that the estimate about whether galaxies reach the Eddington limit includes uncertainties in the dust-to-gas ratio and the CO/H$_2$-HCN conversion factors. Recent numerical simulations also showed that dust-driven winds can be launched even in an initially sub-Eddington system (see Section~\ref{section_simulationrad}). Below we first introduce the analytic models for dust-driven shells in Eddington and sub-Eddington systems in Sections~\ref{section_analyticrad1} and~\ref{section_analyticrad2}, then look into the interaction between gas and radiation based on recent radiation hydrodynamic simulations in Section~\ref{section_simulationrad}. The~importance of radiation pressure on star cluster scales are discussed in Section~\ref{section_starcluster}.



\subsection{ Analytical Models for Dusty Shell Acceleration by Radiation Pressure}\label{section_analyticrad1}

Ignoring the SN feedback, the dynamics of properties of dust-driven winds can be studied from first principles. The momentum equation of a dust-driven wind is  (see more details in~\citep{Lamers99}, Chapter 7)
\begin{equation}
\rho \frac{dv}{dt}+\frac{dP}{dt}+\rho \frac{GM}{r^2}=\rho\frac{\kappa F}{c}\label{equ_radmomentum1},
\end{equation}
where $M$ is the total mass inside radius $r$, $F$ is the radiation flux, and $\kappa$ is the dust opacity. For a spherical wind, the flux can be written as a function of the total luminosity
\begin{equation}
F= \frac{L}{4\pi r^2 c},
\end{equation}
and we introduce the Eddington limit for dust $\Gamma_{\rm d}=\kappa L/(4\pi cG M)$. For steady-state wind the Eulerian description gives $\partial v/\partial t=0$, and Equation (\ref{equ_radmomentum1}) is rewritten as
\begin{equation}
v \frac{dv}{dr} + \frac{dP}{dr} = \frac{\kappa L}{4\pi r^2}\left(1-\frac{1}{\Gamma_{\rm d}}\right)\label{equ_radmomentum2}.
\end{equation}

Integrating Equation (\ref{equ_radmomentum2})  over $dm = 4\rho \pi r^2 dr$, and note that the total radiation force is
\begin{equation}
\int_{V}\frac{\kappa \rho L}{4\pi c r^2}dV = \int_{R}^{\infty} \frac{\kappa \rho L}{c}dr = \tau \frac{L}{c},
\end{equation}
where $\tau = \int_{R}^{\infty} \kappa \rho dr$ is the radial optical depth, and $R$ is the starburst region where the dusty wind is launched. As a result, the integrated momentum equation of the wind is
\begin{equation}
\dot{M} v_{\infty} = \tau \frac{L}{c} \left(1-\frac{1}{\Gamma_{\rm d}}\right)\label{equ_radmomentum3},
\end{equation}
where $\dot{M}$ is the mass-loading rate, which is assumed to be a constant for spherical wind, and $v_{\infty}$ is the asymptotic velocity of the dusty wind. For Eddington ratio $\Gamma_{\rm d} >1$, the dusty gas is unstable to driving an unbound outflow, but for $\Gamma_{\rm d}<1$, only “fountain flow” can be produced. Ignoring the gravity, we have $\dot{M}v_{\infty} \approx \tau L/c$. For UV photons, we take the single-scattering limit, i.e., all UV photons only scatter once, thus $\tau_{\rm UV}\leq 1$. On the other hand, infrared (IR) can be scattered many times and $\tau_{\rm IR}\gg1$ is possible. Theoretically, the momentum of the dusty wind is boosted from $L/c$ to $\tau_{\rm IR}L/c \gg L/c$ for systems which are optical thick to IR.

Now we consider a dusty shell accelerated by radiation pressure. Assuming the mass of the shell is $M_{\rm sh}$, and the shell has a radius of $r$ and thickness of $\Delta r$. The momentum Equation (\ref{equ_radmomentum1}) can be rewritten as
\begin{equation}
M_{\rm sh}\frac{dv}{dt}+\frac{GM}{r^2}M_{\rm sh}=\tau\frac{L}{c}\label{equ_radmomentum4}.
\end{equation}

Here, we take $\tau=\kappa \rho \Delta r$ as the optical depth of the shell. If we use an isothermal sphere to model the gas  as a function of radius $\rho(r)=f_{g}\sigma^2/(2\pi G r^2)$ with $f_{g}$ being the mass fraction of gas and $\sigma$ being the velocity dispersion, and assume the mass $M_{\rm sh}$ is from the swept-up gas inside radius $r$, thus, we~ have $M_{\rm sh}$  as a function of $r$ that $M_{\rm sh}=2f_g \sigma^2 r/G$. Assuming the single-scattering limit $\tau=1$, we~can use Equation (\ref{equ_radmomentum4}) to solve the velocity of wind as a function of $r$:
\begin{equation}
v(r)=2\sigma\sqrt{\left(\frac{L}{L_M}-1\right) \ln \left(\frac{r}{R}\right)}\label{equ_vel1}.
\end{equation}

This solution (\ref{equ_vel1}) is given by Murray~et al.~\citep{Murray05}. 
Here $L_{M}$ is the Eddington luminosity for single-scattering limit that $L_{M}=G M_{\rm sh}M c/r^2$, or in particular, $L_M = 4 f_g c \sigma^4/G$ for an isothermal sphere. An unbound dust-driven wind requires its host galaxy to have a luminosity $L>L_M$, i.e.,  a~ super-Eddington luminosity. On the other hand, the $L_M \propto \sigma^4$ may be related to the Faber-Jackson relation~\citep{Faber76}. Murray~et al.~\citep{Murray05} argued that a starburst that reaches $L_M$ moderates its star formation rate and its luminosity not to increase significantly further. Elliptical galaxies reach $L_M$ during their growth at $z\gtrsim1$, and that this is the origin of the Faber-Jackson relation.

A more generalized equation for a dusty shell includes both UV and IR opacities. \mbox{Thompson~et al.~\citep{Thompson15}} (see also~\citep{HT17}) modified Equation (\ref{equ_radmomentum4}) as
\begin{equation}
M_{\rm sh}\frac{dv}{dt}+\frac{GM}{r^2}M_{\rm sh}=(1+\tau_{\rm IR}+{\rm e}^{-\tau_{\rm UV}})\frac{L}{c}\label{equ_radmomentum5}.
\end{equation}

For a system with $\tau_{\rm UV}\leq 1$, the optical depth term $1+\tau_{\rm IR}+{\rm e}^{-\tau_{\rm UV}}$ goes to $\tau_{\rm UV}$. For $\tau_{\rm UV} \gg 1$ the optical depth term becomes $1+\tau_{\rm IR}$. The generalized Eddington limit for a shell is defined as
\begin{equation}
L_{\rm Edd} =\frac{GM M_{\rm sh}c}{R^2}(1+\tau_{\rm IR}-{\rm e}^{-\tau_{\rm UV}})^{-1},
\end{equation}
and the Eddington radio $\Gamma = L/L_{\rm Edd}$ can be reduced to $\Gamma \rightarrow \Gamma_{\rm IR} =\kappa_{\rm IR}L/(4\pi GMc)$ for $\tau_{\rm IR}\gg 1$, and $\Gamma \rightarrow \Gamma_{\rm UV} = \kappa_{\rm UV} L/(4\pi GMc)$ for $\tau_{\rm UV}\ll 1$. The Eddington limit and Eddington ratio for the single-scattering limit are $L_{\rm single}=GMM_{\rm sh}c/R^2$ and $\Gamma_{\rm single} =LR^2/(GMM_{\rm sh}c)$ respectively. Please note that $L_{\rm single}$ is basically the same as $L_M$ in Equation (\ref{equ_vel1}), but $L_M$ is for an isothermal sphere and $L_{\rm single}$ is for a point source.

For geometrically thin dusty shell with a fixed mass, the opacity of the shell $\propto r^{-2}$ and eventually becomes optical thin for UV photons. For $\Gamma_{\rm IR}+\Gamma_{\rm single}>1$ and $\Gamma_{\rm UV}>1$, the shell can be accelerated first by IR radiation, then by UV radiation. Following Thompson~et al.~\citep{Thompson15}, the asymptotic velocity of the shell is
\begin{equation}
v_{\infty} \approx \left(\frac{2 R_{\rm UV}L}{M_{\rm sh}c}\right)^{1/2} \approx 600\,{\rm km}\,{\rm s}^{-1}\,L_{12}^{1/2}\kappa_{\rm UV,3}^{1/4}M_{\rm sh,9}^{-1/4},
\end{equation}
where $R_{\rm UV}=(\kappa_{\rm UV}M_g/4\pi)^{1/2}$ is the critical radius where the shell becomes optical thin to the incident UV radiaiton, $L_{12}=L/10^{12}\,L_{\odot}$, $\kappa_{\rm UV,3}=\kappa_{\rm UV}/10^3\,$cm$^2$\,g$^{-1}$, and $M_{\rm sh,9}=M_{\rm sh}/10^9\,M_{\odot}$.

There are many simplifications in the picture of shell acceleration. Firstly, the SN feedback is ignored. A more realistic model should include the SN-driven wind combined to the radiation-pressure-driven wind, and the dusty shell may exist in the hot flow. Secondly, the shell is assumed to have a fixed mass. In general, a massive shell sweeps up mass of the ISM and CGM, the~mass of the shell increases, and the assumption of a constant $M_{\rm sh}$ breaks down. If the swept-up mass is accumulated in the shell, the shell may be decelerated. Thirdly,  the hydrodynamic instabilities are ignored in the analytic models. As discussed in Section~\ref{section_acceleration1}, a hot wind may shred clouds in the ISM, and similar instabilities may occur due to the radiation-shell or shell-ISM/CGM interaction. Also,~the radiatively driven instabilities may suppress the coupling between radiation and the dusty gas. Recently, more detailed analytic models of momentum-driven winds have been developed~\citep{Lochhaas18}. In addition, multidimensional MHD and radiation hydrodynamic simulations need to  be carried out to further explore the dynamics of the momentum-driven shells.

\subsection{ Self-gravitating Luminous Disks}\label{section_analyticrad2}

In contrast to the spherical systems, cylindrical disk systems show different properties of dust-driven winds. The Eddington ratio is no longer a constant but varies along the height above the disks. Considering a disk with uniform brightness and surface density  $I(r\leq r_{\rm rad})=I$ and $\Sigma(r\leq r_D)=\Sigma$, where $r_{\rm rad}$ and $r_D$ are the outer radius of the luminous and gravitating portion of the disk. Then the gravity along the polar $z$-axis is (more details see Zhang~et al.~\citep{Zhang12})
\begin{equation}
f_{\rm grav}(z) =-2\pi G \Sigma \left(1-\frac{z}{\sqrt{z^2+r_D^2}}\right)
\end{equation}
and the vertical radiation force is
\begin{equation}
f_{\rm rad}(z)=\frac{2\pi \kappa I}{c}\int_{0}^{r_{\rm rad}}\frac{z^{2}rdr}{(r^{2}+z^{2})^{2}}
=\frac{\pi\kappa I}{c}\frac{r_{\rm rad}^{2}}{z^{2}+r_{\rm rad}^{2}}.
\end{equation}

Then the Eddington ratio along the pole is a function of $z$
\begin{equation}
\Gamma(z)=\Gamma_{0}\left(\frac{r_{\rm rad}}{r_{D}}\right)^{2}
\left(\frac{z^{2}+r_{D}^{2}}{z^{2}+r_{\rm rad}^{2}}+\frac{z\sqrt{z^{2}+r_{D}^{2}}}{z^{2}+r_{\rm rad}^{2}}\right),\label{ratio}
\end{equation}
where $\Gamma_0$ is the Eddington ratio at the center of the disk. For $r_D \approx r_{\rm rad}$, we have $\Gamma_{\infty} = 2\,\Gamma_0$, \mbox{and $\Gamma_{\infty} =2$} if the disk is at Eddington limit. The bright self-gravitating disk is unstable to driving dusty winds. This result is obviously different from that for a spherical system which has a constant $\Gamma$ and $v_{\infty} \propto (\Gamma-1)$.

For a test particle along the pole, the equation of the particle is
\begin{eqnarray}
v^{2}=v_{0}^{2} + 4\pi G\Sigma r_{D}\left[\hat{r}\Gamma_{0}\arctan\left(\frac{\hat{z}}{\hat{r}}\right)
-\left(1+\hat{z}-\sqrt{1+\hat{z}^{2}}\right)\right],\label{orbit01}
\end{eqnarray}
where $\hat{r}=r_{\rm rad}/r_D$ and $\hat{z}=z/r_D$. For $\Gamma_0 =1$ and $r_D = r_{\rm rad}$, the asymptotic terminal velocity of the test particle is
\begin{equation}
v_{\infty}=v_{c}\sqrt{\pi/2-1}\simeq380\,\textrm{km}\,\textrm{s}^{-1}\Sigma_{0}^{1/2}r_{D,\rm 1kpc}^{1/2},\label{terminal01}
\end{equation}
where the characteristic velocity of the wind $v_c = \sqrt{4\pi G \Sigma r_D} \simeq 500\,\textrm{km}\,\textrm{s}^{-1}\Sigma_{0}^{1/2}r_{D,\rm 1kpc}^{1/2}$, \mbox{$\Sigma_0 = \Sigma/$g\,cm$^{-2}$}, and $r_{D,\rm 1kpc} = r_D/1\,$kpc. Equation (\ref{terminal01}) gives the maximum asymptotic velocity of a test particle from a disk at the Eddington limit $\Gamma_0 =1$. Moreover, using the Schmidt law $\Sigma_{\rm SFR}\propto \Sigma_{\rm gas}^{1.4}$~\citep{Kennicutt98a,Kennicutt98b}, we have
\begin{eqnarray}
v_{\infty}\sim400\,\textrm{km}\,\textrm{s}^{-1}f_{g,0.5}^{-0.5}\left(\frac{\rm SFR}{50M_{\odot}\textrm{yr}^{-1}}\right)^{0.36}
r_{D,\rm 1kpc}^{-0.21},\label{terminal02}
\end{eqnarray}
where the gas fraction is $f_g =0.5f_{g,0.5}$. The $v_{\infty} \propto {\rm SFR}^{0.36}$ law may be consistent with some observations in ULIRGs and star-forming galaxies~\citep{Martin05,Weiner09}.

\subsection{ Coupling between Radiation Field and Dusty Gas}\label{section_simulationrad}

The optical depth in some dense star cluster and many rapidly star-forming galaxies are thick to the IR radiation, and the dust reprocessed IR radiation may be absorbed and re-emitted multiple times in these systems. Equation (\ref{equ_radmomentum3}) shows that the momentum of a dust-driven wind is $\dot{M} v_{\infty} \approx \tau_{\rm IR}L/c$, where $L/c$ is the initial momentum provided by the radiation field. However, this is only an analytic estimate, in which we assume the wind is spherical and the radiation field is isotropic along each direction. Roth~et al.~\citep{Roth12} found that the coupling between radiation and gas may be significantly reduced due to the geometry effect: the IR photons may escape along sightline of lower column density of a non-spherical torus, and the momentum transfer from a radiation field to the gas is limited to $\sim$1--$5\,L/c$ even $\tau_{\rm IR}$ is big along some special directions.

Importantly, radiatively driven instability e.g.,~\citep{Jacquet11,Jiang13} may create low-density channels in the gas, and suppress the momentum coupling between radiation and gas because a fraction of the radiation escapes from these channels. It has been argued that the rate of momentum deposition will never exceed a few $L/c$ due to the radiative Rayleigh-Taylor instability (RRTI,~\citep{Krumholz09a,Krumholz09b}). To better understand the dynamics of radiation-gas interaction, a series of multidimensional radiation hydrodynamic (RHD) simulations have been carried out. The results depend on the algorithms for solving the RHD equations and may also depend on the spatial resolution of the simulations.

We can simplify a problem as follows: an IR radiation field is injected into a computational box vertically at the bottom of the box and interacting with a shell of dusty gas with a vertically stratified structure, and the gravity is also along the vertical direction. A key question for this problem is: what is the efficiency of momentum transfer from the radiation field to the gas. In other words, the momentum of the gas in the local box is written as
\begin{equation}
\dot{M}v_{\infty} = (1+\eta \tau_{\rm IR})\frac{L}{c},
\end{equation}
what is the efficiency $\eta$ for a local box simulation? And is it possible to drive an unbound dusty wind by radiation pressure? The results depend on the numerical algorithms. Therefore, we need to briefly look into the methods to solve the RHD equations.

The time-dependent radiative transfer equation is
\begin{equation}
\frac{1}{c}\frac{\partial I_\nu}{\partial t} + \mathbf{n} \cdot \mathbf{\nabla} I_\nu = \eta_\nu-\kappa_\nu \rho I_\nu \label{RT1}
\end{equation}
where $\nu$ is the frequency, $I_{\nu}$ is the radiation intensity, $\eta_{\nu}$ is the emissivity, and $\kappa_{\nu}$ is the opacity. Integrating Equation (\ref{RT1}) over a solid angle and the frequency, we obtain the zeroth and first momentum equations
\begin{eqnarray}
\pdif{E_r}{t} + \mathbf{\nabla} \cdot \mathbf{F}_r & = & cS_r(E),\label{eq:radenergy} \\
\frac{1}{c^2}\pdif{\mathbf{F}_r}{t}+\mathbf{\nabla} \cdot{\sf P}_r & = & \mathbf{S}_r(\mathbf{P}),\label{eq:radmomentum}
\end{eqnarray}
where
\begin{eqnarray}
E_{\rm r}  &=& \frac{1}{c} \int \int I_\nu(\hat{n}) d\Omega d\nu, \label{eq:J} \\
\mathbf{F}_{\rm r} &=& \int \int I_\nu(\hat{n}) \mu_i d\Omega d\nu, \label{eq:H}\\
{\sf P}_{\rm r}  &=& \frac{1}{c}  \int \int I_\nu(\hat{n}) \mu_i \mu_j d\Omega d\nu, \label{eq:K}
\end{eqnarray}
are the frequency-integrated gray energy, flux, and tensor of the radiation field. Also, the frequency-integrated energy and momentum source term $S_r(E)$ and $S_r(\mathbf{P})$ can be added to the hydrodynamic energy- and momentum-conservation equations respectively (see~\citep{Lowrie99,Jiang12} for more details. The hydrodynamic equations coupled with radiation, combined with the radiative transfer Equation (\ref{RT1}) or the zeroth and first momentum Equations (\ref{eq:radenergy}) and (\ref{eq:radmomentum}) give the set of RHD equations. Since the total number of unknown fluid and radiation variables are more than the number of equations, we need other inputs, i.e., the so-called closure relations relating $\mathbf{P}$ to $E_r$ to solve all the variables \citep{Mihalas84}.

A widely used closure relation is the flux-limited-diffusion (FLD) approximation. Neglecting the first momentum Equation (\ref{eq:radmomentum}), and assume the radiation field is locally isotropic, so  $\mathbf{P}=E_r \mathbb I/3$, and
\begin{equation}
\mathbf{F}_r=-\frac{c \lambda}{\kappa_{R}\rho} \nabla E_r,\label{eq:fld}
\end{equation}
where $\lambda(E)$ is the flux-limiter to prevent the flux becoming unphysical~\citep{Levermore81}. As a consequence, the~FLD approximation (\ref{eq:fld}) with the zeroth momentum equation closes the RHD equations.

A better treatment than the FLD approximation is the M1 closure method. The Eddington tensor is defined as ${\sf f} \equiv {\sf P}_r/E_r$. The key technique is to calculate the Eddington tensor and close the set of RHD with both the zeroth and first radiation momentum equations. The M1 closure method calculates the Eddington tensor as
\begin{equation}
{\sf f} = \frac{1-\chi}{2} {\mathbb I}
+ \frac{3\chi-1}{2}~{\bf n}\otimes {\bf n},
\end{equation}
where ${\bf n}=\mathbf{F}_r/|\mathbf{F}_r|$, also $\chi$ depends only on the reduced flux and can be calculated following~\citep{Levermore84}.

A more elaborate but more expensive closure method is called the variable Eddington tensor (VET) method, in which the Eddington tensor is direct calculated by the definition in each timestep:
\begin{eqnarray}
{\sf f} =\frac{{\sf P}_r}{E_r}=\frac{\int I_\nu \hat{\bold{n}}\hat{\bold{n}}d\Omega}{\int I_\nu d\Omega},
\label{Edd}
\end{eqnarray}
and the intensity $I$ is solved based on the time-independent radiative transfer equation
\begin{equation}
\mathbf{n} \cdot \mathbf{\nabla} I = \eta-\kappa_R \rho I
\end{equation}
using the short characteristic method~\citep{Olson87,Kunasz88,Stone92b,Sekora10,Davis12}.

Although there are several other algorithms~\citep{Jiang14,Rosen17}, the FLD, M1 and VET methods provide the major closure methods which are applied to the RHD simulations of radiation-pressure-driven gas~\citep{KT12,KT13,Davis14,Rosdahl15,Tsang15,ZD17,Kannan18}.

Krumholz \& Thompson~\citep{KT12} used the FLD algorithm to investigate the efficiency of momentum coupling between IR radiation and vertically stratified gas in a local 2D box and found that the radiation pressure on dust drives RRTI and supersonic turbulence. Although the radiation trapping factor $\eta$ can be big at the base of the system, a system which is initially sub-Eddington for dust can never launch an unbound wind. In the following work, Krumholz \& Thompson~\citep{KT13} turned off the gravity and prolong the 2D computational box to investigate the maximum efficiency of radiation-gas coupling using the FLD algorithm. Without gravity, a wind can be launched and accelerated, but the RRTI-like instability (without gravity, only drives by the radiation pressure) forces the gas to spread out and forms many low-density channels, which limit the momentum transfer from the radiation to a few $L/c$, regardless of the optical depth of the system. Thus, the efficiency $\eta$ is low, and lower for higher $\tau_{\rm IR}$. They concluded that an initially sub-Eddington system can never launch and accelerate winds--since most ULIRGs are below the IR Eddington limit for dust, radiation pressure is not likely to be crucial to galactic winds.

The problems studied by Krumholz \& Thompson~\citep{KT12} (hereafter the KT problems) have been revisited by many others using different algorithms based on various RHD codes. \mbox{Rosdahl \& Teyssier~\citep{Rosdahl15}} adopted the M1 closure method to study the KT problems. They found that the dusty gas receives a larger acceleration than in the FLD simulations, but the acceleration is not sufficient to overcome the gravity, and the gas eventually settles down at the bottom of the system, similar to the results in Krumholz \& Thompson~\citep{KT12}. Kannan~et al.~\citep{Kannan18} also used the M1 closure method to re-investigate the KT problems and found similar results.

\begin{figure}[H]
\centering
\includegraphics[width=16cm]{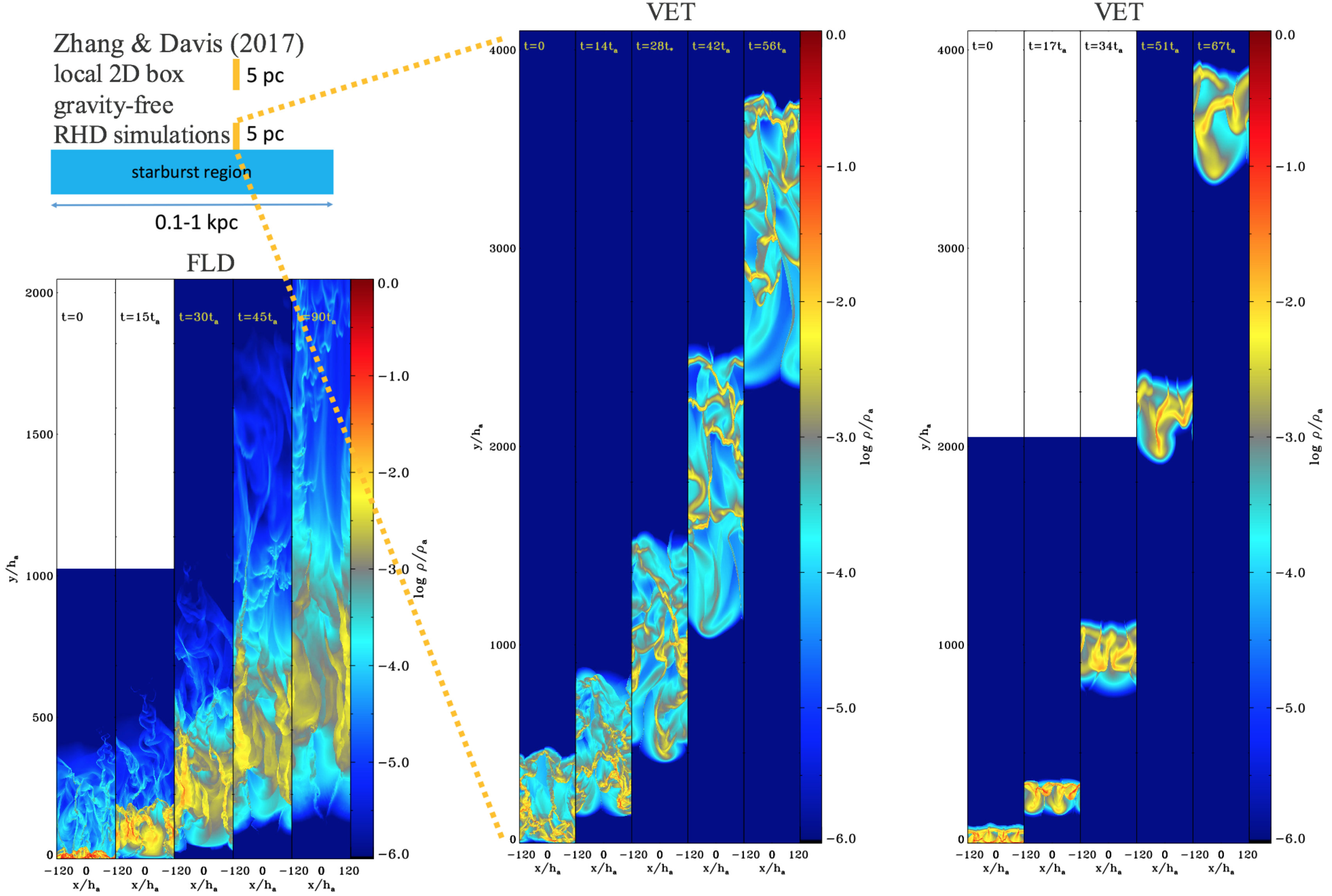}
\caption{Density snapshots in 2D radiation hydrodynamic simulations of dusty shells accelerated by radiation pressure on dust using two algorithms: the flux-limited-diffusion (FLD) method (\textbf{left panel}), the VET method (\textbf{middle and right panels}). The initial IR optical depth for dust is $\tau_{\rm IR}=3$ (\textbf{left and middle panels}), and $\tau_{\rm IR}=1$ (\textbf{right panel}), with the same radiation field injected at the bottom boundaries. The FLD run shows that radiatively driven instability creates low-density channels and the dusty gas spreads out of the entire box. The VET runs show that the dusty gas is still accelerated as a whole with more IR flux trapped in the gas, and more efficient momentum coupling between radiation and gas than that in the FLD run.}\label{fig_shell}
\end{figure}

Using the VET algorithm, Davis~et al.~\citep{Davis14} also revisited the KT problems. In contrast to the results using the FLD and M1 closure method, Davis~et al. showed a stronger momentum coupling between radiation and dusty gas. Importantly, although the RRTI develops and limits the radiation-gas interaction, an initial sub-Eddington system can turn to a super-Eddington system due to the radiation heating on dust, and the gas can still be accelerated upward by radiation and produce an unbound outflow. In the absence of gravity, Zhang~et al.~\citep{ZD17} using the VET method showed that the gas spreads out along the vertical direction, but the gas is still accelerated as a whole. In contrast to the result in Krumholz \& Thompson~\citep{KT13}, Zhang~et al.~found that the momentum transfer from the radiation to the gas is not merely a few $L/c$, but the efficiency $\eta$ $\sim$ 0.5--0.9 for a large range of $\tau_{\rm IR}$. Therefore, the moderate result  $\dot{M}v_{\infty}$ $\sim$ $[1+ (0.5-0.9)\tau_{\rm IR}]L/c$ is between the analytic estimate for spherical wind $\eta = 1$ and the FLD result $\eta\ll 1$. Figure~\ref{fig_shell} shows some detailed comparison between a FLD run and two VET runs. In summary, the RHD simulations using the VET algorithm showed more efficient momentum boost than those using the FLD and M1 closure method. Also, similar to the VET simulations, the RHD simulations based on the Monte Carlo radiation transfer scheme shows similar results~\citep{Tsang15}. Different from Krumholz \& Thompson~\citep{KT13}, Zhang~et al. proposed that radiation pressure is still important to driving dusty winds in most rapidly star-forming galaxies which are initially sub-Eddington but optically thick to the IR radiation.

In addition to the algorithm-dependent differences, the behavior of the radiation-pressure-driven dusty shells also depends on the spatial resolution of the simulations. The KT problems focus on the resolution of the pressure scale height with a resolved length to $h_* = c_s^2/g$ with $c_s$ being the sound speed in the gas, and the resolution $\sim$$h_*$ is far below the pc scales. Recently, Krumholz~\citep{Krumholz18b} studied the direct stellar radiation interacting with the dusty accretion flows and found that whether radiation feedback is able to reverse an accreting/falling flow depends on the spatial resolution of the simulations. They argued that the IR radiation pressure on dust may also be overestimated by low-resolution simulations. On the other hand, it has been tested that the behavior of a dusty shell on the scales of pressure height $h_*$ is very different from that on galaxy scales with lower resolution. The~study on the effects of spatial resolution is still underway.

\subsection{ Radiation Feedback in Star Clusters}\label{section_starcluster}


Even galaxies are below the average Eddington limit for dust, their star clusters which are much denser than the surrounding medium  may still reach or exceed the Eddington limit on small scales, e.g.,~\citep{Leroy18}. The models and calculations in Sections~\ref{section_analyticrad1}--\ref{section_starcluster} can be applied to star clusters. \mbox{Murray~et al.~\citep{Murray11}} proposed that the radiation pressure from clusters with mass $\gtrsim$$10^6\,M_{\odot}$ is able to expel the surrounding dusty gas out of the clusters and the galactic disk with velocities \mbox{$\sim$\,a few$\times$\,100\,km\,s$^{-1}$}. Thus, the cold/cool outflowing gas can be accelerated by radiation from the disk, and travel to a distance of $\sim$$50$--$100\,$kpc. These outflows may be observed as the Mg II or Na D absorbers. Zhang~et al.~\citep{Zhang18} carried out RHD simulations to model the dynamics of dusty clouds accelerated by a radiation field, and found that the dusty clouds can be accelerated to hundreds of km\,s$^{-1}$ with a significant longer lifetime compared to those entrained in a hot flow where the momentum injection is comparable to that in the radiation field.

In addition to driving galactic winds, radiation pressure from star clusters also disrupts the parent GMCs, drives turbulence and affects star formation before the explosion of the first SN. In particular, multidimensional RHD simulations have been carried out to study this problem. Skinner \& Ostriker~\citep{Skinner15}  performed simulations using the M1 closure method: the IR flux is injected from super star clusters with $\tau_{\rm IR}>1$ and interacts with a turbulence GMC. They found that radiation pressure reduces the star formation efficiency only if the dust opacity $\kappa_{\rm IR}\gtrsim 15\,{\rm cm^{-2}}\,{\rm g}$. Tsang \& Milosavljevi\'{c}~\citep{Tsang18} did similar simulations using the Monte Carlo radiation code and basically reaffirm the main conclusions in Skinner \& Ostriker~\citep{Skinner15}. Raskutti~et al.~\citep{Raskutti16,Raskutti17} also used the M1 closure method and found that the direct FUV radiation pressure from star clusters with $\tau_{\rm IR}<1$ is also important to suppressing the star formation efficiency in GMCs.

\section{Galactic Winds Driven by Cosmic Rays}\label{section_CR}
\vspace{-6pt}
\subsection{Some Early Work and Analytic Work}\label{section_CRearly}

Stellar feedback also includes cosmic rays (CRs). The shocks of SN remnants are efficient accelerators of CRs by diffusive shock acceleration (first-order Fermi acceleration)~\citep{Fermi49,Axford77,Krymskii77,Bell78a,Bell78b,Blandford78,Schlickeiser89a,Schlickeiser89b}. About $\sim$$10\%$ of the SN energy ($\sim$$10^{50}\,$ergs) can be converted into non-thermal CR energy~\citep{Helder12,Morlino12,Ackermann13}, most of which is deposited in protons with their energies following a power-law distribution peaked at $\sim\,$GeV~\citep{Strong07,Zweibel13,Grenier15,Zweibel17}. The local energy density of CRs is $\sim$$1.8\,{\rm eV}\,{\rm cm^{-3}}$~\citep{Webber98}, and the total energy of CRs in our Galaxy is $\sim$$10^{56}\,{\rm ergs}$. It appears that the GeV CRs are confined to the Galaxy for about $2\times 10^{7}\,$yr ~\citep{Garcia77,Yanasak01}, therefore the CR luminosity is $\sim$$10^{41}\,{\rm ergs}\,{\rm s^{-1}}$ in our Galaxy. It has long been proposed that CRs coupled to the gas may also drive large-scale galactic winds in the Galaxy.

Ipavich~\citep{Ipavich75} first studied the CR-driven galactic winds. He found that if CRs travel through a magnetized plasma medium faster than the Alfv\'en speed, they can be coupled to the plasma by the emission of MHD waves. The mass-loading rate of CR-driven winds can reach $\sim$$1$--$10\,M_{\odot}\,{\rm yr^{-1}}$. Breitschwerdt~et al.~\citep{Breitschwerdt91} studied the interaction of CRs and hot gas along large-scale magnetic fields and found that the combination of CR and thermal effects can drive a supersonic galactic wind with a mass loss rate of $\sim$$1\,M_{\odot}\,{\rm yr^{-1}}$ in the Galaxy. Breitschwerdt~et al.'s work was further extended by their group~\citep{Breitschwerdt93,Zirakashvili96,Ptuskin97,Breitschwerdt02}. For example, Zirakashvili~et al.~\citep{Zirakashvili96} suggested that the coupling between CRs and plasma  is provided by the resonant excitation of small-scale magnetic field fluctuations generated by CR streaming instability~\citep{Kulsrud69,Wentzel74}, and the nonlinear Alfv\'en wave damping is sufficiently strong to heat the plasma. They investigated the winds from a rotating disk galaxy and found that the mass loss rate is larger compared to that from a non-rotating and undamped model. The winds also take away a large fraction of angular momentum. Ptuskin~et al.~\citep{Ptuskin97} further explored CR streaming instability and the transport of relativistic nucleons in CR-driven winds. Everett~et al.~\citep{Everett08,Everett10a} considered galactic winds driven by the combination of both CRs and thermal gas pressure under Milky Way conditions and found that both CRs and thermal gas pressure are essential to explaining the observed Galactic diffuse soft X-ray and radio emission.

For star-forming and starburst galaxies, the total energy injected by CR protons is~\citep{Thompson07,Socrates08}
\begin{eqnarray}
L_{\rm CR} \approx 3.2\times 10^{41}\,{\rm ergs}\,{\rm s^{-1}}\,e_{51}\mu_{100}{\rm SFR_{1}}\approx 6\times 10^{-4}\,e_{51}\mu_{100}\epsilon_{-3}^{-1}L,
\end{eqnarray}
where $e$ and $\mu$ have been given in Section~\ref{section_CC85}, and $L$ is the galactic luminosity. Although $L_{\rm CR} \ll L$, the~momentum injection by CRs may still be significant. CRs scatter many times due to the interstellar magnetic irregularities. The CR escaping timescale in the Galaxy is about $2\times 10^{7}\,$yr, thus the mean free path of the GeV CRs is $\lambda_{\rm CR}$ $\sim$ $1\,$pc, corresponding to a diffusivity of $10^{28}$--$10^{29}\,{\rm cm^{2}}\,{\rm s^{-1}}$. Some~other work suggested that $\lambda_{\rm CR}$ $\sim$ $0.1\,$pc~\citep{Kulsrud05}. For a typical scale height of winds $H$ $\sim$ $1\,$kpc, the CR scattering optical depth is $\tau_{\rm CR}$ $\sim$ $H/\lambda_{\rm CR}$ $\sim$ $ 10^3-10^{4}$, and the momentum deposited by CRs is
\begin{equation}
P_{\rm CR} \sim \tau_{\rm CR}\frac{L_{\rm CR}}{c}\sim 0.6\,e_{51}\mu_{100}\epsilon_{-3}^{-1}\left(\frac{\tau_{\rm CR}}{10^3}\right)\left(\frac{L}{c}\right)\label{equ:Pcr},
\end{equation}
which is comparable to the momentum injected by the direction radiation. Socrates~et al.~\citep{Socrates08} argued that the CR feedback may be important in star-forming galaxies and starbursts to driving galactic winds. However, this estimate includes several uncertainties. The mechanism responsible for the mean free path $\lambda_{\rm CR}$ and CR diffusivity in our Galaxy are not fully understood, and $\lambda_{\rm CR}$ in other galaxies are not known. CRs are destroyed by interacting with ambient protons and producing pions with a timescale of $t_{\rm pp}$ $\sim$ $7\times 10^{7}\,{\rm yr}/(n_{\rm H}/\,{\rm cm^{-3}})$~\citep{Schlickeiser02,Thompson07}. For starbursts with $n_{\rm H}$ $\sim$ $10^{3}-10^{4}\,$cm$^{-3}$, $t_{\rm pp}$ $\sim$ $7\times 10^{3}-10^{4}\,$yr may be significantly shorter than the CR diffusion time, therefore the hadronic losses cannot be ignored. More recently, it has been found that the momentum injection from an individual SNR to the ISM can significant increase once the effect of CRs produced by the SNR is also included~\citep{Diesing18}. Numerical simulations to explore stellar feedback with the combination of SNRs and CRs are underway. In addition, how to model the impact of CRs in star-forming galaxies and starbursts is still an open question.

\subsection{Recent Numerical Simulations}

Recently, 3D hydrodynamic and MHD simulations have been performed to study the CR-driven galactic winds, with various approximations on CR transport~\citep{Uhlig12,Booth13,Hanasz13,Salem14,Fujita18,Girichidis16a,Pakmor16a,Pakmor16b,Simpson16,Pfrommer17,Ruszkowski17a,Wiener17b,Jacob18,Girichidis18,Jiang18,Farber18,Thomas18,Holguin18}. Since the properties of the CR-driven winds sensitively depend on the details of CR transport, we first look to the set of two-fluid MHD equations which describe the winds
\begin{eqnarray}
&&\frac{\partial \rho}{\partial t} + \bnabla \cdot (\rho {\bf u})  =  \dot{m}_{\rm other},\label{equ:massCR}\\
&&\frac{\partial \rho {\bf u}}{\partial t} + \nabla \cdot \left( \rho {\bf u}{\bf u} + P_{\rm tot}{\bf I}-\frac{{\bf B}{\bf B}}{4\pi} \right)   =  \rho {\bf g} + {\dot P}_{\rm other}, \\
&&\frac{\partial {\bf B}}{\partial t}--\bnabla \times ({\bf u}\times {\bf B})   =  0, \\
&&\frac{\partial e}{\partial t}  + \bnabla \cdot \left[ {\bf u}(e_{\rm cr}+P)--\frac{({\bf u} \cdot {\bf B}){\bf B}}{4\pi} \right] \nonumber \\
&& \quad\quad\quad\quad =  \rho {\bf u}_{g}\cdot {\bf g}--{\bf u} \cdot [\bnabla(P_{\rm cr}+P_{\rm A})]+S_{\rm A}+ \dot{u}_{\rm other}, \\
&&\frac{\partial e_{\rm cr}}{\partial t} + \bnabla \cdot [{\bf u}(e_{\rm cr}+P_{\rm cr}) + {\bf F}_{\rm cr}]
= {\bf u} \cdot \bnabla P_{\rm cr}+ Q_{\rm cr} + Q_{\rm had}+Q_{\rm Coul}+\left. \frac{\partial e_{\rm cr}}{\partial t}  \right\rvert_{\rm scatt} \label{equ:energyCR},
\end{eqnarray}
where $\rho,{\rm u}$, ${\rm B}$ and ${\bf g}$ are the gas density, velocity, the local magnetic field and gravity, respectively. The~total pressure, MHD pressure and the total MHD density are
\begin{eqnarray}
&& P_{\rm tot} = P_{\rm th}+\frac{{\bf B}{\bf B}}{8\pi}+P_{\rm cr}+P_{\rm A}, \\
&& P= P_{\rm th}+ \frac{{\bf B}{\bf B}}{8\pi}, \\
&& e = \frac{1}{2}\rho {\bf u}{\bf u}+ e_{\rm gas}+ \frac{{\bf B}{\bf B}}{8\pi}
\end{eqnarray}
with $P_{\rm th}$, $P_{\rm cr}$ being the thermal and CR pressure, and $P_{\rm A}$ is contributed by the Alfv\'en wave. The CR energy source term $Q_{\rm cr}$ is the heating rate from SNe ($\sim$$10\%$ of the SN energy), $Q_{\rm had}$ is the hadronic losses~\citep{Pfrommer17,Girichidis18}, and $Q_{\rm Coul}$ is the Coulomb losses due to the CRs interaction with the ambient gas~\citep{Guo08}. Other variables are discussed as follows:
\begin{enumerate}

\item[(1)] {\bf Other Sources.} The mass, momentum and energy injection rates by other sources depend on the models of galaxies. Here $\dot{m}_{\rm other}$ is the combination of the mass injection rate by sources including jets, stellar and SN-driven winds subtracting the mass used for star formation e.g.,~\citep{Farber18}. The~momentum injection rate $\dot{P}_{\rm other}$ can be provided by SNe or radiation in the absence of AGNs. The energy rate $\dot{u}_{\rm other}$ is also given by SNe, and radiative cooling needs to be subtracted from this term.

\item[(2)] {\bf Alfv\'en Wave Damping.} As discussed by early theoretical work e.g.,~\citep{Breitschwerdt91,Zirakashvili96,Ptuskin97,Everett08, Everett10a}, the energy source term $S_{\rm A}$ is caused by Alfv\'en wave damping, which is still not completely understood~\citep{Kulsrud05, Zweibel17,Thomas18}. The damping mechanisms include the ion-neutral damping, which~is caused by the friction between ions and neutrals~\citep{DePontieu01,Khodachenko04,Khodachenko06,Soler16}, linear and nonlinear Landau damping~\citep{Malmberg67,Hollweg71,Lee73,McKenzie83,Miller91,Mouhot09}, and turbulent damping~\citep{Farmer04,Yan04,Mouhot09,Zweibel13,Lazarian16,Zweibel17}. Everett \& Zweibel~\citep{Everett11} considered both ion-neutral and nonlinear Landau damping, and found that these damping mechanisms are only important if the magnetic fields are above $\gtrsim$$10\,\mu$G, or high CR pressure ($\sim$$10^{-11}\,{\rm ergs} \,{\rm cm^{-3}}$). The CR diffusion also depends on the wave damping.

\item[(3)]  {\bf CR Diffusion.} The scattering term $\left. {\partial e_{\rm cr}}/{\partial t}  \right\rvert_{\rm scatt}$ in Equation (\ref{equ:energyCR}) includes CR diffusion and streaming. The diffusion term is usually written as $\bnabla \cdot (\bkappa \cdot \bnabla e_{\rm cr})$, where $\bkappa$ is the diffusion coefficient. For isotropic CR diffusion, the diffusion term can be rewritten as~\citep{Yang12}
\begin{equation}
\bnabla \cdot (\bkappa \cdot \bnabla e_{\rm cr}) = \bnabla \cdot (\kappa_{\parallel} {\bf b}{\bf b} \cdot e_{\rm cr})+  \bnabla \cdot [\kappa_{\perp} ({\bf I}-{\bf b}{\bf b}) \cdot e_{\rm cr}].
\end{equation}
where $\bkappa$ is the diffusion tensor, the  $\kappa_{\parallel}$ and $\kappa_{\perp}$ are the diffusion coefficients parallel and perpendicular to the magnetic field, and ${\bf b}={\bf B}/|{\bf B}|$ is the unity vector along the magnetic~field.
\end{enumerate}

If magnetic field lines are sufficiently tangled on small scales, CR diffusion can be approximated  to be isotropic~\citep{Yang12}. As mentioned in Section~\ref{section_CRearly}, the value of diffusion coefficient in our Galaxy can be estimated by $\kappa$ $\sim$ $H^2/t_{\rm CR}$ with $H$ $\sim$ $1\,$kpc and the CR traveling time  $t$ $\sim$ $10^7\,$yr so $\kappa$ $\sim$ $3\times 10^{28}\,{\rm cm^{2}}\,{\rm g^{-1}}$. However, CR diffusion is considered to be anisotropic in most cases, and the coefficient $\kappa_{\perp}$ is found to be much lower than the parallel one $\kappa_{\parallel} \gg \kappa_{\perp}$. The diffusivity also depends on the rigidity and may vary in the multiphase medium. The rigidity is $R_{\rm i} =r_{\rm L}B c$ with $r_{\rm L}$ being the gyroradius, \mbox{and $\kappa_{\parallel} \propto R_{\rm i}^{0.6}$~\citep{Swordy90,Shalchi05}.} Farber~et al.~\citep{Farber18} suggested that $\kappa$ depends on the gas temperature and used $\kappa_{\parallel}=10^{29}\,{\rm cm^{2}}\,{\rm s^{-1}}$ for $T<10^{4}\,{\rm K}$, and $\kappa_{\parallel}=3\times10^{27}\,{\rm cm^{2}}\,{\rm s^{-1}}$ for $T\gtrsim 10^{4}\,{\rm K}$.

On the other hand, it is unclear if the Galactic diffusivity can be applied to other galaxies. One example is NGC 253. Heesen~et al.~\citep{Heesen09a,Heesen09b,Heesen11} investigated the magnetic fields in NGC 253 and estimated that the average diffusion coefficient is $\kappa$ $\sim$ $2\times 10^{29}\,{\rm cm^{2}}\,{\rm s^{-1}}$, \mbox{and $\kappa_{\perp}\approx 1.5\times 10^{28}\,{\rm cm^{2}}\,{\rm s^{-1}}E({\rm GeV})^{0.5\pm 0.7}$,} which seems to be one order of magnitude higher than the values in our Galaxy. The comparison between the observed values for NGC 253 and the theoretical model in~\citep{Shalchi10} was discussed in~\citep{Buffie13}. The estimate of the CR diffusivity in other galaxies has also been explored~\citep{Heesen16}.

\begin{enumerate}

\item[(4)] {\bf CR Streaming.} A widely used CR streaming model for galactic wind simulations calculates the streaming velocity as ${\bf v}_{\rm st} =--{\bf v}_{\rm A}({\bf B} \cdot \bnabla P_{\rm cr})/|{\bf B} \cdot \bnabla P_{\rm cr}|$ with ${\bf v}_{\rm A}$ being the Alfv\'en velocity, and the energy term $\left. {\partial e_{\rm cr}}/{\partial t}  \right\rvert_{\rm scatt}$ contributed by streaming is $-|{\bf v}_{\rm A} \cdot \bnabla P_{\rm cr}|$. The CR energy flux due to streaming is given by ${\bf F}_{\rm cr} = {\bf v}_{\rm st}(E_{\rm cr}+P_{\rm cr})$, and the total CR energy flux including the diffusive flux along the direction of local magnetic field is\mbox{ ${\bf F}_{\rm cr}=(E_{\rm cr}+P_{\rm cr})({\bf v}+{\bf v}_{\rm st})-{\bf b} \kappa({\bf b} \cdot \bnabla P_{\rm cr})$}. In the self-confinement scenario, in which CRs scatter on waves excited by the stream\mbox{ instability~\citep{Kulsrud69,Wentzel74,Kulsrud05,Zweibel13},} an effective draft speed $f{\bf v}_{\rm st}$ is used to replace $v_{\rm st}$, and the factor $f$ can be calculated by balancing the wave growth rate with the wave damping rate~\citep{Ruszkowski17a,Wiener13,Wiener17a}.
\end{enumerate}

Another way to calculate $F_{\rm cr}$ is to directly solve it by adding a new equation and some closure relations to the set of Equations (\ref{equ:massCR}) to (\ref{equ:energyCR}). Before we introduce the more sophisticated algorithms to solve the equations for CR-driven winds, we briefly review a series of recent 3D global and local stratified numerical simulations in the literature. All the simulations include some physical processes introduced above but ignored others. Most simulations showed positive results that the CR feedback drives galactic winds in various galaxies. 
Uhlig~et al.~\citep{Uhlig12} studied the impact of CR streaming at the local sound speed, and found that powerful winds are observed in dwarf galaxies ($\sim$$10^{9}\,M_{\odot}$) and low-mass galaxies ($\lesssim$$10^{11}M_{\odot}$), while fountain flows are driven in larger halos ($\gtrsim$$10^{11}M_{\odot}$). This halo-dependent conclusion is confirmed by Jacobi~et al.~\citep{Jacob18}.\mbox{ Hanasz~et al.~\citep{Hanasz13}} considered anisotropic diffusion of CRs in star-forming ($40\,M_{\odot}\,{\rm yr^{-1}}$) disk galaxies with gas surface density $\Sigma_{\rm g}$ $\sim$ $100\,M_{\odot}\,{\rm pc^{-2}}$, and found that CRs alone can trigger the formation of strong winds. \mbox{Booth~et al.~\citep{Booth13}} assumed isotropic diffusion and compare their simulations to Milky Way and Small Magellanic Cloud (SMC)-like galaxies, and found that the mass-loading rate is up to $\beta_{\rm CR}$ $\sim$ $10$ in dwarf systems, in which warm gas $\sim$$10^{4}\,$K is dominated the winds. Salem \& Bryan~\citep{Salem14} also assumed isotropic diffusion and found stable winds with $\beta_{\rm CR}$ $\sim$ $0.3$ from $\sim$$10^{12}M_{\odot}$ halos. Interestingly, F\mbox{ujita~et al.~\citep{Fujita18}} did similar simulations as in Salem \& Bryan~\citep{Salem14} but found no significant difference in mass-loading rates of SN-driven winds with or without CR pressure. Pakmor~et al.~\citep{Pakmor16a,Pakmor16b} argued that anisotropic diffusion is more reliable than the isotropic approximation for galaxies with $10^{11}\,M_{\odot}$ halos. Girichidis~et al.~\citep{Girichidis16a} performed the first MHD simulations in stratified boxes to model CR-driven winds from the Milky-Way-like disks $\Sigma_{\rm g}$ $\sim$ $10\,M_{\odot}\,{\rm pc^{-2}}$ with SNe, assuming anisotropic diffusion for CRs. They also included hadronic cooling in~\citep{Girichidis18} and found that $5$--$25\%$ of the injected CR energy cools via hadronic losses. Ruszkowski~et al.~\citep{Ruszkowski17a} included both the CR (self-confinement) streaming and anisotropic diffusion in their MHD model and found that the presence of moderately super-Alfv\'en CR streaming enhances the efficiency of galactic winds from galaxies with $10^{12}\,M_{\odot}$ halos. Farber~et al.~\citep{Farber18} extended Ruszkowski~et al.~\citep{Ruszkowski17a}'s work by considering temperature-dependent anisotropic diffusion and observed the decoupling of CRs in the cold/cool neutral ISM (<$10^{4}\,$K). CRs propagate faster in the cold/cool gas than in the ionized medium. Holguin~et al.~\citep{Holguin18} further developed a more elaborate model on CR streaming ${\bf u}_{\rm st}=f{\bf u}_{\rm A}$ with $f$ depending on the turbulent Mach number, and investigated the effects of turbulence damping. Wiener~et al.~\citep{Wiener17b} investigated the relative importance of CR diffusion vs streaming and found significant differences between these two mechanisms in dwarf galaxies ($\sim$$10^{10}\,M_{\odot}$). On the other hand, the heating by CR-driven winds in CGM and clusters of galaxies has also been explored e.g.,~\citep{Guo08,Salem16,Ruszkowski17b,Butsky18}

We expect that future numerical simulations of the CR feedback may use more sophisticated algorithms for CR transport. Returning to the CR Vlasov equation, the advection-diffusion equation for CR distribution function is~\citep{Skilling71,Jiang18}
\begin{equation}
\frac{\partial  f_{\rm p}}{\partial t}+({\bf v}+{\bf v}_{\rm st})\cdot\bnabla  f_{\rm p}=
\bnabla\cdot(\kappa_{\rm p} {\bf b}{\bf b}\cdot\bnabla  f_{\rm p})
+\frac{1}{3}p\frac{\partial  f_{\rm p}}{\partial p} \bnabla\cdot ({\bf v}+{\bf v}_s)+ Q.
\label{eqn:crevol}
\end{equation}

Equation (\ref{eqn:crevol}) is similar to the Radiative Transfer (RT) equation. Jiang \& Oh~\citep{Jiang18} adopted a method similar to the two-moment methods for solving the RT equation. The zeroth momentum equation of Equation (\ref{eqn:crevol}) is the CR energy Equation (\ref{equ:energyCR}), and the first momentum equation using the reduced speed of light approximation is
\begin{equation}
\frac{1}{\tilde{c}^2}\frac{\partial {\bf F}_{\rm cr}}{\partial t} +\bnabla P_{\rm cr} = -\frac{1}{\kappa} [{\bf F}_{\rm cr}-{\bf u}(e_{\rm cr}+P_{\rm cr})]\label{equ:momentCR}.
\end{equation}

Please note that we have reduced their ${\sf P}_{\rm cr}$ tensor to $P_{\rm cr}$, $\kappa$ is the “equivalent” diffusion coefficient, and~the technique to use the reduced speed of light ($\tilde{c}<c$) can be also found in  RT \mbox{algorithms~\citep{Skinner13,Jiang14,Zhang18}}. As a result, the CR streaming energy flux no longer holds ${\bf F}_{\rm cr}=(E_{\rm cr}+P_{\rm cr})({\bf v}+{\bf v}_{\rm st})$, \mbox{but Equation (\ref{equ:momentCR})} is added to the original set of MHD equations to solve the CR transport. Jiang \& Oh~\citep{Jiang18} found that their two-moment method is more stable and robust to handle CR streaming. Recently, \mbox{Thomas \& Pfrommer~\citep{Thomas18}} used the Eddington approximation of RT to develop another two-moment algorithm of CT transport in the self-confinement picture, including both the CR streaming, diffusion, and other damping mechanisms. The CR energy density, flux and Alfv\'en wave density can be solved together by the set of MHD equations. Compared to Jiang \&  Oh~\citep{Jiang18}'s method which is correct to the order of $v_{\rm A}/c$, the CR transport algorithm by Thomas \& Pfrommer~\citep{Thomas18} reaches the order of $(v_{\rm A}/c)^2$. These two-moment algorithms are potentially better to model the CR feedback and CR-driven winds.


One may ask whether CRs can accelerate cool/cloud clouds~\citep{Everett11,Wiener17a,Mao18}. It is found that the CR pressure gradient can push the cool/cloud clouds outward, but CRs can be decoupled with cool/cold gas. It seems that the velocities gained by the pressure gradient is not sufficient to match the observed velocities~\citep{Wiener17a}. Please note that most of the CR-driven wind simulations are on galactic scales, it is also important to perform small-scale simulations to better understand the interaction between CRs and multiphase gas especially cool/cold clouds.

\section{Conclusions and Discussion}\label{section_conclusions}

I have reviewed the theory of galactic winds driven by stellar feedback processes, including~supernova (SN) explosions, radiation pressure from starlight on dust grains, \mbox{and cosmic rays.}

The observed galactic winds are multiphase, which can be divided into very hot ($\sim$$10^8\,$K), hot~($\sim$$10^6$--$10^7\,$K), warm ionized ($\sim$$10^{4}\,$K), neutral atomic ($\sim$$10^{3}\,$K) and cold molecular and dust ($\lesssim$$100\,$K) phases. The comparison between theory and growing observations is still incomplete.

Core-collapse SN explosions have long been considered as the primary energy and momentum sources of galactic winds. A SN-driven large-scale galactic winds can be approximated described by the 1D spherical solution of the CC85 model~\citep{CC85}, which is controlled by two key parameters: the thermalization efficiency that measures the thermal energy converted from the net SN explosions to the wind, and the mass-loading rate. These two parameters can be constrained using the X-ray data from star-forming and starburst galaxies, or by hydrodynamic simulations. Please note that the CC85 model is for adiabatic flows, radiative cooling may significantly change the properties of winds above a distance of $\sim$$1$--$10\,$kpc.

A prevailing scenario of the multiphase galactic winds is that the multiphase material is advected into a SN-driven hot wind and accelerated by the ram pressure of the hot wind to a velocity of a few hundred km\,s$^{-1}$. This scenario is supported by some global simulations on galactic scales and some observations of the association between multiphase gas, but recent numerical simulations on smaller scales found that clouds entrained in a hot flow are completely shredded via hydrodynamic instabilities long before being fully accelerated by the hot wind. Magnetic fields may be important to prolonging the lifetime of clouds, but more work needs to be done to explore the cloud dynamics in hot flows as well as large-scale interaction between hot flows and the multiphase ISM in a wide range of galaxies with more realistic structure of magnetic fields. Another possible origin of warm/cool outflowing gas arises from thermal instability of the hot winds.

On the other hand, radiation pressure from starlight on dust grains can be important in rapidly star-forming and starburst galaxies. Massive stars emit UV photons, which are absorbed by dust and re-radiated in IR from dust grains. Many galaxies are above the Eddington limit for UV photons, but~only galaxies with flux $\gtrsim$$10^{13}\,L_{\odot}\,$kpc$^{-2}$ may reach the Eddington limit for IR photons. The~estimate of the Eddington limit is uncertain in the dust-to-gas ratio. However, even a galaxy is globally sub-Eddington for dust, its star clusters which are much denser than the surrounding medium may still reach the Eddington limit, and drive dusty outflows from them. The dusty gas from star clusters may be eventually pushed out of the galaxies by the momentum injection of SNe and radiation pressure. The analytic models have been proposed to describe the geometrically thin dusty shells pushed by radiation pressure, while the behavior of dusty gas from a self-gravitating disk is different from that from a spherical system. The momentum coupling between the dust and UV photons can be modeled using the single-scattering limit, in which each photon is only scattered or absorbed once. The coupling between the dust and IR photons can be given by $P_{\rm rad}\approx (1+\eta \tau_{\rm IR})L/c$, with $\eta\sim$$0.5$--$0.9$, obtained~from recent radiation hydrodynamic simulations using the state-of-the-art algorithms. According to the simulations, an initially sub-Eddington system may still launch an unbound wind by radiation pressure on dust. In addition to driving galactic winds, radiation pressure can also disrupt GMCs, which is beyond the scope of this review article.

Whether CRs are important to driving winds in star-forming and starburst galaxies is less unclear, as most of the CR knowledge is gained from our Galaxy. For example, the CR diffusivity may be different between our Galaxy and starbursts such as NGC 253. Taking the CR properties in Milky Way as the fiducial values, CRs are no less important than SNe or radiation pressure to providing momentum injection and coupling to the gas in various galaxies. Cool/cold gas may be decoupled with CRs but still be accelerated by the pressure gradient generated by CR transport. Recently, a series of hydrodynamic and MHD simulations have been carried out to explore the impact of CR streaming and diffusion, primarily focusing on Milky Way-like galaxies, then extending to star-forming and high-redshift galaxies. Two-moment algorithms used for radiative transfer may also be applied to solve the CR transport equations. We expect the two-moment algorithms can be used to model the CR-driven winds in the near future. Also, the impact of CRs on rapidly star-forming and starburst galaxies need to be explored more detailed.

A more realistic model for galaxy evolution and galactic winds needs to combine all feedback mechanisms including stellar and black hole/AGN feedback together, as many galaxies include both rapidly star-forming regions as well as central AGNs. For stellar feedback alone, some work has already combined SN and radiation pressure feedback, or SN and CR feedback together, and attempted to build a unified model for stellar-driven galactic winds. However, the details of galactic winds need to be investigated systematically and compared to the multiwavelength observations more carefully. For~example, can the momentum flux from a momentum-driven wind be boosted by an optically thick system with multiple scattering? Although we found that numerical algorithm is important to model the dynamics and thermal properties of momentum-driven winds, recently new evidence showed that spatial resolution may be crucial to modeling SN and radiation pressure, and sub-grid models need to be adopted for numerical simulations on galactic scales. Also, geometry effect may be important: feedback simulation in a vertically stratified box shows different result from that in a global simulation with the same setup. It has been found that the momentum injection by stellar feedback may drive turbulence as well as launching and accelerating winds, so how much momentum is transferred to turbulence and how much to galactic winds? Among SNe, radiation feedback and gravity, which is the most important mechanism to drive turbulence and regulate star formation on various scales? These~ questions need to be further studied in the near future.

\vspace{6pt}

\funding
{This research was supported by NSF grant AST-1616171.}

\acknowledgments
{I would like to thank the referees for giving useful comments which improved the paper. I also thank Shane Davis, Fulai Guo for helpful discussion. }

\conflictsofinterest
{The authors declare no conflict of interest.}


\reftitle{References}



\begin{thebibliography}{999}
\bibitem[Lynds \& Sandage(1963)]{Lynds63}Lynds, C.R.; Sandage, A.R. Evidence for an Explosion in the Center of the Galaxy M82. {\em Astrophys. J.} {\bf 1963}, {\em 137}, 1005--1021, doi:10.1086/147579.

\bibitem[Burbidge~et al.(1964)]{Burbidge64}Burbidge, E.M.; Burbidge, G.R.; Rubin, V.C. A Study of the Velocity Field in M82 and its Bearing on Explosive Phenomena in that Galaxy. {\em Astrophys. J.} {\bf 1964}, {\em 140}, 942--968, doi:10.1086/147997.

\bibitem[Schmidt(1963)]{Schmidt63}Schmidt, M. 3C 273: A Star-Like Object with Large Red-Shift. {\em Nature} {\bf 1963}, {\em 197}, 1040, doi:10.1038/1971040a0.

\bibitem[Hoyle \& Fowler(1963a)]{Hoyle63a}Hoyle, F.; Fowler, W.A. Nature of Strong Radio Sources. {\em Nature} {\bf 1963}, {\em 197}, 533--535, doi:10.1038/197533a0.

\bibitem[Hoyle \& Fowler(1963b)]{Hoyle63b}Hoyle, F.; Fowler, W.A. On the nature of strong radio sources. {\em Mon.~Not.~R.~Astron.~Soc.} {\bf 1963}, {\em 125}, 169--176, doi:10.1093/mnras/125.2.169.

\bibitem[Salpeter(1964)]{Salpeter64}Salpeter, E.E. Accretion of Interstellar Matter by Massive Objects. {\em Astrophys.~J.} {\bf 1964}, {\em 140}, 796--800, doi:10.1086/147973.

\bibitem[Zel'dovich(1964)]{Zeldovich64}Zel'dovich, Y.B. The Fate of a Star and the Evolution of Gravitational Energy Upon Accretion. {\em Dokl.~Phys.} {\bf 1964}, {\em 9}, 195.

\bibitem[Lynden-Bell(1969)]{Lynden69}Lynden-Bell, D. Galactic Nuclei as Collapsed Old Quasars. {\em Nature} {\bf 1969}, {\em 223}, 690--694, doi:10.1038/223690a0.

\bibitem[Lynden-Bell \& Rees(1971)]{Lynden71}Lynden-Bell, D.; Rees, M.J. On quasars, dust and the galactic centre. {\em Mon.~Not.~R.~Astron.~Soc.} {\bf 1971}, {\em 152}, 461--475, doi:10.1093/mnras/152.4.461.

\bibitem[Veilleux~et al.(2005)]{Veilleux05}Veilleux, S.; Cecil, G.; Bland-Hawthorn, J. Galactic Wind. {\em Annu. Rev. Astron. Astrophys.} {\bf 2005}, {\em 43}, 769--826, doi:10.1146/annurev.astro.43.072103.150610.

\bibitem[Heckman \& Thompson(2017)]{HT17}Heckman, T.M.; Thompson, T.A. A Brief Review of Galactic Winds. {\em arxiv} {\bf 2017}, arXiv:1701.09062

\bibitem[Naab \& Ostriker(2017)]{Naab17}Naab, T.; Ostriker, J.P. Theoretical Challenges in Galaxy Formation. {\em Annu. Rev. Astron. Astrophys.} {\bf 2017}, {\em 55}, 59--109, doi:10.1146/annurev-astro-081913-040019.

\bibitem[Dekel \& Silk(1986)]{DS86}Dekel, A.; Silk, J. The origin of dwarf galaxies, cold dark matter, and biased galaxy formation. {\em Astrophys.~J.} {\bf 1986}, {\em 303}, 39--55, doi:10.1086/164050.

\bibitem[Leitherer~et al.(1992)]{Leitherer92}Leitherer, C.; Robert, C.; Drissen, L. Deposition of mass, momentum, and energy by massive stars into the interstellar medium. {\em Astrophys.~J.} {\bf 1992}, {\em 401}, 596--617, doi:10.1086/172089.

\bibitem[Heckman~et al.(1993)]{Heckman93}Heckman, T.M.; Lehnert M.D.; Armus L. \emph{Galactic Superwinds}; Shull, J.M., Thronson, H.A., Jr., Eds; The Environment and Evolution of GalaxiesL Kluwer: Dordrecht, The Netherlands, {2017}; {Volume 188}, p. 455, doi:10.1007/978-94-011-1882-8\_25.

\bibitem[Tomisaka \& Bregman(1993)]{Tomisaka93}Tomisaka, K.; Bregman, J.N. Extended hot-gas halos around starburst galaxies. {\em Publ.~Astron.~Soc.~Jpn.} {\bf 1993}, {\em 45}, 513--528.

\bibitem[Suchkov~et al.(1994)]{Suchkov94}Suchkov, A.A.; Balsara, D.S.; Heckman, T.M.; Leitherer, C. Dynamics and X-ray emission of a galactic superwind interacting with disk and halo gas. {\em Astrophys.~J.} {\bf 1994}, {\em 430}, 511--532, doi:10.1086/174427.

\bibitem[Leitherer \& Heckman(1995)]{Leitherer95}Leitherer, C.; Heckman, T.M. Synthetic properties of starburst galaxies. {\em Astrophys. J. Suppl. Ser.} {\bf 1995}, {\em 96}, 9--38, doi:10.1086/192112.

\bibitem[Heckman~et al.(2000)]{Heckman00}Heckman, T.M.; Lehnert, M.D.; Strickland, D.K.; Armus, L. Absorption-Line Probes of Gas and Dust in Galactic Superwinds. {\em Astrophys. J. Suppl. Ser.} {\bf 2000}, {\em 129}, 493--516, doi:10.1086/313421.

\bibitem[Magorrian~et al.(1998)]{Magorrian98}Magorrian, J.; Tremaine, S.; Richstone, D. The Demography of Massive Dark Objects in Galaxy Centers. {\em Astron. J. } {\bf 1998}, {\em 115}, 2285--2305, doi:10.1086/300353.

\bibitem[Silk \& Rees(1998)]{Silk98}Silk, J.; Rees, M.J. Quasars and galaxy formation. {\em Astron.~Astrophys.} {\bf 1998}, {\em 331}, L1--L4.

\bibitem[Haehnelt~et al.(1998)]{Haehnelt98}Haehnelt, M.G.; Natarajan, P.; Rees, M.J. High-redshift galaxies, their active nuclei and central black holes. {\em Mon.~Not.~R.~Astron.~Soc.} {\bf 1998}, {\em 300}, 817--827, doi:10.1046/j.1365-8711.1998.01951.x.

\bibitem[Ferrarese \& Merritt(2000)]{Ferrarese00}Ferrarese, L.; Merritt, D. A Fundamental Relation between Supermassive Black Holes and Their Host Galaxies. {\em Astrophys. J. Lett.} {\bf 2000}, {\em 539}, L9--L12, doi:10.1086/312838.

\bibitem[Gebhardt~et al.(2000)]{Gebhardt00}Gebhardt, K.; Bender, R.; Bower, G. A Relationship between Nuclear Black Hole Mass and Galaxy Velocity Dispersion. {\em Astrophys. J. Lett.} {\bf 2000}, {\em 539}, L13--L16, doi:10.1086/312840.


\bibitem[Fabian(2012)]{Fabian12}Fabian, A.C. Observational Evidence of Active Galactic Nuclei Feedback. {\em Annu. Rev. Astron. Astrophys.} {\bf 2012}, {\em 50}, 455--489, doi:10.1146/annurev-astro-081811-125521.

\bibitem[Kormendy \& Ho(2013)]{Kormendy13}Kormendy, J.; Ho, L.C. Coevolution (Or Not) of Supermassive Black Holes and Host Galaxies. {\em Annu. Rev. Astron. Astrophys.} {\bf 2013}, {\em 51}, 511--653, doi:10.1146/annurev-astro-082708-101811.

\bibitem[Heckman \& Best(2014)]{Heckman14}Heckman, T.M.; Best, P.N. The Coevolution of Galaxies and Supermassive Black Holes: Insights from Surveys of the Contemporary Universe. {\em Annu. Rev. Astron. Astrophys.} {\bf 2014}, {\em 52}, 589--660, doi:10.1146/annurev-astro-081913-035722.

\bibitem[King \& Pounds(2015)]{King15}King, A.; Pounds, K. Powerful Outflows and Feedback from Active Galactic Nuclei. {\em Annu. Rev. Astron.~Astrophys.} {\bf 2015}, {\em 53}, 115--154, doi:10.1146/annurev-astro-082214-122316.

\bibitem[Di Matteo~et al.(2005)]{Matteo05}Di Matteo, T.; Springel, V.; Hernquist, L. Energy input from quasars regulates the growth and activity of black holes and their host galaxies. {\em Nature} {\bf 2005}, {\em 433}, 604--607, doi:10.1038/nature03335.

\bibitem[Zubovas~et al.(2013)]{Zubovas13}Zubovas, K.; Nayakshin, S.; King, A.; Wilkinson, M. AGN outflows trigger starbursts in gas-rich galaxies. {\em Mon.~Not.~R.~Astron.~Soc.} {\bf 2013}, {\em 433}, 3079--3090, doi:10.1093/mnras/stt952.

\bibitem[Bieri~et al.(2017)]{Bieri17}Bieri, R.; Dubois, Y.; Rosdahl, J.; Wagner, A.; Silk, J.; Mamon, G.A. Outflows driven by quasars in high-redshift galaxies with radiation hydrodynamics. {\em Mon.~Not.~R.~Astron.~Soc.} {\bf 2017}, {\em 464}, 1854--1873, doi:10.1093/mnras/stw2380.


\bibitem[Croton~et al.(2006)]{Croton06}Croton, D.J.; Springel, V.; White, S.D.M. The many lives of active galactic nuclei: Cooling flows, black holes and the luminosities and colours of galaxies. {\em Mon.~Not.~R.~Astron.~Soc.} {\bf 2006}, {\em 365}, 11--28, doi:10.1111/j.1365-2966.2005.09675.x.

\bibitem[Bower~et al.(2006)]{Bower06}Bower, R.G.; Benson, A.J.; Malbon, R. Breaking the hierarchy of galaxy formation. {\em Mon.~Not.~R.~Astron.~Soc.} {\bf 2006}, {\em 370}, 645--655, doi:10.1111/j.1365-2966.2006.10519.x.

\bibitem[McNamara \& Nulsen(2007)]{McNamara07}McNamara, B.R.; Nulsen, P.E.J. Heating Hot Atmospheres with Active Galactic Nuclei. {\em Annu. Rev. Astron.~Astrophys.} {\bf 2007}, {\em 45}, 117--175, doi:10.1146/annurev.astro.45.051806.110625.

\bibitem[Okamoto~et al.(2008)]{Okamoto08}Okamoto, T.; Nemmen, R.S.; Bower, R.G. The impact of radio feedback from active galactic nuclei in cosmological simulations: Formation of disc galaxies. {\em Mon.~Not.~R.~Astron.~Soc.} {\bf 2008}, {\em 385}, 161--180, doi:10.1111/j.1365-2966.2008.12883.x.

\bibitem[Gaibler~et al.(2011)]{Gaibler11}Gaibler, V.; Khochfar, S.; Krause, M. Asymmetries in extragalactic double radio sources: Clues from 3D simulations of jet-disc interaction. {\em Mon.~Not.~R.~Astron.~Soc.} {\bf 2011}, {\em 411}, 155--161, doi:10.1111/j.1365-2966.2010.17674.x.

\bibitem[Guo \& Mathews(2012)]{Guo12}Guo, F.; Mathews, W.G. The Fermi Bubbles. I. Possible Evidence for Recent AGN Jet Activity in the Galaxy. {\em Astrophys.~J.} {\bf 2012}, {\em 756}, 181, doi:10.1088/0004-637X/756/2/181.

\bibitem[Guo(2016)]{Guo16}Guo, F. On the Importance of Very Light Internally Subsonic AGN Jets in Radio-mode AGN Feedback. {\em Astrophys.~J.} {\bf 2016}, {\em 826}, 17, doi:10.3847/0004-637X/826/1/17.

\bibitem[Guo~et al.(2018)]{Guo18}Guo, F.; Duan, X.; Yuan, Y.-F. Reversing cooling flows with AGN jets: Shock waves, rarefaction waves and trailing outflows. {\em Mon.~Not.~R.~Astron.~Soc.} {\bf 2018}, {\em 473}, 1332--1345, doi:10.1093/mnras/stx2404.

\bibitem[Duan~et al.(2018)]{Duan18}Duan, X.; Guo, F. Metal-rich Trailing Outflows Uplifted by AGN Bubbles in Galaxy Clusters. {\em Astrophys.~J.} {\bf 2018}, {\em 861}, 106, doi:10.3847/1538-4357/aac9ba.


\bibitem[Salom\'{e}~et al.(2006)]{Salome06}Salom\'{e}, P.; Combes, F.; Edge, A.C. Cold molecular gas in the Perseus cluster core. Association with X-ray cavity, H$\alpha$ filaments and cooling flow. {\em Astron.~Astrophys.} {\bf 2006}, {\em 454}, 437--445.

\bibitem[Guo \& Oh(2009)]{Guo09}Guo, F.; Oh, S.P. Could AGN outbursts transform cool core clusters? {\em Mon.~Not.~R.~Astron.~Soc.} {\bf 2009}, {\em 400}, 1992--1999, doi:10.1111/j.1365-2966.2009.15592.x.

\bibitem[McNamara \& Nulsen(2012)]{McNamara12}McNamara, B.R.; Nulsen, P.E.J. Mechanical feedback from active galactic nuclei in galaxies, groups and clusters. {\em New J. Phys.} {\bf 2012}, {\em 14}, 055023, doi:10.1088/1367-2630/14/5/055023.

\bibitem[Gaspari~et al.(2012)]{Gaspari12}Gaspari, M.; Ruszkowski, M.; Sharma, P. Cause and Effect of Feedback: Multiphase Gas in Cluster Cores Heated by AGN Jets. {\em Astrophys. J.} {\bf 2012}, {\em 746}, 94, doi:10.1088/0004-637X/746/1/94.

\bibitem[Ciotti~et al.(2017)]{Ciotti17}Ciotti, L.; Pellegrini, S.; Negri, A.; Ostriker, J.P. The Effect of the AGN Feedback on the Interstellar Medium of Early-Type Galaxies:2D Hydrodynamical Simulations of the Low-Rotation Case. {\em Astrophys. J.} {\bf 2017}, {\em 835}, 15, doi:10.3847/1538-4357/835/1/15.


\bibitem[Reeves~et al.(2009)]{Reeves09}Reeves, J.N.; O'Brien, P.T.; Braito, V. A Compton-thick Wind in the High-luminosity Quasar, PDS 456. {\em Astrophys. J.} {\bf 2009}, {\em 701}, 493--507, doi:10.1088/0004-637X/701/1/493.

\bibitem[Tombesi~et al.(2011)]{Tombesi11}Tombesi, F.; Cappi, M.; Reeves, J.N.; Palumbo, G.G.C.; Braito, V.; Dadina, M. Evidence for Ultra-fast Outflows in Radio-quiet Active Galactic Nuclei. II. Detailed Photoionization Modeling of Fe K-shell Absorption Lines. {\em Mon.~Not.~R.~Astron.~Soc.} {\bf 2011}, {\em 742}, 44, doi:10.1088/0004-637X/742/1/44.

\bibitem[Tombesi~et al.(2012)]{Tombesi12}Tombesi, F.; Cappi, M.; Reeves, J.N.; Braito, V. Evidence for ultrafast outflows in radio-quiet AGNs--III. Location and energetics. {\em Mon.~Not.~R.~Astron.~Soc.~Lett.} {\bf 2012}, {\em 422}, L1--L5, doi:10.1111/j.1745-3933.2012.01221.x.


\bibitem[Mullaney~et al.(2013)]{Mullaney13}Mullaney, J.R.; Alexander, D.M.; Fine, S.; Goulding, A.D.; Harrison, C.M.; Hickox, R.C. Narrow-line region gas kinematics of 24,264 optically selected AGN: The radio connection. {\em Mon.~Not.~R.~Astron.~Soc.} {\bf 2013}, {\em 433}, 622--638, doi:10.1093/mnras/stt751.

\bibitem[Cicone~et al.(2014)]{Cicone14}Cicone, C.; Maiolino, R.; Sturm, E. Massive molecular outflows and evidence for AGN feedback from CO observations. {\em Astron.~Astrophys.} {\bf 2014}, {\em 562}, A21, doi:10.1051/0004-6361/201322464.

\bibitem[Zakamska \& Greene(2014)]{Zakamska14}Zakamska, N.L.; Greene, J.E. Quasar feedback and the origin of radio emission in radio-quiet quasars. {\em Mon.~Not.~R.~Astron.~Soc.} {\bf 2014}, {\em 442}, 784--804, doi:10.1093/mnras/stu842.

\bibitem[Morganti(2017)]{Morganti17}Morganti, R. The many routes to AGN feedback. {\em arxiv} {\bf 2018}, arXiv:1712.05301.

\bibitem[Rupke \& Veilleux(2011)]{Rupke11}Rupke, D.S.N.; Veilleux, S. Integral Field Spectroscopy of Massive, Kiloparsec-scale Outflows in the Infrared-luminous QSO Mrk 231. {\em Astrophys. J. Lett.} {\bf 2011}, {\em 729}, L27, doi:10.1088/2041-8205/729/2/L27.

\bibitem[Feruglio~et al.(2015)]{Feruglio15}Feruglio, C.; Fiore, F.; Carniani, S. The multi-phase winds of Markarian 231: From the hot, nuclear, ultra-fast wind to the galaxy-scale, molecular outflow. {\em Astron. Astrophys.} {\bf 2015}, {\em 583}, A99, doi:10.1051/0004-6361/201526020.

\bibitem[Morganti~et al.(2016)]{Morganti16}Morganti, R.; Veilleux, S.; Oosterloo, T.; Teng, S.H.; Rupke, D. Another piece of the puzzle: The fast H I outflow in Mrk 231. {\em Astron. Astrophys.} {\bf 2016}, {\em 593}, A30, doi:10.1051/0004-6361/201628978.

\bibitem[Morganti~et al.(1998)]{Morganti98}Morganti, R.; Oosterloo, T.; Tsvetanov, Z. A Radio Study of the Seyfert Galaxy IC 5063: Evidence for Fast Gas Outflow. {\em Astron. J.} {\bf 1998}, {\em 115}, 915--927, doi:10.1086/300236.

\bibitem[Dasyra~et al.(2016)]{Dasyra16}Dasyra, K.M.; Combes, F.; Oosterloo, T.; Oonk, J.B.R.; Morganti, R.; Salom\'{e}, P.; Vlahakis, N. ALMA reveals optically thin, highly excited CO gas in the jet-driven winds of the galaxy IC 5063. {\em Astron. Astrophys.} {\bf 2016}, {\em 595}, L7, doi:10.1051/0004-6361/201629689.

\bibitem[Oosterloo~et al.(2017)]{Oosterloo17}Oosterloo, T.; Raymond O.J.B.; Morganti, R. Properties of the molecular gas in the fast outflow in the Seyfert galaxy IC 5063. {\em Astron. Astrophys.} {\bf 2017}, {\em 608}, A38, doi:10.1051/0004-6361/201731781.



\bibitem[Benson~et al(2003)]{Benson03}Benson, A.J.; Bower, R.G.; Frenk, C.S.; Lacey, C.G.; Baugh, C.M.; Cole, S. What Shapes the Luminosity Function of Galaxies? {\em Astrophys.~J.} {\bf 2003}, {\em 599}, 38--49, doi:10.1086/379160.

\bibitem[Baldry~et al.(2008)]{Baldry08}Baldry, I.K.; Glazebrook, K.; Driver, S.P. On the galaxy stellar mass function, the mass-metallicity relation, and the implied baryonic mass function. {\em Mon.~Not.~R.~Astron.~Soc.} {\bf 2008}, {\em 388}, 945--959, doi:10.1111/j.1365-2966.2008.13348.x.

\bibitem[Li \& White(2009)]{Li09}Li, C.; White, S.D.M. The distribution of stellar mass in the low-redshift Universe. {\em Mon.~Not.~R.~Astron.~Soc.} {\bf 2009}, {\em 398}, 2177--2187, doi:10.1111/j.1365-2966.2009.15268.x.


\bibitem[Somerville~et al.(2008)]{Somerville08}Somerville, R.S.; Hopkins, P.F.; Cox, T.J.; Robertson, B.E.; Hernquist, L. A semi-analytic model for the co-evolution of galaxies, black holes and active galactic nuclei. {\em Mon.~Not.~R.~Astron.~Soc.} {\bf 2008}, {\em 391}, 481--506, doi:10.1111/j.1365-2966.2008.13805.x.

\bibitem[Vogelsberger~et al.(2014)]{Vogel14}Vogelsberger, M.; Genel, S.; Springel, V. Properties of galaxies reproduced by a hydrodynamic simulation. {\em Nature} {\bf 2014}, {\em 509}, 177--182, doi:10.1038/nature13316.


\bibitem[Mo~et al.(2004)]{Mo04}Mo, H.J.; Yang, X.; van den Bosch, F.C.; Jing, Y.P. The dependence of the galaxy luminosity function on large-scale environment. {\em Mon.~Not.~R.~Astron.~Soc.} {\bf 2004}, {\em 349}, 205--212, doi:10.1111/j.1365-2966.2004.07485.x.

\bibitem[Puchwein \& Springel(2013)]{Puchwein13}Puchwein, E.; Springel, V. Shaping the galaxy stellar mass function with supernova- and AGN-driven winds. {\em Mon.~Not.~R.~Astron.~Soc.} {\bf 2013}, {\em 428}, 2966--2979, doi:10.1093/mnras/sts243.

\bibitem[Schaye~et al.(2015)]{Schaye15}Schaye, J.; Crain, R.A.; Bower, R.G. The EAGLE project: Simulating the evolution and assembly of galaxies and their environments. {\em Mon.~Not.~R.~Astron.~Soc.} {\bf 2015}, {\em 446}, 521--554, doi:10.1093/mnras/stu2058.

\bibitem[Dav\'{e}~et al.(2017)]{Dave17}Dav\'{e}, R.; Rafieferantsoa, M.H.; Thompson, R.J.; Hopkins, P.F. MUFASA: Galaxy star formation, gas, and~metal properties across cosmic time. {\em Mon.~Not.~R.~Astron.~Soc.} {\bf 2017}, {\em 467}, 115--132, doi:10.1093/mnras/stx108.


\bibitem[Tremonti~et al.(2004)]{Tremonti04}Tremonti, C.A.; Heckman, T.M.; Kauffmann, G. The Origin of the Mass-Metallicity Relation: Insights from 53,000 Star-forming Galaxies in the Sloan Digital Sky Survey. {\em Astrophys. J.} {\bf 2004}, {\em 613}, 898--913, doi:10.1086/423264.

\bibitem[Finlator \& Dav\'{e}(2008)]{Finlator08}Finlator, K.; Dav\'{e}, R. The Origin of the Galaxy Mass-Metallicity Relation and Implications for Galactic Outflows. {\em Mon.~Not.~R.~Astron.~Soc.} {\bf 2008}, {\em 385}, 2181--2204, doi:10.1111/j.1365-2966.2008.12991.x.

\bibitem[Peeples \& Shankar(2011)]{Peeples11}Peeples, M.S.; Shankar, F. Constraints on Star-Formation Driven Galaxy Winds from the Mass-Metallicity Relation at z = 0. {\em Mon.~Not.~R.~Astron.~Soc.} {\bf 2011}, {\em 417}, 2962--2981, doi:10.1111/j.1365-2966.2011.19456.x.

\bibitem[Andrews \& Martini(2013)]{Andrews13}Andrews, B.H.; Martini, P. The Mass-Metallicity Relation with the Direct Method on Stacked Spectra of SDSS Galaxies. {\em Astrophys.~J.} {\bf 2013}, {\em 765}, 140, doi:10.1088/0004-637X/765/2/140.

\bibitem[Sanders~et al.(2015)]{Sanders15}Sanders, R.L.; Shapley, A.E.; Kriek, M. The MOSDEF Survey: Mass, Metallicity, and Star-formation Rate at $z$ $\sim$ $2.3$. {\em Astrophys.~J.} {\bf 2015}, {\em 799}, 138, doi:10.1088/0004-637X/799/2/138.

\bibitem[S\'{a}nchez~et al.(2017)]{Sanchez17}S\'{a}nchez, S.F.; Barrera-Ballesteros, J.K.; S\'{a}nchez-Menguiano, L. The mass-metallicity relation revisited with CALIFA. {\em Mon.~Not.~R.~Astron.~Soc.} {\bf 2017}, {\em 469}, 2121--2140, doi:10.1093/mnras/stx808.

\bibitem[Aguirre~et al.(2001a)]{Aguirre01a}Aguirre, A.; Hernquist, L.; Katz, N.; Gardner, J.; Weinberg, D.H. Enrichment of the Intergalactic Medium by Radiation Pressure-driven Dust Efflux. {\em Astrophys.~J.~Lett.} {\bf 2001}, {\em 556}, 11--15, doi:10.1086/322860.

\bibitem[Aguirre~et al.(2001b)]{Aguirre01b}Aguirre, A.; Hernquist, L.; Schaye, J.; Weinberg, D.H.; Katz, N.; Gardner, J. Metal Enrichment of the Intergalactic Medium at z = 3 by Galactic Winds. {\em Astrophys.~J.} {\bf 2001}, {\em 560}, 599--605, doi:10.1086/323070.

\bibitem[Aguirre~et al.(2008)]{Aguirre08}Aguirre, A.; Dow-Hygelund, C.; Schaye, J.; Theuns, T. Metallicity of the Intergalactic Medium Using Pixel Statistics. IV. Oxygen. {\em Astrophys.~J.} {\bf 2008}, {\em 689}, 851--864, doi:10.1086/592554.

\bibitem[Prochaska~et al.(2017)]{Prochaska17}Prochaska, J.X.; Werk, J.K.; Worseck, G. The COS-Halos Survey: Metallicities in the Low-redshift Circumgalactic Medium. {\em Astrophys.~J.} {\bf 2017}, {\em 837}, 169, doi:10.3847/1538-4357/aa6007.

\bibitem[Springel \& Hernquist(2008a)]{SH08a}Springel, V.; Hernquist, L. Cosmological smoothed particle hydrodynamics simulations: A hybrid multiphase model for star formation. {\em Mon.~Not.~R.~Astron.~Soc.} {\bf 2003}, {\em 339}, 289--311, doi:10.1046/j.1365-8711.2003.06206.x.

\bibitem[Springel \& Hernquist(2008b)]{SH08b}Springel, V.; Hernquist, L. The history of star formation in a $\Lambda$ cold dark matter universe. {\em Mon.~Not.~R.~Astron.~Soc.} {\bf 2003}, {\em 339}, 312--334, doi:10.1046/j.1365-8711.2003.06207.x.


\bibitem[Schaye~et al.(2010)]{Schaye10}Schaye, J.; Dalla Vecchia, C.; Booth, C.M. The physics driving the cosmic star formation history. {\em Mon.~Not.~R.~Astron.~Soc.} {\bf 2010}, {\em 402}, 1536--1560, doi:10.1111/j.1365-2966.2009.16029.x.

\bibitem[Muratov~et al.(2015)]{Muratov15}Muratov, A.L.; Kere\u{s}, D.; Faucher-Gigu\`{e}re, C.-A.; Hopkins, P.F.; Quataert, E.; Murray, N. Gusty, gaseous flows of FIRE: Galactic winds in cosmological simulations with explicit stellar feedback. {\em Mon.~Not.~R.~Astron.~Soc.} {\bf 2015}, {\em 454}, 2691--2713, doi:10.1093/mnras/stv2126.



\bibitem[Bertone~et al.(2007)]{Bertone07}Bertone, S.; De Lucia, G.; Thomas, P.A. The recycling of gas and metals in galaxy formation: Predictions of a dynamical feedback model. {\em Mon.~Not.~R.~Astron.~Soc.} {\bf 2007}, {\em 379}, 1143--1154, doi:10.1111/j.1365-2966.2007.11997.x.

\bibitem[Chisholm \& Matsushita(2016)]{Chisholm16}Chisholm, J.; Matsushita, S. The Molecular Baryon Cycle of M82. {\em Astrophys.~J.} {\bf 2016}, {\em 830}, 72, doi:10.3847/0004-637X/830/2/72.


\bibitem[van de Voort(2017)]{Voort17}Van de Voort, F. The Effect of Galactic Feedback on Gas Accretion and Wind Recycling. In {\em Gas Accretion onto Galaxies}; Fox, A.J., Dav\'e, R., Eds.; Astrophysics and Space Science Library, Springer International Publishing~AG: New York, NY, USA, 2017; pp. 301--321, ISBN 978-3-319-52511-2.

\bibitem[Tumlinson~et al.(2017)]{Tumlinson17}Tumlinson, J.; Peeples, M.S.; Werk, J.K. The Circumgalactic Medium. {\em Annu. Rev. Astron. Astrophys.} {\bf 2017}, {\em 55}, 389--432, doi:10.1146/annurev-astro-091916-055240.

\bibitem[Krumholz~et al.(2017)]{Krumholz17}Krumholz, M.R.; Kruijssen, J.M.D.; Crocker, R.M. A dynamical model for gas flows, star formation and nuclear winds in galactic centres. {\em Mon.~Not.~R.~Astron.~Soc.} {\bf 2017}, {\em 466}, 1213--1233, doi:10.1093/mnras/stw3195.


\bibitem[\"{U}bler~et al.(2014)]{Ubler14}\"{U}bler, H.; Naab, T.; Oser, L.; Aumer, M.; Sales, L.V.; White, S.D.M. Why stellar feedback promotes disc formation in simulated galaxies. {\em Mon.~Not.~R.~Astron.~Soc.} {\bf 2014}, {\em 443}, 2092--2111, doi:10.1093/mnras/stu1275.

\bibitem[DeFelippis~et al.(2017)]{DeFelippis17}DeFelippis, D.; Genel, S.; Bryan, G.L.; Fall, S.M. The Impact of Galactic Winds on the Angular Momentum of Disk Galaxies in the Illustris Simulation. {\em Astrophys.~J.} {\bf 2017}, {\em 841}, 16, doi:10.3847/1538-4357/aa6dfc.




\bibitem[Strickland \& Stevens(2000)]{SS00}Strickland, D.K.; Stevens, I.R. Starburst-driven galactic winds--I. Energetics and intrinsic X-ray emission. {\em Mon.~Not.~R.~Astron.~Soc.} {\bf 2000}, {\em 314}, 511--545, doi:10.1046/j.1365-8711.2000.03391.x.

\bibitem[McDowell~et al.(2003)]{McDowell03}McDowell, J.C.; Clements, D.L.; Lamb, S.A. Chandra Observations of Extended X-Ray Emission in Arp 220. {\em Astrophys.~J.} {\bf 2003}, {\em 591}, 154--166, doi:10.1086/375289.

\bibitem[Strickland \& Heckman(2007)]{SH07}Strickland, D.K.; Heckman, T.M. Iron Line and Diffuse Hard X-Ray Emission from the Starburst Galaxy M82. {\em Astrophys.~J.} {\bf 2007}, {\em 658}, 258--281, doi:10.1086/511174.

\bibitem[Strickland \& Heckman(2009)]{SH09}Strickland, D.K.; Heckman, T.M. Supernova Feedback Efficiency and Mass Loading in the Starburst and Galactic Superwind Exemplar M82. {\em Astrophys.~J.} {\bf 2009}, {\em 697}, 2030--2056, doi:10.1088/0004-637X/697/2/2030.

\bibitem[Bravo-Guerrero \& Stevens(2017)]{Bravo17}Bravo-Guerrero, J.; Stevens, I.R. Superwind evolution: The young starburst-driven wind galaxy NGC 2782. {\em Mon.~Not.~R.~Astron.~Soc.} {\bf 2017}, {\em 467}, 3788--3800, doi:10.1093/mnras/stx327.


\bibitem[Shopbell \& Bland-Hawthorn(1998)]{SB98}Shopbell, P.L.; Bland-Hawthorn, J. The Asymmetric Wind in M82. {\em Astrophys.~J.} {\bf 1998}, {\em 493}, 129--153, doi:10.1086/305108.
\bibitem[Cecil~et al.(2002)]{Cecil02}Cecil, G.; Bland-Hawthorn, J.; Veilleux, S. Tightly Correlated X-ray/H$\alpha$-emitting Filaments in the Superbubble and Large-Scale Superwind of NGC 3079. {\em Astrophys.~J.} {\bf 2002}, {\em 576}, 745--752, doi:10.1086/341861.
\bibitem[Weiner~et al.(2009)]{Weiner09}Weiner, B.J.; Coil, A.L.; Prochaska, J.X. Ubiquitous Outflows in DEEP2 Spectra of Star-Forming Galaxies at $z = 1.4$. {\em Astrophys.~J.} {\bf 2009}, {\em 692}, 187--211, doi:10.1088/0004-637X/692/1/187.
\bibitem[Martin~et al.(2013)]{Martin13}Martin, C.L.; Shapley, A.E.; Coil, A.L.; Kornei, K.A.; Murray, N.; Pancoast, A. Scattered Emission from $z$ $\sim$ $1$ Galactic Outflows. {\em Astrophys.~J.} {\bf 2013}, {\em 770}, 41, doi:10.1088/0004-637X/770/1/41.
\bibitem[Tang~et al.(2014)]{Tang14}Tang, Y.; Giavalisco, M.; Guo, Y.; Kurk, J. Probing Outflows in $z = 1$ $\sim$ $2$ Galaxies through Fe II/Fe II* Multiplets. {\em Astrophys.~J.} {\bf 2014}, {\em 793}, 92, doi:10.1088/0004-637X/793/2/92.
\bibitem[Rubin~et al.(2014)]{Rubin14}Rubin, K.H.R.; Prochaska, J.X.; Koo, D.C.; Phillips, A.C.; Martin, C.L.; Winstrom, L.O. Evidence for Ubiquitous Collimated Galactic-scale Outflows along the Star-forming Sequence at $z$ $\sim$ $0.5$. {\em Astrophys.~J.} {\bf 2014}, {\em 794}, 156, doi:10.1088/0004-637X/794/2/156.
\bibitem[Heckman~et al.(2015)]{Heckman15}Heckman, T.M.; Alexandroff, R.M.; Borthakur, S.; Overzier, R.; Leitherer, C. The Systematic Properties of the Warm Phase of Starburst-Driven Galactic Winds. {\em Astrophys.~J.} {\bf 2015}, {\em 809}, 147, doi:10.1088/0004-637X/809/2/147.
\bibitem[Du~et al.(2016)]{Du16}Du, X.; Shapley, A.E.; Martin, C.L.; Coil, A.L. The Kinematics of C IV in Star-forming Galaxies at $z\approx1.2$. {\em Astrophys.~J.} {\bf 2016}, {\em 829}, 64, doi:10.3847/0004-637X/829/2/64.
\bibitem[Chisholm~et al.(2017)]{Chisholm17}Chisholm, J.; Tremonti, C.A.; Leitherer, C.; Chen, Y. The mass and momentum outflow rates of photoionized galactic outflows. {\em Mon.~Not.~R.~Astron.~Soc.} {\bf 2017}, {\em 469}, 4831--4849, doi:10.1093/mnras/stx1164.



\bibitem[Rupke~et al.(2002)]{Rupke02}Rupke, D.S.; Veilleux, S.; Sanders, D.B. Keck Absorption-Line Spectroscopy of Galactic Winds in Ultraluminous Infrared Galaxies. {\em Astrophys.~J.} {\bf 2002}, {\em 570}, 588--609, doi:10.1086/339789.
\bibitem[Martin(2005)]{Martin05}Martin, C.L. Mapping Large-Scale Gaseous Outflows in Ultraluminous Galaxies with Keck II ESI Spectra: Variations in Outflow Velocity with Galactic Mas. {\em Astrophys.~J.} {\bf 2005}, {\em 621}, 227--245, doi:10.1086/427277.

\bibitem[Rupke~et al.(2005a)]{Rupke05a}Rupke, D.S.; Veilleux, S.; Sanders, D.B. Outflows in Infrared-Luminous Starbursts at z < 0.5. I. Sample, Na I D Spectra, and Profile Fitting. {\em Astrophys. J. Suppl. Ser.} {\bf 2005}, {\em 160}, 87--114, doi:10.1086/432886.

\bibitem[Rupke~et al.(2005b)]{Rupke05b}Rupke, D.S.; Veilleux, S.; Sanders, D.B. Outflows in Infrared-Luminous Starbursts at z < 0.5. II. Analysis and Discussion. {\em Astrophys. J. Suppl. Ser.} {\bf 2005}, {\em 160}, 115--148, doi:10.1086/432889.

\bibitem[Rupke~et al.(2005c)]{Rupke05c}Rupke, D.S.; Veilleux, S.; Sanders, D.B. Outflows in Active Galactic Nucleus/Starburst-Composite Ultraluminous Infrared Galaxies. {\em Astrophys.~J.} {\bf 2005}, {\em 632}, 751--780, doi:10.1086/444451.

\bibitem[Chen~et al.(2010)]{Chen10}Chen, Y.-M.; Tremonti, C.A.; Heckman, T.M. Absorption-line Probes of the Prevalence and Properties of Outflows in Present-day Star-forming Galaxies. {\em Astrophys.~J.} {\bf 2010}, {\em 140}, 445--461, doi:10.1088/0004-6256/140/2/445.
\bibitem[Kornei~et al.(2013)]{Kornei13}Kornei, K.A.; Shapley, A.E.; Martin, C.L.; Coil, A.L.; Lotz, J.M.; Weiner, B.J. Fine-structure Fe II* Emission and Resonant Mg II Emission in $z \sim 1$ Star-forming Galaxies. {\em Astrophys.~J.} {\bf 2013}, {\em 774}, 50--74, doi:10.1088/0004-637X/774/1/50.
\bibitem[Rupke \& Veilleux(2015)]{Rupke15}Rupke, D.S.N.; Veilleux, S. Spatially Extended Na I D Resonant Emission and Absorption in the Galactic Wind of the Nearby Infrared-Luminous Quasar F05189-2524. {\em Astrophys.~J.} {\bf 2015}, {\em 801}, 126, doi:10.1088/0004-637X/801/2/126.


\bibitem[Sakamoto~et al.(1999)]{Sakamoto99}Sakamoto, K.; Okumura, S.K.; Ishizuki, S.; Scoville, N.Z. Bar-driven Transport of Molecular Gas to Galactic Centers and Its Consequences. {\em Astrophys.~J.} {\bf 1999}, {\em 525}, 691--701, doi:10.1086/307910.



\bibitem[Veilleux~et al.(2009)]{Veilleux09}Veilleux, S.; Rupke, D.S.N.; Swaters, R. Warm Molecular Hydrogen in the Galactic Wind of M82. {\em Astrophys.~J.~Lett.} {\bf 2009}, {\em 700}, 149--153, doi:10.1088/0004-637X/700/2/L149.

\bibitem[Westmoquette(2013)]{Westmoquette13}Westmoquette, M. Astrophysics: How to catch a galactic wind. {\em Nature} {\bf 2013}, {\em 499}, 416--417, doi:10.1038/499416a.

\bibitem[Bolatto~et al.(2013)]{Bolatto13}Bolatto, A.D.; Warren, S.R.; Leroy, A.K. The Starburst-Driven Molecular Wind in NGC 253 and the Suppression of Star Formation. {\em Nature} {\bf 2013}, {\em 499}, 450--453, doi:10.1038/nature12351.

\bibitem[Salak~et al.(2013)]{Salak13}Salak, D.; Nakai, N.; Miyamoto, Y.; Yamauchi, A.; Tsuru, T.G. Large-Field CO(J = 1$\rightarrow$0) Observations of the Starburst Galaxy M82. {\em Publ. Astron. Soc. Jpn.} {\bf 2013}, {\em 65}, 66, doi:10.1093/pasj/65.3.66.

\bibitem[Meier~et al.(2015)]{Meier15}Meier, D.S.; Walter, F.; Bolatto, A.D. ALMA Multi-line Imaging of the Nearby Starburst NGC 253. {\em Astrophys.~J.} {\bf 2015}, {\em 801}, 63, doi:10.1088/0004-637X/801/1/63.

\bibitem[Leroy~et al.(2015)]{Leroy15}Leroy, A.K.; Walter, F.; Martini, P. The Multi-phase Cold Fountain in M82 Revealed by a Wide, Sensitive Map of the Molecular Interstellar Medium. {\em Astrophys.~J.} {\bf 2015}, {\em 814}, 83, doi:10.1088/0004-637X/814/2/83.

\bibitem[Walter~et al.(2017)]{Walter17}Walter, F.; Bolatto, A.D.; Leroy, A.K. Dense Molecular Gas Tracers in the Outflow of the Starburst Galaxy NGC 253. {\em Astrophys.~J.} {\bf 2017}, {\em 835}, 265, doi:10.3847/1538-4357/835/2/265.




\bibitem[Roussel~et al.(2010)]{Roussel10}Roussel, H.; Wilson, C.D.; Vigroux, L. SPIRE imaging of M 82: Cool dust in the wind and tidal streams. {\em Astron.~Astrophys.~Lett.} {\bf 2010}, {\em 518}, L66, doi:10.1051/0004-6361/201014567.

\bibitem[Hutton~et al.(2014)]{Hutton14}Hutton, S.; Ferreras, I.; Wu, K. A panchromatic analysis of starburst galaxy M82: Probing the dust properties. {\em Mon.~Not.~R.~Astron.~Soc.} {\bf 2014}, {\em 440}, 150--160, doi:10.1093/mnras/stu185.

\bibitem[Mel\'{e}ndez~et al.(2015)]{Melendez15}Mel\'{e}ndez, M.; Veilleux, S.; Martin, C. Exploring the Dust Content of Galactic Winds with Herschel. I. NGC 4631. {\em Astrophys.~J.} {\bf 2015}, {\em 804}, 46, doi:10.1088/0004-637X/804/1/46.

\bibitem[Hodges-Kluck~et al.(2016)]{Hodges16}Hodges-Kluck, E.; Cafmeyer, J.; Bregman, J.N. Ultraviolet Halos around Spiral Galaxies. I. Morphology. {\em Astrophys.~J.} {\bf 2016}, {\em 833}, 58, doi:10.3847/1538-4357/833/1/58.

\bibitem[McCormick~et al.(2018)]{McCormick18}McCormick, A.; Veilleux, S.; Mel\'{e}ndez, M. Exploring the dust content of galactic winds with Herschel--II. Nearby dwarf galaxies. {\em Mon.~Not.~R.~Astron.~Soc.} {\bf 2018}, {\em 477}, 699--726, doi:10.1093/mnras/sty634.

\bibitem[Jones~et al.(2018)]{Jones18}Jones, T.; Stark, D.P.; Ellis, R.S. Dust in the Wind: Composition and Kinematics of Galaxy Outflows at the Peak Epoch of Star Formation. {\em arxiv} {\bf 2018}, arXiv:1805.01484.

\bibitem[McKeith~et al.(1995)]{McKeith95}McKeith, C.D.; Greve, A.; Downes, D.; Prada, F. The outflow in the halo of M 82. {\em Atron.~Astrophys.} {\bf 1995}, {\em 293}, 703--709.



\bibitem[Walter~et al.(2002)]{Walter02}Walter, F.; Weiss, A.; Scoville, N. Molecular Gas in M82: Resolving the Outflow and Streamers. {\em Astrophys.~J.~Lett.} {\bf 2002}, {\em 580}, L21--L25, doi:10.1086/345287.

\bibitem[Scannapieco \& Br\"{u}ggen(2015)]{SB15}Scannapieco, E.; Br\"{u}ggen, M. The Launching of Cold Clouds by Galaxy Outflows. I. Hydrodynamic Interactions with Radiative Cooling. {\em Astrophys.~J.} {\bf 2015}, {\em 805}, 158, doi:10.1088/0004-637X/805/2/158.

\bibitem[Br\"{u}ggen \& Scannapieco(2016)]{BS16}Br\"{u}ggen, M.; Scannapieco, E. The Launching of Cold Clouds by Galaxy Outflows. II. The Role of Thermal Conduction. {\em Astrophys.~J.} {\bf 2016}, {\em 822}, 31, doi:10.3847/0004-637X/822/1/31.

\bibitem[Zhang~et al.(2017)]{Zhang17}Zhang, D.; Thompson, T.A.; Quataert, E.; Murray, N. Entrainment in trouble: Cool cloud acceleration and destruction in hot supernova-driven galactic winds. {\em Mon.~Not.~R.~Astron.~Soc.} {\bf 2017}, {\em 468}, 4801--4814, doi:10.1093/mnras/stx822.

\bibitem[Schneider \& Robertson(2017)]{SR17}Schneider, E.E.; Robertson, B.E. Hydrodynamical Coupling of Mass and Momentum in Multiphase Galactic Winds. {\em Astrophys.~J.} {\bf 2017}, {\em 834}, 144, doi:10.3847/1538-4357/834/2/144.
\bibitem[McCourt~et al.(2018)]{McCourt18}McCourt, M.; Oh, S.P.; O'Leary, R.; Madigan, A.-M. A characteristic scale for cold gas. {\em Mon.~Not.~R.~Astron.~Soc.} {\bf 2018}, {\em 473}, 5407--5431, doi:10.1093/mnras/stx2687.



\bibitem[Ostriker \& McKee(1988)]{OM88}Ostriker, J.P.; McKee, C.F. Astrophysical Blastwaves. {\em Rev.~Mod.~Phys.} {\bf 1988}, {\em 60}, 1--68, doi:10.1103/RevModPhys.60.1.
\bibitem[Sedov(1959)]{Sedov59}Sedov, L.I. \emph{Similarity and Dimensional Methods in Mechanics}; Academic Press: New York, NY, USA, {1959}.

\bibitem[Taylor(1950)]{Taylor50}Taylor, G. The Formation of a Blast Wave by a Very Intense Explosion. I. Theoretical Discussion. {\em Proc.~R.~Soc.~A.} {\bf 1950}, {\em 201}, 175--186, doi:10.1098/rspa.1950.0050.

\bibitem[Draine(2011)]{Draine11}Draine, B.T. \emph{Physics of the Interstellar and Intergalactic Medium}; Princeton University Press: Princeton, NJ, USA, {2011}; ISBN 978-0-691-12214-4.

\bibitem[Kim \& Ostriker(2015)]{Kim15}Kim, C.-G.; Ostriker, E.C. Momentum Injection by Supernovae in the Interstellar Medium. {\em Astrophys.~J.} {\bf 2015}, {\em 802}, 99, doi:10.1088/0004-637X/802/2/99.

\bibitem[Zhang \& Chevalier(2018)]{ZC18}Zhang, D.; Chevalier, R. Numerical simulations of supernova remnants in turbulent molecular clouds. {\em arxiv} {\bf 2018}, arXiv:1807.06603.

\bibitem[Chevalier \& Gardner(1974)]{Chevalier74}Chevalier, R.A.; Gardner, J. The Evolution of Supernova Remnants. II. Models of an Explosion in a Plane-Stratified Medium. {\em Astrophys.~J.} {\bf 1974}, {\em 192}, 457--463, doi:10.1086/153077.

\bibitem[Cioffi~et al.(1988)]{Cioffi88}Cioffi, D.F.; McKee, C.F.; Bertschinger, E. Dynamics of Radiative Supernova Remnants. {\em Astrophys.~J.} {\bf 1988}, {\em 334}, 252--265, doi:10.1086/166834.

\bibitem[Thornton~et al.(1998)]{Thornton98}Thornton, K.; Gaudlitz, M.; Janka, H.-Th.; Steinmetz, M. Energy Input and Mass Redistribution by Supernovae in the Interstellar Medium. {\em Astrophys.~J.} {\bf 1998}, {\em 500}, 95--119, doi:10.1086/305704.

\bibitem[Blondin~et al.(1998)]{Blondin98}Blondin, J.M.; Wright, E.B.; Borkowski, K.J.; Reynolds, S.P. Transition to the Radiative Phase in Supernova Remnants. {\em Astrophys.~J.} {\bf 1998}, {\em 500}, 342--354, doi:10.1086/305708.

\bibitem[Martizzi~et al.(2015)]{Martizzi15}Martizzi, D.; Faucher-Gigu\`{e}re, C.-A.; Quataert, E. Supernova feedback in an inhomogeneous interstellar medium. {\em Mon.~Not.~R.~Astron.~Soc.} {\bf 2015}, {\em 450}, 504--522, doi:10.1093/mnras/stv562.

\bibitem[Walch \& Naab(2015)]{Walch15}Walch, S.; Naab, T. The energy and momentum input of supernova explosions in structured and ionized molecular clouds. {\em Mon.~Not.~R.~Astron.~Soc.} {\bf 2015}, {\em 451}, 2757--2771, doi:10.1093/mnras/stv1155.


\bibitem[Elmegreen \& Scalo(2004)]{Elmegreen04}Elmegreen, B.G.; Scalo, J. Interstellar Turbulence I: Observations and Processes. {\em Annu. Rev. Astron. Astrophys.} {\bf 2004}, {\em 42}, 211--273, doi:10.1146/annurev.astro.41.011802.094859.


\bibitem[Iffrig \& Hennebelle(2015)]{Iffrig15}Iffrig, O.; Hennebelle, P. Mutual influence of supernovae and molecular clouds. {\em Astron.~Astrophs.} {\bf 2015}, {\em 576}, A95, doi:10.1051/0004-6361/201424556.


\bibitem[Lemaster \& Stone(2009)]{Lemaster09}Lemaster, M.N.; Stone, J.M. Dissipation and Heating in Supersonic Hydrodynamic and MHD Turbulence. {\em Astrophys.~J.} {\bf 2009}, {\em 691}, 1092--1108, doi:10.1088/0004-637X/691/2/1092.

\bibitem[Ostriker~et al.(2001)]{Ostriker01}Ostriker, E.C.; Stone, J.M.; Gammie, C.F. Density, Velocity, and Magnetic Field Structure in Turbulent Molecular Cloud Models. {\em Astrophys.~J.} {\bf 2001}, {\em 546}, 980--1005, doi:10.1086/318290.


\bibitem[Field(1965)]{Field65}Field, G.B. Thermal Instability. {\em Astrophys.~J.} {\bf 1965}, {\em 142}, 531--567, doi:10.1086/148317.

\bibitem[Pikel'Ner(1968)]{Pikel68}Pikel'Ner, S.B. Structure and Dynamics of the Interstellar Medium. {\em Annu. Rev. Astron. Astrophys.} {\bf 1968}, {\em 6}, 165--194, doi:10.1146/annurev.aa.06.090168.001121.

\bibitem[Field, Goldsmith \& Habing(1969)]{Field69}Field, G.B.; Goldsmith, D.W.; Habing, H.J. Cosmic-Ray Heating of the Interstellar Gas. {\em Astrophys.~J.~Lett.} {\bf 1969}, {\em 155}, 149--154, doi:10.1086/180324.

\bibitem[Reynolds~et al.(1974)]{Reynolds74}Reynolds, R.J.; Roesler, F.L.; Scherb, F. The Intensity Distribution of Diffuse Galactic H$\alpha$ Emission. {\em Astrophys.~J.~Lett.} {\bf 1974}, {\em 192}, 53--56, doi:10.1086/181589.

\bibitem[Haffner~et al.(2009)]{Haffner09}Haffner, L.M.; Dettmar, R.-J.; Beckman, J.E. The warm ionized medium in spiral galaxies. {\em Rev.~Mod.~Phys.} {\bf 2009}, {\em 81}, 969--997, doi:10.1103/RevModPhys.81.969.

\bibitem[Cox \& Smith(1974)]{Cox74}Cox, D.P.; Smith, B.W. Large-Scale Effects of Supernova Remnants on the Galaxy: Generation and Maintenance of a Hot Network of Tunnels. {\em Astrophys.~J.~Lett.} {\bf 1974}, {\em 189}, 105--108, doi:10.1086/181476.

\bibitem[McKee \& Ostriker(1977)]{McKee77}McKee, C.F.; Ostriker, J.P. A theory of the interstellar medium--Three components regulated by supernova explosions in an inhomogeneous substrate. {\em Astrophys.~J.} {\bf 1977}, {\em 218}, 148--169, doi:10.1086/155667.

\bibitem[Bowyer~et al.(1995)]{Bowyer95}Bowyer, S.; Lieu, R.; Sidher, S.D.; Lampton, M.; Knude, J. Evidence for a large thermal pressure imbalance in the local interstellar medium. {\em Nature} {\bf 1995}, {\em 375}, 212--214, doi:10.1038/375212a0.

\bibitem[Cox(2005)]{Cox05}Cox, D.P. The Three-Phase Interstellar Medium Revisited. {\em Annu. Rev. Astron. Astrophys.} {\bf 2005}, {\em 43}, 337--385, doi:10.1146/annurev.astro.43.072103.150615.



\bibitem[Hennebelle \& Iffrig(2014)]{Hennebelle14}Hennebelle, P.; Iffrig, O. Simulations of magnetized multiphase galactic disc regulated by supernovae explosions. {\em Astron.~Astrophys.} {\bf 2014}, {\em 570}, 81, doi:10.1051/0004-6361/201423392.

\bibitem[Gatto~et al.(2015)]{Gatto15}Gatto, A.; Walch, S.; Mac Low, M.-M. Modelling the supernova-driven ISM in different environments. {\em Mon.~Not.~R.~Astron.~Soc.} {\bf 2015}, {\em 449}, 1057--1075, doi:10.1093/mnras/stv324.

\bibitem[Kim \& Ostriker(2017)]{Kim17b}Kim, C.-G.; Ostriker, E.C. Three-phase Interstellar Medium in Galaxies Resolving Evolution with Star Formation and Supernova Feedback (TIGRESS): Algorithms, Fiducial Model, and Convergence. {\em Astrophys.~J.} {\bf 2017}, {\em 846}, 133, doi:10.3847/1538-4357/aa8599.

\bibitem[Hill~et al.(2018)]{Hill18}Hill, A.S.; Mac Low, M.-M.; Gatto, A.; Ib\'{a}\~{n}ez-Mej\'{i}a, J.C. Effect of the Heating Rate on the Stability of the Three-phase Interstellar Medium. {\em Astrophys.~J.} {\bf 2018}, {\em 862}, 55, doi:10.1086/181476.


\bibitem[Chevalier \& Clegg(1985)]{CC85}Chevalier, R.A.; Clegg, A.W. Wind from a starburst galaxy nucleus. {\em Nature} {\bf 1985}, {\em 317}, 44--45, doi:10.1038/317044a0.

\bibitem[Parker(1965)]{Parker65}Parker, E.N. Dynamical Theory of the Solar Wind. {\em Space~Sci.~Rev.} {\bf 1965}, {\em 4}, 666--708, doi:10.1007/BF00216273.

\bibitem[Leitherer~et al.(1999)]{Leitherer99}Leitherer, C.; Schaerer, D.; Goldader, J.D. Starburst99: Synthesis Models for Galaxies with Active Star Formation. {\em Astrophys.~J.~Suppl.~Ser.} {\bf 1999}, {\em 123}, 3--40, doi:10.1086/313233.

\bibitem[Cant\'{o}~et al.(2000)]{Canto00}Cant\'{o}, J.; Raga, A.C.; Rodr\'{i}guez, L.F. The Hot, Diffuse Gas in a Dense Cluster of Massive Stars. {\em Astrophys.~J.} {\bf 2000}, {\em 536}, 896--901, doi:10.1086/308983.

\bibitem[Fielding~et al.(2017)]{Fielding17}Fielding, D.; Quataert, E.; Martizzi, D.; Faucher-Gigu\`{e}re, C.-A. How supernovae launch galactic winds? {\em Mon.~Not.~R.~Astron.~Soc.~Lett.} {\bf 2017}, {\em 470}, 39--43, doi:10.1093/mnrasl/slx072.

\bibitem[Schneider~et al.(2018)]{Schneider18}Schneider, E.E.; Robertson, B.E.; Thompson, T.A. Production of Cool Gas in Thermally-Driven Outflows. {\em Astrophys.~J.} {\bf 2018}, {\em 862}, 56, doi:10.3847/1538-4357/aacce1.


\bibitem[Zhang~et al.(2014)]{Zhang14}Zhang, D.; Thompson, T.A.; Murray, N.; Quataert, E. Hot Galactic Winds Constrained by the X-Ray Luminosities of Galaxies. {\em Astrophys.~J.} {\bf 2014}, {\em 784}, 93--101, doi:10.1088/0004-637X/784/2/93.

\bibitem[Grimm~et al.(2003)]{Grimm03}Grimm, H.-J.; Gilfanov, M.; Sunyaev, R. High-mass X-ray binaries as a star formation rate indicator in distant galaxies. {\em Mon.~Not.~R.~Astron.~Soc.} {\bf 2003}, {\em 339}, 793--809, doi:10.1046/j.1365-8711.2003.06224.x.

\bibitem[Ranalli~et al.(2003)]{Ranalli03}Ranalli, P.; Comastri, A.; Setti, G. The 2-10 keV luminosity as a Star Formation Rate indicator. {\em Astron.~Astrophys.} {\bf 2003}, {\em 399}, 39--50, doi:10.1051/0004-6361:20021600.

\bibitem[Gilfanov~et al.(2004)]{Gilfanov04}Gilfanov, M.; Grimm, H.-J.; Sunyaev, R. $L_X-$SFR relation in star-forming galaxies. {\em Mon.~Not.~R.~Astron.~Soc.} {\bf 2004}, {\em 347}, L57--L60, doi:10.1111/j.1365-2966.2004.07450.x.

\bibitem[Dijkstra~et al.(2012)]{Dijkstra12}Dijkstra, M.; Gilfanov, M.; Loeb, A.; Sunyaev, R. Constraints on the redshift evolution of the $L_X-$SFR relation from the cosmic X-ray backgrounds. {\em Mon.~Not.~R.~Astron.~Soc.} {\bf 2012}, {\em 421}, 213--223, doi:10.1111/j.1365-2966.2011.20292.x.

\bibitem[Mineo~et al.(2012a)]{Mineo12a}Mineo, S.; Gilfanov, M.; Sunyaev, R. X-ray emission from star-forming galaxies--I. High-mass X-ray binaries. {\em Mon.~Not.~R.~Astron.~Soc.} {\bf 2012}, {\em 419}, 2095--2115, doi:10.1111/j.1365-2966.2011.19862.x.

\bibitem[Mineo~et al.(2012b)]{Mineo12b}Mineo, S.; Gilfanov, M.; Sunyaev, R. X-ray emission from star-forming galaxies--II. Hot interstellarmedium. {\em Mon.~Not.~R.~Astron.~Soc.} {\bf 2012}, {\em 426}, 1870--1883, doi:10.1111/j.1365-2966.2012.21831.x.

\bibitem[Mineo~et al.(2014)]{Mineo14}Mineo, S.; Gilfanov, M.; Lehmer, B.D.; Morrison, G.E.; Sunyaev, R. X-ray emission from star-forming galaxies--III. Calibration of the LX-SFR relation up to redshift $z \approx 1.3$. {\em Mon.~Not.~R.~Astron.~Soc.} {\bf 2014}, {\em 437}, 1698--1707, doi:10.1093/mnras/stt1999.

\bibitem[Oppenheimer \& Dav\'{e}(2006)]{Oppen06}Oppenheimer, B.D.; Dav\'{e}, R. Cosmological simulations of intergalactic medium enrichment from galactic outflows. {\em Mon.~Not.~R.~Astron.~Soc.} {\bf 2006}, {\em 373}, 1265--1292, doi:10.1111/j.1365-2966.2006.10989.x.

\bibitem[Oppenheimer \& Dav\'{e}(2008)]{Oppen08}Oppenheimer, B.D.; Dav\'{e}, R. Mass, metal, and energy feedback in cosmological simulations. {\em Mon.~Not.~R.~Astron.~Soc.} {\bf 2008}, {\em 387}, 577--600, doi:10.1111/j.1365-2966.2008.13280.x.

\bibitem[Bower~et al.(2012)]{Bower12}Bower, R.G.; Benson, A.J.; Crain, R.A. What shapes the galaxy mass function? Exploring the roles of supernova-driven winds and active galactic nuclei. {\em Mon.~Not.~R.~Astron.~Soc.} {\bf 2012}, {\em 422}, 2816--2840, doi:10.1111/j.1365-2966.2012.20516.x.


\bibitem[Bustard~et al.(2016)]{Bustard16}Bustard, C.; Zweibel, E.G.; D'Onghia, E.A. Versatile Family of Galactic Wind Models. {\em Astrophys.~J.} {\bf 2016}, {\em 819}, 29, doi:10.3847/0004-637X/819/1/29.




\bibitem[Creasey~et al.(2013)]{Creasey13}Creasey, P.; Theuns, T.; Bower, R.G. How supernova explosions power galactic winds. {\em Mon.~Not.~R.~Astron.~Soc.} {\bf 2013}, {\em 429}, 1922--1948, doi:10.1093/mnras/sts439.

\bibitem[Girichidis~et al.(2016b)]{Girichidis16b}Girichidis, P.; Walch, S.; Naab, T. The SILCC (SImulating the LifeCycle of molecular Clouds) project--II. Dynamical evolution of the supernova-driven ISM and the launching of outflows. {\em Mon.~Not.~R.~Astron.~Soc.} {\bf 2016}, {\em 456}, 3432--3455, doi:10.1093/mnras/stv2742.

\bibitem[Martizzi~et al.(2016)]{Martizzi16}Martizzi, D.; Fielding, D.; Faucher-Gigu\`{e}re, C.-A.; Quataert, E. Supernova feedback in a local vertically stratified medium: Interstellar turbulence and galactic winds. {\em Mon.~Not.~R.~Astron.~Soc.} {\bf 2016}, {\em 459}, 2311--2326, doi:10.1093/mnras/stw745.

\bibitem[Gatto~et al.(2017)]{Gatto17}Gatto, A.; Walch, S.; Naab, T. The SILCC project--III. Regulation of star formation and outflows by stellar winds and supernovae. {\em Mon.~Not.~R.~Astron.~Soc.} {\bf 2017}, {\em 466}, 1903--1924, doi:10.1093/mnras/stw3209.

\bibitem[Kim~et al.(2017)]{Kim17a}Kim, C.-G.; Ostriker, E.C.; Raileanu, R. Superbubbles in the Multiphase ISM and the Loading of Galactic Winds. {\em Astrophys.~J.} {\bf 2017}, {\em 834}, 25, doi:10.3847/1538-4357/834/1/25.

\bibitem[Li~et al.(2017)]{Li17}Li, M.; Bryan, G.L.; Ostriker, J.P. Quantifying Supernovae-driven Multiphase Galactic Outflows. {\em Astrophys.~J.} {\bf 2017}, {\em 841}, 101, doi:10.3847/1538-4357/aa7263.

\bibitem[Kim \& Ostriker(2018)]{Kim18}Kim, C.-G.; Ostriker, E.C. Numerical Simulations of Multiphase Winds and Fountains from Star-forming Galactic Disks. I. Solar Neighborhood TIGRESS Model. {\em Astrophys.~J.} {\bf 2018}, {\em 853}, 173, doi:10.3847/1538-4357/aaa5ff.



\bibitem[Thompson~et al.(2016)]{Thompson16}Thompson, T.A.; Quataert, E.; Zhang, D.; Weinberg, D.H. An origin for multiphase gas in galactic winds and haloes. {\em Mon.~Not.~R.~Astron.~Soc.} {\bf 2016}, {\em 455}, 1830--1844, doi:10.1093/mnras/stv2428.

\bibitem[Wang(1995a)]{Wang95a}Wang, B. Ly-alpha Forests and Cooling Outflow from Dwarf Galaxies. {\em Astrophys.~J.~Lett.} {\bf 1995}, {\em 444}, 17--20, doi:10.1086/187849.

\bibitem[Wang(1995b)]{Wang95b}Wang, B. Cooling Gas Outflows from Galaxies. {\em Astrophys.~J.} {\bf 1995}, {\em 444}, 590--609, doi:10.1086/175633.

\bibitem[Efstathiou(2000)]{Efstathiou00}Efstathiou, G. A model of supernova feedback in galaxy formation. {\em Mon.~Not.~R.~Astron.~Soc.} {\bf 2000}, {\em 317}, 697--719, doi:10.1046/j.1365-8711.2000.03665.x.

\bibitem[Silich~et al.(2003)]{Silich03}Silich, S.; Tenorio-Tagle, G.; Mu'~{n}oz-Tu'~{n}\'{o}n, C. On the Rapidly Cooling Interior of Supergalactic Winds. {\em Astrophys.~J.} {\bf 2003}, {\em 590}, 791--796, doi:10.1086/375133.

\bibitem[Silich~et al.(2004)]{Silich04}Silich, S.; Tenorio-Tagle, G.; Rodr\'{i}guez-Gonz\'{a}lez, A. Winds Driven by Super Star Clusters: The Self-Consistent Radiative Solution. {\em Astrophys.~J.} {\bf 2004}, {\em 610}, 226--232, doi:10.1086/421702.

\bibitem[Tenorio-Tagle~et al.(2007)]{Tenorio07}Tenorio-Tagle, G.; W\"{u}nsch, R.; Silich, S.; Palou\v{s}, J. Hydrodynamics of the Matter Reinserted within Super Stellar Clusters. {\em Astrophys.~J.} {\bf 2007}, {\em 658}, 1196--1202, doi:10.1086/511671.

\bibitem[W\"{u}nsch~et al.(2011)]{Wunsch11}W\"{u}nsch, R.; Silich, S.; Palou\v{s}, J. Evolution of Super Star Cluster Winds with Strong Cooling. {\em Astrophys.~J.} {\bf 2011}, {\em 740}, 75, doi:10.1088/0004-637X/740/2/75.

\bibitem[Scannapieco(2017)]{Scannapieco17}Scannapieco, E. The Production of Cold Gas Within Galaxy Outflows. {\em Astrophys.~J.} {\bf 2017}, {\em 837}, 28, doi:10.3847/1538-4357/aa5d0d.



\bibitem[Heckman~et al.(1990)]{Heckman90}Heckman, T.M.; Armus, L.; Miley, G.K. On the nature and implications of starburst-driven galactic superwinds. {\em Astrophys.~J.~Suppl.} {\bf 1990}, {\em 74}, 833--868, doi:10.1086/191522.


\bibitem[Cooper~et al.(2008)]{Cooper08}Cooper, J.L.; Bicknell, G.V.; Sutherland, R.S.; Bland-Hawthorn, J. Three-Dimensional Simulations of a Starburst-driven Galactic Wind. {\em Astrophys.~J.} {\bf 2008}, {\em 674}, 157--171, doi:10.1086/524918.

\bibitem[Fujita~et al.(2009)]{Fujita09}Fujita, A.; Martin, C.L.; Mac Low, M.-M.; New, K.C.B.; Weaver, R. The Origin and Kinematics of Cold Gas in Galactic Winds: Insight from Numerical Simulations. {\em Astrophys.~J.} {\bf 2009}, {\em 698}, 693--714, doi:10.1088/0004-637X/698/1/693.


\bibitem[Cowie \& McKee(1977)]{CM77}Cowie, L.L.; McKee, C.F. The evaporation of spherical clouds in a hot gas. I--Classical and saturated mass loss rates. {\em Astrophys.~J.} {\bf 1977}, {\em 211}, 135--146, doi:10.1086/154911.

\bibitem[McKee~et al.(1978)]{McKee78}McKee, C.F.; Cowie, L.L.; Ostriker, J.P. The acceleration of high-velocity clouds in supernova remnants. {\em Astrophys.~J.~Lett.} {\bf 1978}, {\em 219}, L23--L28, doi:10.1086/182599.

\bibitem[Klein, McKee \& Colella(1994)]{Klein94}Klein, R.I.; McKee, C.F.; Colella, P. On the hydrodynamic interaction of shock waves with interstellar clouds. I: Nonradiative shocks in small clouds. {\em Astrophys.~J.} {\bf 1994}, {\em 420}, 213--236, doi:10.1086/173554.


\bibitem[Stone \& Norman(1992)]{Stone92}Stone, J.M.; Norman, M.L. The three-dimensional interaction of a supernova remnant with an interstellar cloud. {\em Astrophys.~J.~Lett.} {\bf 1992}, {\em 390}, L17--L19, doi:10.1086/186361.

\bibitem[Mac Low~et al.(1994)]{MacLow94}Mac Low, M.-M.; McKee, C.F.; Klein, R.I.; Stone, J.M.; Norman, M.L. Shock interactions with magnetized interstellar clouds. I: Steady shocks hitting nonradiative clouds. {\em Astrophys.~J.} {\bf 1994}, {\em 433}, 757--777, doi:10.1086/174685.

\bibitem[Schiano~et al.(1995)]{Schiano95}Schiano, A.V.R.; Christiansen, W.A.; Knerr, J.M. Interstellar clouds in high-speed, supersonic flows: Two-dimensional simulations. {\em Astrophys.~J.} {\bf 1995}, {\em 439}, 237--255, doi:10.1086/175167.

\bibitem[Vietri~et al.(1997)]{Vietri97}Vietri, M.; Ferrara, A.; Miniati, F. The Survival of Interstellar Clouds against Kelvin-Helmholtz Instabilities. {\em Astrophys.~J.} {\bf 1997}, {\em 483}, 262--273, doi:10.1086/304202.

\bibitem[Marcolini~et al.(2005)]{Marcolini05}Marcolini, A.; Strickland, D.K.; D'Ercole, A.; Heckman, T.M.; Hoopes, C.G. The dynamics and high-energy emission of conductive gas clouds in supernova-driven galactic superwinds. {\em Mon.~Not.~R.~Astron.~Soc.} {\bf 2005}, {\em 362}, 626--648, doi:10.1111/j.1365-2966.2005.09343.x.

\bibitem[Orlando~et al.(2005)]{Orlando05}Orlando, S.; Peres, G.; Reale, F.; Bocchino, F.; Rosner, R.; Plewa, T.; Siegel, A. Crushing of interstellar gas clouds in supernova remnants. I. The role of thermal conduction and radiative losses. {\em Astron.~Astrophys.} {\bf 2005}, {\em 444}, 505--519, doi:10.1051/0004-6361:20052896.

\bibitem[Nakamura~et al.(2006)]{Nakamura06}Nakamura, F.; McKee, C.F.; Klein, R.I.; Fisher, R.T. On the Hydrodynamic Interaction of Shock Waves with Interstellar Clouds. II. The Effect of Smooth Cloud Boundaries on Cloud Destruction and Cloud Turbulence. {\em Astrophys. J. Suppl. Ser.} {\bf 2006}, {\em 164}, 477--505, doi:10.1086/501530.

\bibitem[Orlando~et al.(2006)]{Orlando06}Orlando, S.; Bocchino, F.; Peres, G.; Reale, F.; Plewa, T.; Rosner, R. Crushing of interstellar gas clouds in supernova remnants. II. X-ray emission. {\em Astron.~Astrophys.} {\bf 2006}, {\em 457}, 545--552, doi:10.1051/0004-6361:20065652.

\bibitem[Orlando~et al.(2008)]{Orlando08}Orlando, S.; Bocchino, F.; Reale, F.; Peres, G.; Pagano, P. The Importance of Magnetic-Field-Oriented Thermal Conduction in the Interaction of SNR Shocks with Interstellar Clouds. {\em Astrophys.~J.} {\bf 2008}, {\em 678}, 274--286, doi:10.1086/529420.

\bibitem[Shin~et al.(2008)]{Shin08}Shin, M.-S.; Stone, J.M.; Snyder, G.F. The Magnetohydrodynamics of Shock-Cloud Interaction in Three Dimensions. {\em Astrophys.~J.} {\bf 2008}, {\em 680}, 336--348, doi:10.1086/587775.

\bibitem[Cooper~et al.(2009)]{Cooper09}Cooper, J.L.; Bicknell, G.V.; Sutherland, R.S.; Bland-Hawthorn, J. Starburst-Driven Galactic Winds: Filament Formation and Emission Processes. {\em Astrophys.~J.} {\bf 2009}, {\em 703}, 330--347, doi:10.1088/0004-637X/703/1/330.

\bibitem[Pittard~et al.(2009)]{Pittard09}Pittard, J.M.; Falle, S.A.E.G.; Hartquist, T.W.; Dyson, J.E. The turbulent destruction of clouds--I. A $k-\epsilon$ treatment of turbulence in 2D models of adiabatic shock-cloud interactions. {\em Mon.~Not.~R.~Astron.~Soc.} {\bf 2009}, {\em 394}, 1351--1378, doi:10.1111/j.1365-2966.2009.13759.x.

\bibitem[Pittard~et al.(2010)]{Pittard10}Pittard, J.M.; Hartquist, T.W.; Falle, S.A.E.G. The turbulent destruction of clouds--II. Mach number dependence, mass-loss rates and tail formation. {\em Mon.~Not.~R.~Astron.~Soc.} {\bf 2010}, {\em 405}, 821--838, doi:10.1111/j.1365-2966.2010.16504.x.

\bibitem[Al\={u}zas~et al.(2014)]{Aluzas14}Al\={u}zas, R.; Pittard, J.M.; Falle, S.A.E.G.; Hartquist, T.W. Numerical simulations of a shock interacting with multiple magnetized clouds. {\em Mon.~Not.~R.~Astron.~Soc.} {\bf 2014}, {\em 444}, 971--993, doi:10.1093/mnras/stu1501.


\bibitem[McCourt~et al.(2015)]{McCourt15}McCourt, M.; O'Leary, R.M.; Madigan, A.-M.; Quataert, E. Magnetized gas clouds can survive acceleration by a hot wind. {\em Mon.~Not.~R.~Astron.~Soc.} {\bf 2015}, {\em 449}, 2--7, doi:10.1093/mnras/stv355.


\bibitem[Banda-Barrag\'{a}n~et al.(2016)]{Banda16}Banda-Barrag\'{a}n, W.E.; Parkin, E.R.; Federrath, C.; Crocker, R.M.; Bicknell, G.V. Filament formation in wind-cloud interactions--I. Spherical clouds in uniform magnetic fields. {\em Mon.~Not.~R.~Astron.~Soc.} {\bf 2016}, {\em 455}, 1309--1333, doi:10.1093/mnras/stv2405.

\bibitem[Pittard \& Parkin(2016)]{Pittard16}Pittard, J.M.; Parkin, E.R. The turbulent destruction of clouds--III. Three-dimensional adiabatic shock-cloud simulations. {\em Mon.~Not.~R.~Astron.~Soc.} {\bf 2016}, {\em 457}, 4470--4498, doi:10.1093/mnras/stw025.


\bibitem[Banda-Barrag\'{a}n~ et al.(2018)]{Banda18}Banda-Barrag\'{a}n, W.E.; Federrath, C.; Crocker, R.M.; Bicknell, G.V. Filament formation in wind-cloud interactions- II. Clouds with turbulent density, velocity, and magnetic fields. {\em Mon.~Not.~R.~Astron.~Soc.} {\bf 2018}, {\em 473}, 3454--3489, doi:10.1093/mnras/stx2541.


\bibitem[Goldsmith \& Pittard(2018)]{Goldsmith18}Goldsmith, K.J.A.; Pittard, J.M. A comparison of shock-cloud and wind-cloud interactions: Effect of increased cloud density contrast on cloud evolution. {\em Mon.~Not.~R.~Astron.~Soc.} {\bf 2018}, {\em 476}, 2209--2219, doi:10.1093/mnras/sty401.

\bibitem[Faucher-Gigu\`{e}re~et al.(2012)]{Andre12}Faucher-Gigu\`{e}re, C.-A.; Quataert, E.; Murray, N. A physical model of FeLoBALs: Implications for quasar feedback. {\em Mon.~Not.~R.~Astron.~Soc.} {\bf 2012}, {\em 420}, 1347--1354, doi:10.1111/j.1365-2966.2011.20120.x.

\bibitem[Krolik~et al.(1981)]{Krolik81}Krolik, J.H.; McKee, C.F.; Tarter, C.B. Two-phase models of quasar emission line regions. {\em Astrophys.~J.} {\bf 1981}, {\em 249}, 422--442, doi:10.1086/159303.

\bibitem[Adebahr~et al.(2017)]{Adebahr17}Adebahr, B.; Krause, M.; Klein, U.; Heald, G.; Dettmar, R.-J. M 82--A radio continuum and polarisation study. II. Polarisation and rotation measures. {\em Astron.~Astrophys.} {\bf 2017}, {\em 608}, 29, doi:10.1051/0004-6361/201629616.



\bibitem[Steidel~et al.(2010)]{Steidel10}Steidel, C.C.; Erb, D.K.; Shapley, A.E. The Structure and Kinematics of the Circumgalactic Medium from Far-ultraviolet Spectra of $z approx 2-3$ Galaxies. {\em Astrophys.~J.} {\bf 2010}, {\em 717}, 289--322, doi:10.1088/0004-637X/717/1/289.

\bibitem[Prochaska~et al.(2011)]{Prochaska11}Prochaska, J.X.; Weiner, B.; Chen, H.-W.; Mulchaey, J.; Cooksey, K. Probing the Intergalactic Medium/Galaxy Connection. V. On the Origin of Ly$\alpha$ and O VI Absorption at $z < 0.2$. {\em Astrophys.~J.} {\bf 2011}, {\em 740}, 91, doi:10.1088/0004-637X/740/2/91.
\bibitem[Tumlinson~et al.(2013)]{Tumlinson13}Tumlinson, J.; Thom, C.; Werk, J.K. The COS-Halos Survey: Rationale, Design, and a Census of Circumgalactic Neutral Hydrogen. {\em Astrophys.~J.} {\bf 2013}, {\em 777}, 59, doi:10.1088/0004-637X/777/1/59.

\bibitem[Werk~et al.(2014)]{Werk14}Werk, J.K.; Prochaska, J.X.; Tumlinson, J. The COS-Halos Survey: Physical Conditions and Baryonic Mass in the Low-redshift Circumgalactic Medium. {\em Astrophys.~J.} {\bf 2014}, {\em 792}, 8, doi:10.1088/0004-637X/792/1/8.

\bibitem[Werk~et al.(2016)]{Werk16}Werk, J.K.; Prochaska, J.X.; Cantalupo, S. The COS-Halos Survey: Origins of the Highly Ionized Circumgalactic Medium of Star-Forming Galaxies. {\em Astrophys.~J.} {\bf 2016}, {\em 833}, 54, doi:10.3847/1538-4357/833/1/54.

\bibitem[Borthakur~et al.(2016)]{Borthakur16}Borthakur, S.; Heckman, T.; Tumlinson, J. The Properties of the Circumgalactic Medium in Red and Blue Galaxies: Results from the COS-GASS+COS-Halos Surveys. {\em Astrophys.~J.} {\bf 2016}, {\em 833}, 259, doi:10.3847/1538-4357/833/2/259.



\bibitem[Whitworth(1979)]{Whitworth79}Whitworth, A. The erosion and dispersal of massive molecular clouds by young stars. {\em Mon.~Not.~R.~Astron.~Soc.} {\bf 1979}, {\em 186}, 59--67, doi:10.1093/mnras/186.1.59.

\bibitem[McKee~et al.(1984)]{McKee84}McKee, C.F.; van Buren, D.; Lazareff, B. Photoionized Stellar Wind Bubbles in a Cloudy Medium. {\em Astrophys.~J.~Lett.} {\bf 1984}, {\em 278}, L115-L118, doi:10.1086/184237.

\bibitem[McKee(1989)]{McKee89}McKee, C.F. Photoionization-Regulated Star Formation and the Structure of Molecular Clouds. {\em Astrophys.~J.} {\bf 1989}, {\em 345}, 782--801, doi:10.1086/167950.

\bibitem[Matzner(2002)]{Matzner02}Matzner, C.D. On the Role of Massive Stars in the Support and Destruction of Giant Molecular Clouds. {\em Astrophys.~J.} {\bf 2002}, {\em 566}, 302--314, doi:10.1086/338030.

\bibitem[Dale~et al.(2012)]{Dale12}Dale, J.E.; Ercolano, B.; Bonnell, I.A. Ionizing feedback from massive stars in massive clusters--II. Disruption of bound clusters by photoionization. {\em Mon.~Not.~R.~Astron.~Soc.} {\bf 2012}, {\em 424}, 377--392, doi:10.1111/j.1365-2966.2012.21205.x.

\bibitem[Walch~et al.(2012)]{Walch12}Walch, S.K.; Whitworth, A.P.; Bisbas, T.; Wünsch, R.; Hubber, D. Dispersal of molecular clouds by ionizing radiation. {\em Mon.~Not.~R.~Astron.~Soc.} {\bf 2012}, {\em 427}, 625--636, doi:10.1111/j.1365-2966.2012.21767.x.

\bibitem[Emerick~et al.(2018)]{Emerick08}Emerick, A.; Bryan, G.L.; Mac Low, M.-M. Stellar Radiation is Critical for Regulating Star Formation and Driving Outflows in Low Mass Dwarf Galaxies. {\em Astrophys.~J.~Lett.} {\bf 2018}, {\em 865}, L22, doi:10.3847/2041-8213/aae315.



\bibitem[Krumholz \& Matzner(2009)]{Krumholz09b}Krumholz, M.R.; Matzner, C.D. The Dynamics of Radiation-pressure-dominated H II Regions. {\em Astrophys.~J.} {\bf 2009}, {\em 703}, 1352--1362, doi:10.1088/0004-637X/703/2/1352.

\bibitem[Kuiper~et al.(2011)]{Kuiper11}Kuiper, R.; Klahr, H.; Beuther, H.; Henning, T. Three-dimensional Simulation of Massive Star Formation in the Disk Accretion Scenario. {\em Astrophys.~J.} {\bf 2011}, {\em 732}, 20, doi:10.1088/0004-637X/732/1/20.

\bibitem[Kuiper~et al.(2012)]{Kuiper12}Kuiper, R.; Klahr, H.; Beuther, H.; Henning, TH. On the stability of radiation-pressure-dominated cavities. {\em Astron.~Astrophys.} {\bf 2012}, {\em 537}, 122, doi:10.1051/0004-6361/201117808.

\bibitem[Klassen~et al.(2016)]{Klassen16}Klassen, M.; Pudritz, R.E.; Kuiper, R.; Peters, T.; Banerjee, R. Simulating the Formation of Massive Protostars. I. Radiative Feedback and Accretion Disks. {\em Astrophys.~J.} {\bf 2016}, {\em 823}, 28, doi:10.3847/0004-637X/823/1/28.

\bibitem[Kuiper~et al.(2016)]{Kuiper16}Kuiper, R.; Turner, N.J.; Yorke, H.W. Protostellar Outflows and Radiative Feedback from Massive Stars. II. Feedback, Star-formation Efficiency, and Outflow Broadening. {\em Astrophys.~J.} {\bf 2016}, {\em 832}, 40, doi:10.3847/0004-637X/832/1/40.

\bibitem[Rosen~et al.(2016)]{Rosen16}Rosen, A.L.; Krumholz, M.R.; McKee, C.F.; Klein, R.I. An unstable truth: How massive stars get their mass. {\em Mon.~Not.~R.~Astron.~Soc.} {\bf 2016}, {\em 463}, 2553--2573, doi:10.1093/mnras/stw2153.

\bibitem[Crocker~et al.(2018)]{Crocker18}Crocker, R.M.; Krumholz, M.R.; Thompson, T.A.; Baumgardt, H.; Mackey, D. Radiation Pressure Limits on the Star Formation Efficiency and Surface Density of Compact Stellar Systems. {\em Mon.~Not.~R.~Astron.~Soc.} {\bf 2018}, {\em 481}, 4895--4906, doi: 10.1093/mnras/sty2659.


\bibitem[Sales~et al.(2016)]{Sales16}Sales, L.V.; Marinacci, F.; Springel, V.; Petkova, M. Stellar feedback by radiation pressure and photoionization. {\em Mon.~Not.~R.~Astron.~Soc.} {\bf 2014}, {\em 439}, 2990--3006, doi:10.1093/mnras/stu155.

\bibitem[Ishiki \& Okamoto(2017)]{Ishiki17}Ishiki, S.; Okamoto, T. Radiation feedback in dusty clouds. {\em Mon.~Not.~R.~Astron.~Soc.~Lett.} {\bf 2017}, {\em 466}, L123--L127, doi:10.1093/mnrasl/slw253.

\bibitem[Ishiki~et al.(2018)]{Ishiki18}Ishiki, S.; Okamoto, T.; Inoue, A.K. The effect of radiation pressure on spatial distribution of dust inside H II regions. {\em Mon.~Not.~R.~Astron.~Soc.} {\bf 2018}, {\em 474}, 1935--1943, doi:10.1093/mnras/stx2833.

\bibitem[Haid~et al.(2018)]{Haid18}Haid, S.; Walch, S.; Seifried, D.; W\"{u}nsch, R.; Dinnbier, F.; Naab, T. The relative impact of photoionizing radiation and stellar winds on different environments. {\em Mon.~Not.~R.~Astron.~Soc.} {\bf 2018}, {\em 478}, 4799--4815, doi:10.1093/mnras/sty1315.


\bibitem[Harwit(1962)]{Harwit62}Harwit, M. Dust, Radiation Pressure, and Star Formation. {\em Astrophys.~J.} {\bf 1962}, {\em 136}, 832--843, doi:10.1086/147440.

\bibitem[Chiao \& Wickramasinghe(1972)]{Chiao72}Chiao, R.Y.; Wickramasinghe, N.C. Radiation-driven efflux and circula-tion of dust in spiral galaxies. {\em Mon.~Not.~R.~Astron.~Soc.} {\bf 1972}, {\em 159}, 361--373, doi:10.1093/mnras/stx2833.

\bibitem[Chiao \& Wickramasinghe(1973)]{Chiao73}Chiao, R.Y.; Wickramasinghe, N.C. The Expulsion of Dust from Galaxies. {\em Astrophys.~J.~Lett.} {\bf 1973}, {\em 14}, L19--L23.

\bibitem[Ferrara~et al.(1990)]{Ferrara90}Ferrara, A.; Ferrini, F.; Barsella, B.; Aiello, S. Removal of dust from spiral galaxies. {\em Astron.~Astrophys.} {\bf 1990}, {\em 240}, 259--261.

\bibitem[Murray~et al.(2005)]{Murray05}Murray, N.; Quataert, E.; Thompson, T.A. On the Maximum Luminosity of Galaxies and Their Central Black Holes: Feedback from Momentum-driven Winds. {\em Astrophys.~J.} {\bf 2005}, {\em 618}, 569--585, doi:10.1086/426067.

\bibitem[Thompson~et al.(2005)]{Thompson05}Thompson, T.A.; Quataert, E.; Murray, N. Radiation Pressure-supported Starburst Disks and Active Galactic Nucleus Fueling. {\em Astrophys.~J.} {\bf 2005}, {\em 630}, 167--185, doi:10.1086/431923.

\bibitem[Murray~et al.(2011)]{Murray11}Murray, N.; M\'{e}nard, B.; Thompson, T.A. Radiation Pressure from Massive Star Clusters as a Launching Mechanism for Super-galactic Winds. {\em Astrophys.~J.} {\bf 2011}, {\em 735}, 66--77, doi:10.1088/0004-637X/735/1/66.

\bibitem[Hopkins~et al.(2012b)]{Hopkins12b}Hopkins, P.F.; Quataert, E.; Murray, N. Stellar feedback in galaxies and the origin of galaxy-scale winds. {\em Mon.~Not.~R.~Astron.~Soc.} {\bf 2012}, {\em 421}, 3522--3537, doi:10.1111/j.1365-2966.2012.20593.x.


\bibitem[O'dell~et al.(1967)]{Odell67}O'dell, C.R.; York, D.G.; Henize, K.G. Structure of the Barnard Loop Nebula as Determined from Gemini 11 Photographs. {\em Astrophys.~J.} {\bf 1967}, {\em 150}, 835--844, doi:10.1086/149386.

\bibitem[Scoville~et al.(2001)]{Scoville01}Scoville, N.Z.; Polletta, M.; Ewald, S.; Stolovy, S.R.; Thompson, R.; Rieke, M. High-Mass, OB Star Formation in M51: Hubble Space Telescope H$\alpha$ and Pa$\alpha$ Imaging. {\em Astron.~J.} {\bf 2001}, {\em 122}, 3017--3045, doi:10.1086/323445.

\bibitem[Murray~et al.(2010)]{Murray10}Murray, N.; Quataert, E.; Thompson, T.A. The Disruption of Giant Molecular Clouds by Radiation Pressure \& the Efficiency of Star Formation in Galaxies. {\em Astrophys.~J.} {\bf 2010}, {\em 709}, 191--209, doi:10.1088/0004-637X/709/1/191.

\bibitem[Skinner \& Ostriker(2015)]{Skinner15}Skinner, M.A.; Ostriker, E.C. Numerical Simulations of Turbulent Molecular Clouds Regulated by Reprocessed Radiation Feedback from Nascent Super Star Clusters. {\em Astrophys.~J.} {\bf 2015}, {\em 809}, 187, doi:10.1088/0004-637X/809/2/187.

\bibitem[Raskutti~et al.(2016)]{Raskutti16}Raskutti, S.; Ostriker, E.C.; Skinner, M.A. Numerical Simulations of Turbulent Molecular Clouds Regulated by Radiation Feedback Forces. I. Star Formation Rate and Efficiency. {\em Astrophys.~J.} {\bf 2016}, {\em 829}, 130, doi:10.3847/0004-637X/829/2/130.

\bibitem[Gupta~et al.(2016)]{Gupta16}Gupta, S.; Nath, B.B.; Sharma, P.; Shchekinov, Y. How radiation affects superbubbles: Through momentum injection in early phase and photo-heating thereafter. {\em Mon.~Not.~R.~Astron.~Soc.} {\bf 2016}, {\em 462}, 4532--4548, doi:10.1093/mnras/stw1920.

\bibitem[Raskutti~et al.(2017)]{Raskutti17}Raskutti, S.; Ostriker, E.C.; Skinner, M.A. Numerical Simulations of Turbulent Molecular Clouds Regulated by Radiation Feedback Forces. II. Radiation-Gas Interactions and Outflows. {\em Astrophys.~J.} {\bf 2017}, {\em 850}, 112, doi:10.3847/1538-4357/aa965e.

\bibitem[Tsang \& Milosavljevi\'{c}(2018)]{Tsang18}Tsang, B.T.-H.; Milosavljevi\'{c}, M. Radiation pressure in super star cluster formation. {\em Mon.~Not.~R.~Astron.~Soc.} {\bf 2018}, {\em 478}, 4142--4161, doi:10.1093/mnras/sty1217.


\bibitem[Renaud~et al.(2013)]{Renaud13}Renaud, F.; Bournaud, F.; Emsellem, E. A sub-parsec resolution simulation of the Milky Way: Global structure of the interstellar medium and properties of molecular clouds. {\em Mon.~Not.~R.~Astron.~Soc.} {\bf 2013}, {\em 436}, 1836--1851, doi:10.1093/mnras/stt1698.

\bibitem[Kannan~et al.(2014)]{Kannan14}Kannan, R.; Stinson, G.S.; Macci\`{o}, A.V. Galaxy formation with local photoionization feedback--I. Methods. {\em Mon.~Not.~R.~Astron.~Soc.} {\bf 2014}, {\em 437}, 2882--2893, doi:10.1093/mnras/stt2098.

\bibitem[Tremblin~et al.(2014)]{Tremblin14}Tremblin, P.; Anderson, L.D.; Didelon, P. Age, size, and position of H II regions in the Galaxy. Expansion of ionized gas in turbulent molecular clouds. {\em Astron.~Astrophys.} {\bf 2014}, {\em 568}, 4, doi:10.1051/0004-6361/201423959.

\bibitem[Bruzual \& Charlot(2003)]{Bruzual03}Bruzual, G.; Charlot, S. Stellar population synthesis at the resolution of 2003. {\em Mon.~Not.~R.~Astron.~Soc.} {\bf 2003}, {\em 344}, 1000--1028, doi:10.1046/j.1365-8711.2003.06897.x.


\bibitem[Proga~et al.(1998)]{Proga98}Proga, D.; Stone, J.M.; Drew, J.E. Radiation-driven winds from luminous accretion discs. {\em Mon.~Not.~R.~Astron.~Soc.} {\bf 1998}, {\em 295}, 595--617, doi:10.1046/j.1365-8711.2003.06897.x.

\bibitem[Proga(1999)]{Proga99a}Proga, D. Comparison of theoretical radiation-driven winds from stars and discs. {\em Mon.~Not.~R.~Astron.~Soc.} {\bf 1999}, {\em 304}, 938--946, doi:10.1046/j.1365-8711.1999.02408.x.

\bibitem[Proga~et al.(1999)]{Proga99b}Proga, D.; Stone, J.M.; Drew, J.E. Line-driven disc wind models with an improved line force. {\em Mon.~Not.~R.~Astron.~Soc.} {\bf 1999}, {\em 310}, 476--482, doi:10.1046/j.1365-8711.1999.02935.x.

\bibitem[Proga \& Kallman(2002)]{Proga02}Proga, D.; Kallman, T.R. On the Role of the Ultraviolet and X-Ray Radiation in Driving a Disk Wind in X-Ray Binaries. {\em Astrophys.~J.} {\bf 2002}, {\em 565}, 455--470, doi:10.1086/324534.

\bibitem[Proga \& Kallman(2004)]{Proga04}Proga, D.; Kallman, T.R. Dynamics of Line-driven Disk Winds in Active Galactic Nuclei. II. Effects of Disk Radiation. {\em Astrophys.~J.} {\bf 2004}, {\em 616}, 688--695, doi:10.1086/425117.

\bibitem[Risaliti~et al.(2010)]{Risaliti10}Risaliti, G.; Elvis, M. A non-hydrodynamical model for acceleration of line-driven winds in active galactic nuclei. {\em Astron.~Astrophys.} {\bf 2010}, {\em 516}, 89, doi:10.1051/0004-6361/200912579.

\bibitem[Dyda \& Proga(2018a)]{Dyda18a}Dyda, S.; Proga, D. Non-axisymmetric line-driven disc winds--I. Disc perturbations. {\em Mon.~Not.~R.~Astron.~Soc.} {\bf 2018}, {\em 475}, 3786--3796, doi:10.1093/mnras/sty030.

\bibitem[Dyda \& Proga(2018b)]{Dyda18b}Dyda, S.; Proga, D. Non-axisymmetric line-driven disc winds II--full velocity gradient. {\em Mon.~Not.~R.~Astron.~Soc.} {\bf 2018}, {\em 478}, 5006--5016, doi:10.1093/mnras/sty1257.

\bibitem[Springel(2005)]{Springel05}Springel, V. The cosmological simulation code GADGET-2. {\em Mon.~Not.~R.~Astron.~Soc.} {\bf 2005}, {\em 364}, 1105--1134, doi:10.1111/j.1365-2966.2005.09655.x.



\bibitem[Gilman(1972)]{Gilman72}Gilman, R.C. On the Coupling of Grains to the Gas in Circumstellar Envelopes. {\em Astrophys.~J.} {\bf 1972}, {\em 423}, 423--426, doi:10.1086/151800.

\bibitem[Salpeter(1974)]{Salpeter74}Salpeter, E.E. Formation and flow of dust grains in cool stellar atmospheres. {\em Astrophys.~J.} {\bf 1974}, {\em 193}, 585--592, doi:10.1086/153196.

\bibitem[Kwok(1975)]{Kwok75}Kwok, S. Radiation pressure on grains as a mechanism for mass loss in red giants. {\em Astrophys.~J.} {\bf 1975}, {\em 198}, 583--591, doi:10.1086/153637.

\bibitem[Deguchi(1980)]{Deguchi80}Deguchi, S. Grain formation in cool stellar envelopes. {\em Astrophys.~J.} {\bf 1980}, {\em 236}, 567--576, doi:10.1086/157775.

\bibitem[Kozasa~et al.(1984)]{Kozasa84}Kozasa, T.; Hasegawa, H.; Seki, J. Grain formation in the expanding gas flow around cool luminous stars. {\em Astrophys.~Space~Sci.} {\bf 1984}, {\em 98}, 61--79, doi:10.1007/BF00651951.

\bibitem[Gail \& Sedlmayr(1985)]{Gail85}Gail, H.-P.; Sedlmayr, E. Dust formation in stellar winds. II--Carbon condensation in stationary, spherically expanding winds. {\em Astron.~Astrophys.} {\bf 1985}, {\em 148}, 183--190.

\bibitem[Bowen(1988)]{Bowen88}Bowen, G.H. Dynamical Modeling of Long-Period Variable Star Atmospheres. {\em Astrophys.~J.} {\bf 1988}, {\em 329}, 299--317, doi:10.1086/166378.

\bibitem[Fleischer~et al.(1992)]{Fleischer92}Fleischer, A.J.; Gauger, A.; Sedlmayr, E. Circumstellar dust shells around long-period variables. I--Dynamical models of C-stars including dust formation, growth and evaporation. {\em Astron.~Astrophys.} {\bf 1992}, {\em 266}, 321--339.

\bibitem[Sedlmayr \& Dominik(1995)]{Sedlmayr95}Sedlmayr, E.; Dominik, C. Dust Driven Winds. {\em Space~Sci.~Rev.} {\bf 1995}, {\em 73}, 211--272, doi:10.1007/BF00751238.

\bibitem[Krueger \& Sedlmayr(1997)]{Krueger97}Krueger, D.; Sedlmayr, E. Two-fluid models for stationary dust driven winds. II. The grain size distribution in consideration of drift. {\em Astron.~Astrophys.} {\bf 1997}, {\em 321}, 557--567.

\bibitem[Ferrarotti \& Gail(2006)]{Ferrarotti06}Ferrarotti, A.S.; Gail, H.-P. Composition and quantities of dust produced by AGB-stars and returned to the interstellar medium. {\em Astron.~Astrophys.} {\bf 2006}, {\em 447}, 553--576, doi:10.1051/0004-6361:20041198.

\bibitem[Takigawa~et al.(2017)]{Takigawa17}Takigawa, A.; Kamizuka, T.; Tachibana, S.; Yamamura, I. Dust formation and wind acceleration around the aluminum oxide-rich AGB star W Hydrae. {\em Sci.~Adv.} {\bf 2017}, {\em 3}, 2149--2153, doi:10.1126/sciadv.aao2149.

\bibitem[Lamers \& Cassinelli(1999)]{Lamers99}Lamers, H.J.G.L.M.; Cassinelli, J.P. \emph{Introduction to Stellar Winds}; Cambridge University Press: Cambridge, UK, {1999}, ISBN 0-521-59565-7.


\bibitem[Kobayashi~et al.(2009)]{Kobayashi09}Kobayashi, H.; Watanabe, S.-I.; Kimura, H.; Yamamoto, T. Dust ring formation due to sublimation of dust grains drifting radially inward by the Poynting-Robertson drag: An analytical model. {\em Icarus} {\bf 2009}, {\em 201}, 395--405, doi:10.1016/j.icarus.2009.01.002.

\bibitem[Kobayashi~et al.(2011)]{Kobayashi11}Kobayashi, H.; Kimura, H.; Watanabe, S.-I.; Yamamoto, T.; Müller, S. Sublimation temperature of circumstellar dust particles and its importance for dust ring formation. {\em Earth~Planets~Space} {\bf 2011}, {\em 63}, 1067--1075, doi:10.5047/eps.2011.03.012.

\bibitem[Baskin \& Laor(2018)]{Baskin18}Baskin, A.; Laor, A. Dust inflated accretion disc as the origin of the broad line region in active galactic nuclei. {\em Mon.~Not.~R.~Astron.~Soc.} {\bf 2018}, {\em 474},1970--1994, doi:10.1093/mnras/stx2850.

\bibitem[Draine \& Salpeter(1979a)]{Draine79a}Draine, B.T.; Salpeter, E.E. On the physics of dust grains in hot gas. {\em Astrophys.~J.} {\bf 1979}, {\em 231}, 77--94, doi:10.1086/157165.

\bibitem[Draine \& Salpeter(1979b)]{Draine79b}Draine, B.T.; Salpeter, E.E. Destruction mechanisms for interstellar dust. {\em Astrophys.~J.} {\bf 1979}, {\em 231}, 438--455, doi:10.1086/157206.

\bibitem[Covatto \& Aannestad(2000)]{Covatto00}Covatto, C.; Aannestad, P.A. Sputtering in the outflows of cool stars. {\em Mon.~Not.~R.~Astron.~Soc.} {\bf 2000}, {\em 318}, 67--72, doi:10.1046/j.1365-8711.2000.03612.x.

\bibitem[Gray \& Edmunds(2004)]{Gray04}Gray, M.D.; Edmunds, M.G. Modification of dust-grain structure by sputtering. {\em Mon.~Not.~R.~Astron.~Soc.} {\bf 2004}, {\em 349}, 491--502, doi:10.1111/j.1365-2966.2004.07502.x.

\bibitem[Nath~et al.(2008)]{Nath08}Nath, B.B.; Laskar, T.; Shull, J.M. Dust Sputtering by Reverse Shocks in Supernova Remnants. {\em Astrophys.~J.} {\bf 2008}, {\em 682}, 1055--1064, doi:10.1086/589224.

%

\bibitem[Everett \& Churchwell(2010)]{Everett10b}Everett, J.E.; Churchwell, E. Dusty Wind-blown Bubbles. {\em Astrophys.~J.} {\bf 2010}, {\em 713}, 592--602,
doi:10.1088/0004-637X/713/1/592.

\bibitem[Jones \& Nuth(2011)]{Jones11}Jones, A.P.; Nuth, J.A. Dust destruction in the ISM: A re-evaluation of dust lifetimes. {\em Astron.~Astrophys.} {\bf 2011}, {\em 530}, 44, doi:10.1051/0004-6361/201014440.

\bibitem[Draine(2011b)]{Draine11b}Draine, B.T. On Radiation Pressure in Static, Dusty H II Regions. {\em Astrophys.~J.} {\bf 2011}, {\em 732}, 100, doi:10.1088/0004-637X/732/2/100.

\bibitem[Berruyer \& Frisch(1983)]{Berruyer83}Berruyer, N.; Frisch, H. Dust-driven winds. I--A two-fluid model and its numerical solution. {\em Astron.~Astrophys.} {\bf 1983}, {\em 126}, 269--277.

\bibitem[Mastrodemos~et al.(1996)]{Mastrodemos96}Mastrodemos, N.; Morris, M.; Castor, J. On the Stability of the Dust-Gas Coupling in Winds from Late-Type Stars. {\em Astrophys.~J.} {\bf 1996}, {\em 468}, 851--860, doi:10.1086/177741.

\bibitem[Goldsmith(2001)]{Goldsmith01}Goldsmith, P.F. Molecular Depletion and Thermal Balance in Dark Cloud Cores. {\em Astrophys.~J.} {\bf 2001}, {\em 557}, 736--746, doi:10.1086/322255.

\bibitem[Narayanan~et al.(2011)]{Narayanan11}Narayanan, D.; Krumholz, M.R.; Ostriker, E.C.; Hernquist, L. The CO-H$_2$ conversion factor in disc galaxies and mergers. {\em Mon.~Not.~R.~Astron.~Soc.} {\bf 2011}, {\em 418}, 664--679, doi:10.1111/j.1365-2966.2011.19516.x.

\bibitem[Narayanan~et al.(2012)]{Narayanan12}Narayanan, D.; Krumholz, M.R.; Ostriker, E.C.; Hernquist, L. A general model for the CO-H$_2$ conversion factor in galaxies with applications to the star formation law. {\em Mon.~Not.~R.~Astron.~Soc.} {\bf 2012}, {\em 421}, 3127--3146, doi:10.1111/j.1365-2966.2012.20536.x.


\bibitem[Semenov~et al.(2003)]{Semenov03}Semenov, D.; Henning, TH.; Helling, CH.; Ilgner, M.; Sedlmayr, E. Rosseland and Planck mean opacities for protoplanetary discs. {\em Astron.~Astrophys.} {\bf 2003}, {\em 410}, 611--621, doi:10.1051/0004-6361:20031279.

\bibitem[Ferguson~et al.(2005)]{Ferguson05}Ferguson, J.W.; Alexander, D.R.; Allard, F. Low-Temperature Opacities. {\em Astrophys.~J.} {\bf 2005}, {\em 623}, 585--596, doi:10.1086/428642.

\bibitem[Andrews \& Thompson(2011)]{Andrews11}Andrews, B.H.; Thompson, T.A. Assessing Radiation Pressure as a Feedback Mechanism in Star-forming Galaxies. {\em Astrophys.~J.} {\bf 2011}, {\em 727}, 97, doi:10.1088/0004-637X/727/2/97.

\bibitem[Sharp(1992)]{Sharp92}Sharp, C.M. Molecular opacities for solar and enhanced CNO abundances--Relevance for accretion disks. {\em Astron.~Astrophys.~Suppl.~Ser.} {\bf 1992}, {\em 94}, 1--12.

\bibitem[Seaton~et al.(1994)]{Seaton94}Seaton, M.J.; Yan, Y.; Mihalas, D.; Pradhan, A.K. Opacities for Stellar Envelopes. {\em Mon.~Not.~R.~Astron.~Soc.} {\bf 1994}, {\em 266}, 805--828, doi:10.1093/mnras/266.4.805.

\bibitem[Iglesias \& Rogers(1996)]{Iglesias96}Iglesias, C.A.; Rogers, F.J. Updated Opal Opacities. {\em Astrophys.~J.} {\bf 1996}, {\em 464}, 943--953, doi:10.1086/177381.

\bibitem[Bell \& Lin(1994)]{Bell94}Bell, K.R.; Lin, D.N.C. Using FU Orionis outbursts to constrain self-regulated protostellar disk models. {\em Astrophys.~J.} {\bf 1994}, {\em 427}, 987--1004, doi:10.1086/174206.

\bibitem[Pollack~et al.(1994)]{Pollack94}Pollack, J.B.; Hollenbach, D.; Beckwith, S.; Simonelli, D.P.; Roush, T.; Fong, W. Composition and radiative properties of grains in molecular clouds and accretion disks. {\em Astrophys.~J.} {\bf 1994}, {\em 421}, 615--639, doi:10.1086/173677.

\bibitem[Alexander(1975)]{Alexander75}Alexander, D.R. Low-Temperature Rosseland Opacity Tables. {\em Astrophys.~J.~Suppl.~Ser.} {\bf 1975}, {\em 29}, 363--374, doi:10.1086/190349.

\bibitem[Alexander \& Ferguson(1994)]{Alexander94}Alexander, D.R.; Ferguson, J.W. Low-temperature Rosseland Opacities. {\em Astrophys.~J.} {\bf 1994}, {\em 437}, 879--891, doi:10.1086/175039.





\bibitem[Coker~et al.(2013)]{Coker13}Coker, C.T.; Thompson, T.A.; Martini, P. Dust Scattering and the Radiation Pressure Force in the M82 Superwind. {\em Astrophys.~J.} {\bf 2013}, {\em 778}, 79, doi:10.1088/0004-637X/778/1/79.

\bibitem[Wilson~et al.(2014)]{Wilson14}Wilson, C.D.; Rangwala, N.; Glenn, J.; Maloney, P.R.; Spinoglio, L.; Pereira-Santaella, M. Extreme Dust Disks in Arp 220 as Revealed by ALMA. {\em Astrophys.~J.~Lett.} {\bf 2014}, {\em 789}, 36, doi:10.1088/2041-8205/789/2/L36.

\bibitem[Barcos-Mu\~{n}oz~et al.(2018)]{Barcos18}Barcos-Mu\~{n}oz, L.; Aalto, S.; Thompson, T.A. Fast, Collimated Outflow in the Western Nucleus of Arp 220. {\em Astrophys.~J.~Lett.} {\bf 2018}, {\em 853}, 28, doi:10.3847/2041-8213/aaa28d.

\bibitem[Barcos-Mu\~{n}oz~et al.(2017)]{Barcos17}Barcos-Mu\~{n}oz, L.; Leroy, A.K.; Evans, A.S. A 33 GHz Survey of Local Major Mergers: Estimating the Sizes of the Energetically Dominant Regions from High-resolution Measurements of the Radio Continuum. {\em Astrophys.~J.} {\bf 2017}, {\em 843}, 117, doi:10.3847/1538-4357/aa789a.

\bibitem[Zhang \& Davis(2017)]{ZD17}Zhang, D.; Davis, S.W. Radiation Hydrodynamic Simulations of Dust-driven Winds. {\em Astrophys.~J.} {\bf 2017}, {\em 839}, 54, doi:10.3847/1538-4357/aa6935.

\bibitem[Zhang~et al.(2018)]{Zhang18}Zhang, D.; Davis, S.W.; Jiang, Y.-F.; Stone, J.M. Dusty Cloud Acceleration by Radiation Pressure in Rapidly Star-forming Galaxies. {\em Astrophys.~J.} {\bf 2018}, {\em 854}, 110, doi:10.3847/1538-4357/aaa8e4.

\bibitem[Thompson \& Krumholz(2016)]{TK16}Thompson, T.A.; Krumholz, M.R. Sub-Eddington star-forming regions are super-Eddington: Momentum-driven outflows from supersonic turbulence. {\em Mon.~Not.~R.~Astron.~Soc.} {\bf 2016}, {\em 455}, 334--342, doi:10.1093/mnras/stv2331.



\bibitem[Faber \& Jackson(1976)]{Faber76}Faber, S.M.; Jackson, R.E. Velocity dispersions and mass-to-light ratios for elliptical galaxies. {\em Astrophys.~J.} {\bf 1976}, {\em 204}, 668--683, doi:10.1086/154215.

\bibitem[Thompson~et al.(2015)]{Thompson15}Thompson, T.A.; Fabian, A.C.; Quataert, E.; Murray, N. Dynamics of dusty radiation-pressure-driven shells and clouds: Fast outflows from galaxies, star clusters, massive stars, and AGN. {\em Mon.~Not.~R.~Astron.~Soc.} {\bf 2015}, {\em 449}, 147--161, doi:10.1093/mnras/stv246.

\bibitem[Lochhaas~et al.(2018)]{Lochhaas18}Lochhaas, C.; Thompson, T.A.; Quataert, E.; Weinberg, D.H. Fast Winds Drive Slow Shells: A Model for the Circumgalactic Medium as Galactic Wind-Driven Bubbles. {\em Mon.~Not.~R.~Astron.~Soc.} {\bf 2018}, {\em 481}, 1873--1896, doi:10.1093/mnras/sty2421.

\bibitem[Zhang \& Thompson(2012)]{Zhang12}Zhang, D.; Thompson, T.A. Radiation pressure-driven galactic winds from self-gravitating discs. {\em Mon.~Not.~R.~Astron.~Soc.} {\bf 2012}, {\em 424}, 1170--1178, doi:10.1111/j.1365-2966.2012.21291.x.

\bibitem[Kennicutt(1998a)]{Kennicutt98a}Kennicutt, R.C., Jr. The Global Schmidt Law in Star-forming Galaxies. {\em Astrophys.~J.} {\bf 1998}, {\em 498}, 541--552, doi:10.1086/305588.

\bibitem[Kennicutt(1998b)]{Kennicutt98b}Kennicutt, R.C., Jr. Star Formation in Galaxies Along the Hubble Sequence. {\em Annu. Rev. Astron. Astrophys.} {\bf 1998}, {\em 36}, 189--232, doi:10.1146/annurev.astro.36.1.189.


\bibitem[Roth~et al.(2012)]{Roth12}Roth, N.; Kasen, D.; Hopkins, P.F.; Quataert, E. Three-dimensional Radiative Transfer Calculations of Radiation Feedback from Massive Black Holes: Outflow of Mass from the Dusty “Torus”. {\em Astrophys.~J.} {\bf 2012}, {\em 759}, 36, doi:10.1088/0004-637X/759/1/36.

\bibitem[Jacquet \& Krumholz(2011)]{Jacquet11}Jacquet, E.; Krumholz, M.R. Radiative Rayleigh-Taylor Instabilities. {\em Astrophys.~J.} {\bf 2011}, {\em 730}, 116, doi:10.1088/0004-637X/730/2/116.

\bibitem[Jiang~et al.(2013)]{Jiang13}Jiang, Y.-F.; Davis, S.W.; Stone, J.M. Nonlinear Evolution of Rayleigh-Taylor Instability in a Radiation-supported Atmosphere. {\em Astrophys.~J.} {\bf 2013}, {\em 763}, 102, doi:10.1088/0004-637X/763/2/102.



\bibitem[Krumholz~et al.(2009)]{Krumholz09a}Krumholz, M.R.; Klein, R.I.; McKee, C.F.; Offner, S.S.R.; Cunningham, A.J. The Formation of Massive Star Systems by Accretion. {\em Science} {\bf 2009}, {\em 323}, 754--757, doi:10.1126/science.1165857.

\bibitem[Lowrie(1999)]{Lowrie99}Lowrie, R.B.; Morel, J.E.; Hittinger, J.A. The Coupling of Radiation and Hydrodynamics. {\em Astrophys.~J.} {\bf 1999}, {\em 521}, 432--450, doi:10.1086/307515.

\bibitem[Jiang~et al.(2012)]{Jiang12}Jiang, Y.-F.; Stone, J.M.; Davis, S.W. A Godunov Method for Multidimensional Radiation Magnetohydrodynamics Based on a Variable Eddington Tensor. {\em Astrophys.~J.} {\bf 2012}, {\em 199}, 14, doi:10.1088/0067-0049/199/1/14.

\bibitem[Mihalas \& Mihalas(1984)]{Mihalas84}Mihalas, D.; Mihalas, B.W. \emph{Foundations of Radiation Hydrodynamics}; Oxford University Press: Oxford, NY, USA, {1984}.

\bibitem[Levermore \& Pomraning(1981)]{Levermore81}Levermore, C.D.; Pomraning, G.C. A flux-limited diffusion theory. {\em Astrophys.~J.} {\bf 1981}, {\em 248}, 321--334, doi:10.1086/159157.

\bibitem[Levermore(1984)]{Levermore84}Levermore, C.D. Relating Eddington factors to flux limiters. {\em J. Quant. Spectrosc. Radiat. Transf.} {\bf 1984}, {\em 31}, 149--160, doi:10.1016/0022-4073(84)90112-2.

\bibitem[Olson \& Kunasz(1987)]{Olson87}Olson, G.L.; Kunasz, P.B. Short characteristic solution of the non-LTE transfer problem by operator perturbation. I. The one-dimensional planar slab. {\em J. Quant. Spectrosc. Radiat. Transf.} {\bf 1987}, {\em 38}, 325--336, doi:10.1016/0022-4073(87)90027-6.

\bibitem[Kunasz \& Auer(1988)]{Kunasz88}Kunasz, P.; Auer, L.H. Short characteristic integration of radiative transfer problems--Formal solution in two-dimensional slabs. {\em J. Quant. Spectrosc. Radiat. Transf.} {\bf 1988}, {\em 39}, 67--79, doi:10.1016/0022-4073(88)90021-0.

\bibitem[Stone~et al.(1992)]{Stone92b}Stone, J.M.; Mihalas, D.; Norman, M.L. ZEUS-2D: A radiation magnetohydrodynamics code for astrophysical flows in two space dimensions. III--The radiation hydrodynamic algorithms and tests. {\em Astrophys.~J.~Suppl.~Ser.} {\bf 1992}, {\em 80}, 819--845, doi:10.1086/191682.

\bibitem[Sekora \& Stone(2010)]{Sekora10}Sekora, M.D.; Stone, J.M. A hybrid Godunov method for radiation hydrodynamics. {\em J.~Comput.~Phys.} {\bf 2010}, {\em 229}, 6819--6852, doi:10.1016/j.jcp.2010.05.024.

\bibitem[Davis~et al.(2012)]{Davis12}Davis, S.W.; Stone, J.M.; Jiang, Y.-F. A Radiation Transfer Solver for Athena Using Short Characteristics. {\em Astrophys.~J.~Suppl.~Ser.} {\bf 2012}, {\em 199}, 9, doi:10.1088/0067-0049/199/1/9.

\bibitem[Jiang~et al.(2014)]{Jiang14}Jiang, Y.-F.; Stone, J.M.; Davis, S.W. An Algorithm for Radiation Magnetohydrodynamics Based on Solving the Time-dependent Transfer Equation. {\em Astrophys.~J.~Suppl.~Ser.} {\bf 2014}, {\em 213}, 7, doi:10.1088/0067-0049/213/1/7.

\bibitem[Rosen~et al.(2017)]{Rosen17}Rosen, A.L.; Krumholz, M.R.; Oishi, J.S.; Lee, A.T.; Klein, R.I. Hybrid Adaptive Ray-Moment Method (HARM$^2$): A highly parallel method for radiation hydrodynamics on adaptive grids. {\em J.~Comput.~Phys.} {\bf 2017}, {\em 330}, 924--942, doi:10.1016/j.jcp.2016.10.048.

\bibitem[Krumholz \& Thompson(2012)]{KT12}Krumholz, M.R.; Thompson, T.A. Direct Numerical Simulation of Radiation Pressure-driven Turbulence and Winds in Star Clusters and Galactic Disks. {\em Astrophys.~J.} {\bf 2012}, {\em 760}, 155, doi:10.1088/0004-637X/760/2/155.

\bibitem[Krumholz \& Thompson(2013)]{KT13}Krumholz, M.R.; Thompson, T.A. Numerical simulations of radiatively driven dusty winds. {\em Mon.~Not.~R.~Astron.~Soc.} {\bf 2013}, {\em 434}, 2329--2346, doi:10.1093/mnras/stt1174.

\bibitem[Davis~et al.(2014)]{Davis14}Davis, S.W.; Jiang, Y.-F.; Stone, J.M.; Murray, N. Radiation Feedback in ULIRGs: Are Photons Movers and Shakers? {\em Astrophys.~J.} {\bf 2014}, {\em 796}, 107, doi:10.1088/0004-637X/796/2/107.

\bibitem[Rosdahl \& Teyssier(2015)]{Rosdahl15}Rosdahl, J.; Teyssier, R. A scheme for radiation pressure and photon diffusion with the M1 closure in RAMSES-RT. {\em Mon.~Not.~R.~Astron.~Soc.} {\bf 2015}, {\em 449}, 4380--4403, doi:10.1093/mnras/stv567.

\bibitem[Tsang \& Milosavljevi\'{c}(2015)]{Tsang15}Tsang, B.T.-H.; Milosavljevi\'{c}, M. Radiation pressure driving of a dusty atmosphere. {\em Mon.~Not.~R.~Astron.~Soc.} {\bf 2015}, {\em 453}, 1108--1120, doi:10.1093/mnras/stv1707.

\bibitem[Kannan~et al.(2018)]{Kannan18}Kannan, R.; Vogelsberger, M.; Marinacci, F.; McKinnon, R.; Pakmor, R.; Springel, V. AREPO-RT: Radiation hydrodynamics on a moving mesh. {\em arxiv} {\bf 2018}, arXiv:1804.01987.

\bibitem[Krumholz(2018)]{Krumholz18b}Krumholz, M.R. Resolution Requirements and Resolution Problems in Simulations of Radiative Feedback in Dusty Gas. {\em Mon.~Not.~R.~Astron.~Soc.} {\bf 2018}, {\em 480}, 3468--3482, doi: 10.1093/mnras/sty2105.


\bibitem[Leroy~et al.(2018)]{Leroy18}Leroy, A.K.; Bolatto, A.D.; Ostriker, E.C. Forming Super Star Clusters Power the Central Starburst in NGC 253. {\em arxiv} {\bf 2018}, arXiv:1804.02083.



\bibitem[Fermi(1949)]{Fermi49}Fermi, E. On the Origin of the Cosmic Radiation. {\em Phys.~Rev.~Lett.} {\bf 1949}, {\em 75}, 1169--1174, doi:10.1103/PhysRev.75.1169.

\bibitem[Axford~et al.(1977)]{Axford77}Axford, W.I.; Leer, E.; Skadron, G. The acceleration of cosmic rays by shock waves. {\em Int.~Cosmic~Ray~Conf.} {\bf 1977}, {\em 11}, 132--137.

\bibitem[Krymskii(1977)]{Krymskii77}Krymskii, G.F. A regular mechanism for the acceleration of charged particles on the front of a shock wave. {\em Dokl.~Akad.~Nauk~SSSR} {\bf 1977}, {\em 243}, 1306--1308.

\bibitem[Bell(1978a)]{Bell78a}Bell, A.R. The acceleration of cosmic rays in shock fronts. I. {\em Mon.~Not.~R.~Astron.~Soc.} {\bf 1978}, {\em 182}, 147--156, doi:10.1093/mnras/182.2.147.

\bibitem[Bell(1978b)]{Bell78b}Bell, A.R. The acceleration of cosmic rays in shock fronts. II. {\em Mon.~Not.~R.~Astron.~Soc.} {\bf 1978}, {\em 182}, 443--455, doi:10.1093/mnras/182.3.443.

\bibitem[Blandford \& Ostriker(1978)]{Blandford78}Blandford, R.D.; Ostriker, J.P. Particle acceleration by astrophysical shocks. {\em Astrophys.~J.~Lett.} {\bf 1978}, {\em 221}, 29--32, doi:10.1086/182658.

\bibitem[Schlickeiser(1989a)]{Schlickeiser89a}Schlickeiser, R. Cosmic-ray transport and acceleration. I--Derivation of the kinetic equation and application to cosmic rays in static cold media. {\em Astrophys.~J.} {\bf 1989}, {\em 336}, 243--263, doi:10.1086/167009.

\bibitem[Schlickeiser(1989b)]{Schlickeiser89b}Schlickeiser, R. Cosmic-Ray Transport and Acceleration. II. Cosmic Rays in Moving Cold Media with Application to Diffusive Shock Wave Acceleration. {\em Astrophys.~J.} {\bf 1989}, {\em 336}, 264--293, doi:10.1086/182658.

\bibitem[Helder~et al.(2012)]{Helder12}Helder, E.A.; Vink, J.; Bykov, A.M.; Ohira, Y.; Raymond, J.C.; Terrier, R. Observational Signatures of Particle Acceleration in Supernova Remnants. {\em Space~Sci.~Rev.} {\bf 2012}, {\em 173}, 369--431, doi:10.1007/s11214-012-9919-8.

\bibitem[Morlino \& Caprioli(2012)]{Morlino12}Morlino, G.; Caprioli, D. Strong evidence for hadron acceleration in Tycho's supernova remnant. {\em Astron.~Astrophys.} {\bf 2012}, {\em 538}, 81, doi:10.1051/0004-6361/201117855.

\bibitem[Ackermann~et al.(2013)]{Ackermann13}Ackermann, M. Detection of the Characteristic Pion-Decay Signature in Supernova Remnants. {\em Science} {\bf 2013}, {\em 339}, 807--811, doi:10.1126/science.1231160.

\bibitem[Strong~et al.(2007)]{Strong07}Strong, A.W.; Moskalenko, I.V.; Ptuskin, V.S. Cosmic-Ray Propagation and Interactions in the Galaxy. {\em Annu.~Rev. Nucl. Part. Sci.} {\bf 2007}, {\em 57}, 285--327, doi:10.1146/annurev.nucl.57.090506.123011.

\bibitem[Grenier~et al.(2015)]{Grenier15}Grenier, I.A.; Black, J.H.; Strong, A.W. The Nine Lives of Cosmic Rays in Galaxies. {\em Annu. Rev. Astron.~Astrophys.} {\bf 2015}, {\em 53}, 199--246, doi:10.1146/annurev-astro-082214-122457.

\bibitem[Zweibel(2013)]{Zweibel13}Zweibel, E.G. The microphysics and macrophysics of cosmic rays. {\em Phys.~Plasmas} {\bf 2013}, {\em 20}, 055501, doi:10.1063/1.4807033.

\bibitem[Zweibel(2017)]{Zweibel17}Zweibel, E.G. The basis for cosmic ray feedback: Written on the wind. {\em Phys.~Plasmas} {\bf 2017}, {\em 24}, 055402, doi:10.1063/1.4984017.

\bibitem[Webber(1998)]{Webber98}Webber, W.R. A New Estimate of the Local Interstellar Energy Density and Ionization Rate of Galactic Cosmic Cosmic Rays. {\em Astrophys.~J.} {\bf 1998}, {\em 506}, 329--334, doi:10.1086/306222.

\bibitem[Garcia-Munoz~et al.(1977)]{Garcia77}Garcia-Munoz, M.; Mason, G.M.; Simpson, J.A. The age of the galactic cosmic rays derived from the abundance of Be-10. {\em Astrophys.~J.} {\bf 1977}, {\em 217}, 859--877, doi:10.1086/155632.

\bibitem[Yanasak~et al.(2001)]{Yanasak01}Yanasak, N.E. Measurement of the Secondary Radionuclides $^{10}$Be,$^{26}$Al, $^{36}$Cl, $^{54}$Mn, and $^{14}$C and Implications for the Galactic Cosmic-Ray Age. {\em Astrophys.~J.} {\bf 2001}, {\em 563}, 768--792, doi:10.1086/323842.



\bibitem[Ipavich(1975)]{Ipavich75}Ipavich, F.M. Galactic Winds Driven by Cosmic Rays. {\em Astrophys.~J.} {\bf 1975}, {\em 196}, 107--120, doi:10.1086/153397.

\bibitem[Breitschwerdt~et al.(1991)]{Breitschwerdt91}Breitschwerdt, D.; McKenzie, J.F.; Voelk, H.J. Galactic winds. I--Cosmic ray and wave-driven winds from the Galaxy. {\em Astron.~Astrophys.} {\bf 1991}, {\em 245}, 79--98.

\bibitem[Breitschwerdt~et al.(1993)]{Breitschwerdt93}Breitschwerdt, D.; McKenzie, J.F.; Voelk, H.J. Galactic winds. II--Role of the disk-halo interface in cosmic ray driven galactic winds. {\em Astron.~Astrophys.} {\bf 1993}, {\em 269}, 54--66.

\bibitem[Zirakashvili~et al.(1996)]{Zirakashvili96}Zirakashvili, V.N.; Breitschwerdt, D.; Ptuskin, V.S.; Voelk, H.J. Magnetohydrodynamic wind driven by cosmic rays in a rotating galaxy. {\em Astron.~Astrophys.} {\bf 1996}, {\em 311}, 113--126.

\bibitem[Ptuskin~et al.(1997)]{Ptuskin97}Ptuskin, V.S.; Voelk, H.J.; Zirakashvili, V.N.; Breitschwerdt, D. Transport of relativistic nucleons in a galactic wind driven by cosmic rays. {\em Astron.~Astrophys.} {\bf 1997}, {\em 321}, 434--443.

\bibitem[Breitschwerdt~et al.(2002)]{Breitschwerdt02}Breitschwerdt, D.; Dogiel, V.A.; V\"olk, H.J. The gradient of diffuse gamma -ray emission in the Galaxy. {\em Astron.~Astrophys.} {\bf 2002}, {\em 385}, 216--238, doi:10.1051/0004-6361:20020152.

\bibitem[Kulsrud \& Pearce(1969)]{Kulsrud69}Kulsrud, R.; Pearce, W.P. The Effect of Wave-Particle Interactions on the Propagation of Cosmic Rays. {\em Astrophys.~J.} {\bf 1969}, {\em 156}, 445--469, doi:10.1086/149981.

\bibitem[Wentzel(1974)]{Wentzel74}Wentzel, D.G. Cosmic-ray propagation in the Galaxy--Collective effects. {\em Annu. Rev. Astron. Astrophys.} {\bf 1974}, {\em 12}, 71--96, doi:10.1146/annurev.aa.12.090174.000443.

\bibitem[Everett~et al.(2008)]{Everett08}Everett, J.E.; Zweibel, E.G.; Benjamin, R.A.; McCammon, D.; Rocks, L.; Gallagher, J.S., III. The Milky Way's Kiloparsec-Scale Wind: A Hybrid Cosmic-Ray and Thermally Driven Outflow. {\em Astrophys.~J.} {\bf 2008}, {\em 674}, 258--270, doi:10.1086/524766.

\bibitem[Everett~et al.(2010)]{Everett10a}Everett, J.E.; Schiller, Q.G.; Zweibel, E.G. Synchrotron Constraints on a Hybrid Cosmic-ray and Thermally Driven Galactic Wind. {\em Astrophys.~J.} {\bf 2010}, {\em 711}, 13--24, doi:10.1088/0004-637X/711/1/13.



\bibitem[Thompson~et al.(2007)]{Thompson07}Thompson, T.A.; Quataert, E.; Waxman, E. The Starburst Contribution to the Extragalactic $\gamma$-Ray Background. {\em Astrophys.~J.} {\bf 2007}, {\em 654}, 219--225, doi:10.1086/509068.

\bibitem[Socrates~et al.(2008)]{Socrates08}Socrates, A.; Davis, S.W.; Ramirez-Ruiz, E. The Eddington Limit in Cosmic Rays: An Explanation for the Observed Faintness of Starbursting Galaxies. {\em Astrophys.~J.} {\bf 2008}, {\em 687}, 202--215, doi:10.1086/590046.

\bibitem[Kulsrud(2005)]{Kulsrud05}Kulsrud, R.M. \emph{Plasma Physics for Astrophysics}; Princeton University Press: Princeton, NJ, USA, {2005}.

\bibitem[Schlickeiser(2002)]{Schlickeiser02}Schlickeiser, R. \emph{Cosmic Ray Astrophysics}; Springer: Berlin, Germany, {2002}; ISBN 3-540-66465-3.

\bibitem[Diesing \& Caprioli(2018)]{Diesing18}Diesing, R.; Caprioli, D. Effect of Cosmic Rays on the Evolution and Momentum Deposition of Supernova Remnants. {\em Phys.~Rev.~Lett.} {\bf 2018}, {\em 121}, 091101, doi:10.1103/PhysRevLett.121.091101.




\bibitem[Uhlig~et al.(2012)]{Uhlig12}Uhlig, M.; Pfrommer, C.; Sharma, M.; Nath, B.B.; En$\beta$lin, T.A.; Springel, V. Galactic winds driven by cosmic ray streaming. {\em Mon.~Not.~R.~Astron.~Soc.} {\bf 2012}, {\em 423}, 2374--2396, doi:10.1111/j.1365-2966.2012.21045.x.

\bibitem[Booth~et al.(2013)]{Booth13}Booth, C.M.; Agertz, O.; Kravtsov, A.V.; Gnedin, N.Y. Simulations of Disk Galaxies with Cosmic Ray Driven Galactic Winds. {\em Astrophys.~J.~Lett.} {\bf 2013}, {\em 777}, L16, doi:10.1088/2041-8205/777/1/L16.

\bibitem[Hanasz~et al.(2013)]{Hanasz13}Hanasz, M.; Lesch, H.; Naab, T.; Gawryszczak, A.; Kowalik, K.; W\'olta\'nski, D. Cosmic Rays Can Drive Strong Outflows from Gas-rich High-redshift Disk Galaxies. {\em Astrophys.~J.~Lett.} {\bf 2013}, {\em 777}, 38, doi:10.1088/2041-8205/777/2/L38.

\bibitem[Salem \& Bryan(2014)]{Salem14}Salem, M.; Bryan, G.L. Cosmic ray driven outflows in global galaxy disc models. {\em Mon.~Not.~R.~Astron.~Soc.} {\bf 2014}, {\em 437}, 3312--3330, doi:10.1093/mnras/stt2121.

\bibitem[Girichidis~et al.(2016a)]{Girichidis16a}Girichidis, P. Launching Cosmic-Ray-driven Outflows from the Magnetized Interstellar Medium. {\em Astrophys.~J.~Lett.} {\bf 2016}, {\em 816}, L19, doi:10.3847/2041-8205/816/2/L19.

\bibitem[Pakmor~et al.(2016a)]{Pakmor16a}Pakmor, R.; Pfrommer, C.; Simpson, C.M.; Springel, V. Galactic Winds Driven by Isotropic and Anisotropic Cosmic-Ray Diffusion in Disk Galaxies. {\em Astrophys.~J.~Lett.} {\bf 2016}, {\em 824}, L30, doi:10.3847/2041-8205/824/2/L30.

\bibitem[Pakmor~et al.(2016b)]{Pakmor16b}Pakmor, R.; Pfrommer, C.; Simpson, C.M.; Kannan, R.; Springel, V. Semi-implicit anisotropic cosmic ray transport on an unstructured moving mesh. {\em Mon.~Not.~R.~Astron.~Soc.} {\bf 2016}, {\em 462}, 2603--2616, doi:10.1093/mnras/stw1761.

\bibitem[Simpson~et al,(2016)]{Simpson16}Simpson, C.M. The Role of Cosmic-Ray Pressure in Accelerating Galactic Outflows. {\em Astrophys.~J.~Lett.} {\bf 2016}, {\em 827}, L29, doi:10.3847/2041-8205/827/2/L29.

\bibitem[Pfrommer~et al.(2017)]{Pfrommer17}Pfrommer, C.; Pakmor, R.; Schaal, K.; Simpson, C.M.; Springel, V. Simulating cosmic ray physics on a moving mesh. {\em Mon.~Not.~R.~Astron.~Soc.} {\bf 2017}, {\em 465}, 4500--4529, doi:10.1093/mnras/stw2941.

\bibitem[Ruszkowski~et al.(2017a)]{Ruszkowski17a}Ruszkowski, M.; Yang, H.-Y.K.; Zweibel, E. Global Simulations of Galactic Winds Including Cosmic-ray Streaming. {\em Astrophys.~J.} {\bf 2017}, {\em 834}, 208, doi:10.3847/1538-4357/834/2/208.

\bibitem[Wiener~et al.(2017b)]{Wiener17b}Wiener, J.; Pfrommer, C.; Oh, S.P. Cosmic ray-driven galactic winds: Streaming or diffusion? {\em Mon.~Not.~R.~Astron.~Soc.} {\bf 2017}, {\em 467}, 906--921, doi:10.1093/mnras/stx127.

\bibitem[Jacob~et al.(2018)]{Jacob18}Jacob, S.; Pakmor, R.; Simpson, C.M.; Springel, V.; Pfrommer, C. The dependence of cosmic ray-driven galactic winds on halo mass. {\em Mon.~Not.~R.~Astron.~Soc.} {\bf 2018}, {\em 475}, 570--584, doi:10.1093/mnras/stx3221.

\bibitem[Fujita \& Mac Low(2018)]{Fujita18}Fujita, A.; Mac Low, M.-M. Cosmic ray driven outflows in an ultraluminous galaxy. {\em Mon.~Not.~R.~Astron.~Soc.} {\bf 2018}, {\em 477}, 531--538, doi:10.1093/mnras/sty715.

\bibitem[Girichidis~et al.(2018)]{Girichidis18}Girichidis, P.; Naab, T.; Hanasz, M.; Walch, S. Cooler and smoother--the impact of cosmic rays on the phase structure of galactic outflows. {\em Mon.~Not.~R.~Astron.~Soc.} {\bf 2018}, {\em 479}, 3042--3067, doi:10.1093/mnras/sty1653.

\bibitem[Jiang \& Oh(2018)]{Jiang18}Jiang, Y.-F.; Oh, S.P. A New Numerical Scheme for Cosmic-Ray Transport. {\em Astrophys.~J.} {\bf 2018}, {\em 854}, 5, doi:10.3847/1538-4357/aaa6ce.

\bibitem[Farber~et al.(2018)]{Farber18}Farber, R.; Ruszkowski, M.; Yang, H.-Y.K.; Zweibel, E.G. Impact of Cosmic-Ray Transport on Galactic Winds. {\em Astrophys.~J.} {\bf 2018}, {\em 856}, 112, doi:10.3847/1538-4357/aab26d.

\bibitem[Thomas \& Pfrommer(2018)]{Thomas18}Thomas, T.; Pfrommer, C. Cosmic-ray hydrodynamics: Alfv\'en-wave regulated transport of cosmic rays. {\em arxiv} {\bf 2018}, arXiv:1805.11092.

\bibitem[Holguin~et al.(2018)]{Holguin18}Holguin, F.; Ruszkowski, M.; Lazarian, A.; Farber, R.; Yang, H.-Y.K. Role of Cosmic Ray Streaming and Turbulent Damping in Driving Galactic Winds. {\em arxiv} {\bf 2018}, arXiv:1807.05494.

\bibitem[Guo \& Oh(2008)]{Guo08}Guo, F.; Oh, S.P. Feedback heating by cosmic rays in clusters of galaxies. {\em Mon.~Not.~R.~Astron.~Soc.} {\bf 2008}, {\em 384}, 251--266, doi:10.1111/j.1365-2966.2007.12692.x.


\bibitem[De Pontieu~et al.(2001)]{DePontieu01}De Pontieu, B.; Martens, P.C.H.; Hudson, H.S. Chromospheric Damping of Alfv\'en Waves. {\em Astrophys.~J.} {\bf 2001}, {\em 558}, 859--871, doi:10.1086/322408.

\bibitem[Khodachenko~et al.(2004)]{Khodachenko04}Khodachenko, M.L.; Arber, T.D.; Rucker, H.O.; Hanslmeier, A. Collisional and viscous damping of MHD waves in partially ionized plasmas of the solar atmosphere. {\em Astron.~Astrophys.} {\bf 2004}, {\em 422}, 1073--1084, doi:10.1051/0004-6361:20034207.

\bibitem[Khodachenko~et al.(2006)]{Khodachenko06}Khodachenko, M.L.; Rucker, H.O.; Kislyakov, A.G.; Zaitsev, V.V.; Urpo, S. Microwave Diagnostics of Dynamic Processes and Oscillations in Groups of Solar Coronal Magnetic Loops. {\em Adv.~Space~Res.} {\bf 2006}, {\em 122}, 137--148, doi:10.1007/s11214-006-7767-0.

\bibitem[Soler~et al.(2016)]{Soler16}Soler, R.; Terradas, J.; Oliver, R.; Ballester, J.L. The role of Alfv\'en wave heating in solar prominences. {\em Astron.~Astrophys.} {\bf 2016}, {\em 592}, 28, doi:10.1051/0004-6361/201628722.

\bibitem[Malmberg \& Wharton(1967)]{Malmberg67}Malmberg, J.H.; Wharton, C.B. Collisionless Damping of Large-Amplitude Plasma Waves. {\em Phys.~Rev.~Lett.} {\bf 1967}, {\em 19}, 775--777, doi:10.1103/PhysRevLett.19.775.

\bibitem[Hollweg(1971)]{Hollweg71}Hollweg, J.V. Nonlinear Landau Damping of Alfv\'{e}n Waves. {\em Phys.~Rev.~Lett.} {\bf 1971}, {\em 27}, 1349--1352, doi:10.1103/PhysRevLett.27.1349.

\bibitem[Lee \& V\"olk(1973)]{Lee73}Lee, M.A.; V\"olk, H.J. Damping and Non-Linear Wave-Particle Interactions of Alfvén-Waves in the Solar Wind. {\em Astrophys.~Space~Sci.} {\bf 1973}, {\em 24}, 31--49, doi:10.1007/BF00648673.

\bibitem[McKenzie \& Bond(1983)]{McKenzie83}McKenzie, J.F.; Bond, R.A.B. The role of non-linear Landau damping in cosmic ray shock acceleration. {\em Astron.~Astrophys.} {\bf 1983}, {\em 123}, 111--117.

\bibitem[Miller(1991)]{Miller91}Miller, J.A. Magnetohydrodynamic turbulence dissipation and stochastic proton acceleration in solar flares. {\em Astrophys.~J.} {\bf 1991}, {\em 376}, 342--354, doi:10.1086/170284.

\bibitem[Mouhot \& Villani(2009)]{Mouhot09}Mouhot, C.; Villani, C. On Landau Damping. {\em arxiv} {\bf 2009}, arXiv:0904.2760.

\bibitem[Farmer \& Goldreich(2004)]{Farmer04}Farmer, A.J.; Goldreich, P. Wave Damping by Magnetohydrodynamic Turbulence and Its Effect on Cosmic-Ray Propagation in the Interstellar Medium. {\em Astrophys.~J.} {\bf 2004}, {\em 604}, 671--674, doi:10.1086/382040.

\bibitem[Yan \& Lazarian(2004)]{Yan04}Yan, H.; Lazarian, A. Cosmic-Ray Scattering and Streaming in Compressible Magnetohydrodynamic Turbulence. {\em Astrophys.~J.} {\bf 2004}, {\em 614}, 757--769, doi:10.1086/423733.



\bibitem[Lazarian(2016)]{Lazarian16}Lazarian, A. Damping of Alfv\'en Waves by Turbulence and Its Consequences: From Cosmic-ray Streaming to Launching Winds. {\em Astrophys.~J.} {\bf 2016}, {\em 833}, 131, doi:10.3847/1538-4357/833/2/131.

\bibitem[Everett \& Zweibel(2011)]{Everett11}Everett, J.E.; Zweibel, E.G. The Interaction of Cosmic Rays with Diffuse Clouds. {\em Astrophys.~J.} {\bf 2011}, {\em 739}, 60, doi:10.1088/0004-637X/739/2/60.


\bibitem[Yang~et al.(2012)]{Yang12}Yang, H.-Y.K.; Ruszkowski, M.; Ricker, P.M.; Zweibel, E.; Lee, D. The Fermi Bubbles: Supersonic Active Galactic Nucleus Jets with Anisotropic Cosmic-Ray Diffusion. {\em Astrophys.~J.} {\bf 2012}, {\em 761}, 185, doi:10.1088/0004-637X/761/2/185.

\bibitem[Swordy(1990)]{Swordy90}Swordy, S.P.; Mueller, D.; Meyer, P.; L'Heureux, J.; Grunsfeld, J.M. Relative Abundances of Secondary and Primary Cosmic Rays at High Energies. {\em Astrophys.~J.} {\bf 1990}, {\em 349}, 625--633, doi:10.1086/168349.

\bibitem[Shalchi \& Schlickeiser(2005)]{Shalchi05}Shalchi, A.; Schlickeiser, R. Evidence for the Nonlinear Transport of Galactic Cosmic Rays. {\em Astrophys.~J.~Lett.} {\bf 2005}, {\em 626}, 97--99, doi:10.1086/431905.

\bibitem[Heesen~et al.(2009a)]{Heesen09a}Heesen, V.; Beck, R.; Krause, M.; Dettmar, R.-J. Cosmic rays and the magnetic field in the nearby starburst galaxy NGC 253. I. The distribution and transport of cosmic rays. {\em Astron.~Astrophys.} {\bf 2009}, {\em 494}, 563--577, doi:10.1051/0004-6361:200810543.

\bibitem[Heesen~et al.(2009b)]{Heesen09b}Heesen, V.; Krause, M.; Beck, R.; Dettmar, R.-J. Cosmic rays and the magnetic field in the nearby starburst galaxy NGC 253. II. The magnetic field structure. {\em Astron.~Astrophys.} {\bf 2009}, {\em 506}, 1123--1135, doi:10.1051/0004-6361/200911698.

\bibitem[Heesen~et al.(2011)]{Heesen11}Heesen, V.; Beck, R.; Krause, M.; Dettmar, R.-J. Cosmic rays and the magnetic field in the nearby starburst galaxy NGC 253 III. Helical magnetic fields in the nuclear outflow. {\em Astron.~Astrophys.} {\bf 2011}, {\em 535}, 79, doi:10.1051/0004-6361/201117618.

\bibitem[Shalchi~et al.(2010)]{Shalchi10}Shalchi, A.; B\"{u}sching, I.; Lazarian, A.; Schlickeiser, R. Perpendicular Diffusion of Cosmic Rays for a Goldreich-Sridhar Spectrum. {\em Astrophys.~J.} {\bf 2010}, {\em 725}, 2117--2127, doi:10.1088/0004-637X/725/2/2117.

\bibitem[Buffie~et al.(2013)]{Buffie13}Buffie, K.; Heesen, V.; Shalchi, A. Theoretical Explanation of the Cosmic-Ray Perpendicular Diffusion Coefficient in the Nearby Starburst Galaxy NGC 253. {\em Astrophys.~J.} {\bf 2013}, {\em 764}, 37, doi:10.1088/0004-637X/764/1/37.

\bibitem[Heesen~et al.(2016)]{Heesen16}Heesen, V.; Dettmar, R.-J.; Krause, M.; Beck, R.; Stein, Y. Advective and diffusive cosmic ray transport in galactic haloes. {\em Mon.~Not.~R.~Astron.~Soc.} {\bf 2016}, {\em 458}, 332--353, doi:10.1093/mnras/stw360.


\bibitem[Wiener~et al.(2013)]{Wiener13}Wiener, J.; Oh, S.P.; Guo, F. Cosmic ray streaming in clusters of galaxies. {\em Mon.~Not.~R.~Astron.~Soc.} {\bf 2013}, {\em 434}, 2209--2228, doi:10.1093/mnras/stt1163.

\bibitem[Wiener~et al.(2017a)]{Wiener17a}Wiener, J.; Oh, S.P.; Zweibel, E.G. Interaction of cosmic rays with cold clouds in galactic haloes. {\em Mon.~Not.~R.~Astron.~Soc.} {\bf 2017}, {\em 467}, 646--660, doi:10.1093/mnras/stx109.



\bibitem[Salem~et al.(2016)]{Salem16}Salem, M.; Bryan, G.L.; Corlies, L. Role of cosmic rays in the circumgalactic medium. {\em Mon.~Not.~R.~Astron.~Soc.} {\bf 2016}, {\em 456}, 582--601, doi:10.1093/mnras/stv2641.

\bibitem[Ruszkowski~et al.(2017b)]{Ruszkowski17b}Ruszkowski, M.; Yang, H.-Y.K.; Reynolds, C.S. Cosmic-Ray Feedback Heating of the Intracluster Medium. {\em Astrophys.~J.} {\bf 2017}, {\em 844}, 13, doi:10.3847/1538-4357/aa79f8.

\bibitem[Butsky \& Quinn(2018)]{Butsky18}Butsky, I.; Quinn, T.R. The Role of Cosmic Ray Transport in Shaping the Simulated Circumgalactic Medium. {\em arxiv} {\bf 2018}, arXiv:1803.06345.




\bibitem[Skilling(1971)]{Skilling71}Skilling, J. Cosmic Rays in the Galaxy: Convection or Diffusion? {\em Astrophys.~J.} {\bf 1971}, {\em 170}, 256--273, doi:10.1086/151210.

\bibitem[Skiner \& Ostriker(2013)]{Skinner13}Skinner, M.A.; Ostriker, E.C. A Two-moment Radiation Hydrodynamics Module in Athena Using a Time-explicit Godunov Method. {\em Astrophys.~J.~Suppl.~Ser.} {\bf 2013}, {\em 206}, 21, doi:10.1088/0067-0049/206/2/21.




\bibitem[Mao \& Ostriker(2018)]{Mao18}Mao, S.A.; Ostriker, E.C. Galactic Disk Winds Driven by Cosmic Ray Pressure. {\em Astrophys.~J.} {\bf 2018}, {\em 854}, 89, doi:10.3847/1538-4357/aaa88e.










\end{thebibliography}


\end{document}